\documentclass[breaklinks=true]{aa}

\usepackage{silence}
\usepackage{graphicx}
\usepackage{lscape}

\usepackage{txfonts}
\usepackage{stackengine}

\usepackage{subcaption}
\usepackage{geometry}

\usepackage{float}
\usepackage{tabularx}
\usepackage{longtable}
\usepackage{booktabs}
\usepackage{gensymb}
\usepackage{textgreek}
\usepackage{textcomp}
\usepackage{multirow}
\usepackage{orcidlink}

\usepackage{graphicx}
\usepackage{rotating}
\usepackage{multirow}
\usepackage{array}
\usepackage{placeins}

\usepackage{natbib,twoopt}
\bibpunct{(}{)}{;}{a}{}{,}
\hypersetup{
    colorlinks=true,
    linkcolor=red,
    citecolor=blue,
    urlcolor=blue 
}

\makeatletter
\let\linenumbers\relax
\let\endlinenumbers\relax
\let\internallinenumbers\relax
\let\resetlinenumber\relax
\makeatother

\begin{document}

   \title{Spectral modelling of IXPE HSP blazars: exploring one-zone and two-zone leptonic scenarios} 
   
   \titlerunning{Spectral modelling of IXPE HSP blazars}
    \authorrunning{S. Capecchiacci et al.}

\author{
  Sara Capecchiacci \inst{\ref*{physics_crete}, \ref*{forth}}\orcidlink{0009-0007-1918-577X} \and
  Ioannis Liodakis \inst{\ref*{forth}}\orcidlink{0000-0001-9200-4006} \and
  Andrea Tramacere\inst{\ref*{geneva_obs}} \orcidlink{0000-0002-8186-3793} \and
  Axel Arbet-Engels\inst{\ref*{garching_mpp}, \ref*{geneva_obs}} \orcidlink{0000-0001-9076-9582} \and
  Steven R. Ehlert \inst{\ref*{nasa_marshall}}\orcidlink{0000-0003-4420-2838} \and
  Philip Kaaret \inst{\ref*{nasa_marshall}}\orcidlink{0000-0002-3638-0637} \and
  Riccardo Middei \inst{\ref*{inaf_obs_rome}}\orcidlink{0000-0001-9815-9092} \and
  Romana Miku\v{s}incov\'{a} \inst{\ref*{inaf_roma}}\orcidlink{0000-0001-7374-843X} \and
  Michela Negro \inst{\ref*{lousiana_stateuni}}\orcidlink{0000-0002-6548-5622} \and
  Jorge Otero-Santos \inst{\ref*{infn_padova}}\orcidlink{0000-0002-4241-5875} \and
  Alessandro Paggi \inst{\ref*{inaf_bolo}}\orcidlink{0000-0002-5646-2410} \and 
  M. Lynne Saade \inst{\ref*{huntsville}, \ref*{nasa_huntsville}}\orcidlink{0000-0001-7163-7015}
  }
  
\institute{
    Department of Physics, University of Crete, 70013, Heraklion, Greece \label{physics_crete} \and
  Institute of Astrophysics, Foundation for Research and Technology-Hellas, 70013, Heraklion, Greece \label{forth} \and
  University of Geneva, Chemin d’Ecogia 16, CH-1290 Versoix, Switzerland \label{geneva_obs} \and
  Max-Planck-Institut für Physik, Boltzmannstr. 8, DE-85748 Garching, Germany \label{garching_mpp} \and
  NASA Marshall Space Flight Center, Huntsville, AL 35812, USA \label{nasa_marshall} \and
  INAF Osservatorio Astronomico di Roma, Via Frascati 33, 00078 Monte Porzio Catone (RM), Italy \label{inaf_obs_rome} \and
  INAF Istituto di Astrofisica e Planetologia Spaziali, Via del Fosso del Cavaliere 100, 00133 Roma, Italy \label{inaf_roma} \and
  Department of Physics and Astronomy, Louisiana State University, LA, USA \label{lousiana_stateuni} \and
  Istituto Nazionale di Fisica Nucleare, Sezione di Padova, 35131 Padova, Italy \label{infn_padova} \and
  INAF Istituto di Radioastronomia, via Gobetti 101, 40129, Bologna, Italy \label{inaf_bolo} \and
  Science \& Technology Institute, Universities Space Research Association, 320 Sparkman Drive, Huntsville, AL 35805, USA \label{huntsville} \and
  NASA Marshall Space Flight Center, Huntsville, AL 35812, USA \label{nasa_huntsville}
  }

  \abstract{Physical properties of blazar jets can be inferred or constrained through spectral energy distribution (SED) modelling.
  However, different emission models often reproduce observed SEDs equally well while yielding different physical interpretations.
  In this work, we compared one- and two-zone leptonic models to investigate how the inferred jet properties depend on the adopted description of the emitting region.
  We took a sample of high-synchrotron-peaked blazars observed by the Imaging X-ray Polarimetry Explorer (IXPE) and reconstructed simultaneous multiwavelength SEDs for each observation.
  The SEDs were modelled with the one-zone JetSeT framework and compared with an energy-stratified two-zone model.
  We explored correlations between the inferred physical parameters and the observed polarisation degree in the radio, optical, and X-ray bands and estimated the effective number of emitting cells contributing at different frequencies.
  We found that polarisation is linked to selected properties of the emitting regions, with significant correlations involving the magnetic field in the one-zone model and the radius and magnetic field of the compact two-zone region.
  The two models yield broadly consistent estimates of the magnetic field strength and emitting-region size.
  The effective number of emitting cells increases toward lower frequencies, supporting an energy-stratified scenario in which higher-energy emission originates from more compact regions, while remaining compatible with a fully turbulent jet.
  These results highlight the degeneracy of broadband SED modelling and demonstrate the importance of polarisation as a complementary probe of the internal structure and magnetic field properties of blazar jets.

}
   \keywords{Techniques: polarimetric -- galaxies: active -- galaxies: nuclei -- galaxies: jets}

   \maketitle

\section{Introduction}
\label{sect_introduction}

A subclass of active galactic nuclei (AGN) presents relativistic jets of magnetized plasma oriented towards us.
The alignment of the jet with our line of sight makes these astrophysical objects extremely bright due to relativistic beaming effects, and their emission spans from radio wavelengths to extremely high energies.
This type of astrophysical sources is known as blazars.

Their spectral energy distribution (SED) presents two bumps, the low-energy bump, associated with synchrotron emission (the synchrotron bump), and the high-energy bump, which is generally associated with electron inverse Compton emission, although the emission mechanisms taking place are still debated.
The frequency of the low-energy peak allows us to divide these sources into three classes \citep{abdo2010}: low-synchrotron-peaked, intermediate-synchrotron-peaked, and high-synchrotron-peaked blazars (LSP, ISP, and HSP, respectively).
LSP present synchrotron peaks falling into the millimetre-to-optical range ($\nu_{\rm synch}<10^{14} \ \rm Hz$), the HSP peaks fall in the X-ray range ($\nu_{\rm synch}>10^{15} \ \rm Hz$), and the ISP peaks fall in the middle range between these frequencies.

The emission and particle acceleration mechanisms occurring in these powerful sources and the structure of the jet itself are still under investigation.
Modelling the SED of blazars provides a powerful tool for inferring their physical conditions, including the magnetic field strength, the properties of the emitting particle population, the size and location of the emitting region, and the dominant radiation mechanisms.
However, different physical scenarios can often reproduce the same observed SED with comparable accuracy, leading to significant degeneracies in the inferred model parameters \citep{boettcher2012modeling}.
It is therefore essential to compare different emission models to assess the robustness of the derived physical quantities, identify model-dependent assumptions, and determine which observables are most effective at distinguishing between competing scenarios.

Many different models have been proposed, with the main possible scenarios considering emission dominated by hadronic or leptonic processes.
In leptonic scenarios, the high-energy component of the blazar SED is primarily attributed to inverse Compton scattering, in which relativistic electrons transfer energy to lower-energy photons.
The seed photon population can consist of the synchrotron radiation produced by the same electrons, giving rise to synchrotron self-Compton (SSC) emission, or of photons originating outside the jet.
In the latter case, known as external Compton (EC) scattering, the target photons can be supplied by the broad-line region, the dusty torus, or other external radiation fields surrounding the central engine.
From a polarimetric perspective, leptonic models generally predict relatively low polarisation degrees (PD) at high energies because inverse Compton scattering tends to reduce the polarisation of the incident radiation.
As a result, the SSC component is expected to be less polarised than the synchrotron emission from which its seed photons originate, while the EC component is predicted to exhibit little or no polarisation, reflecting the predominantly unpolarised nature of the external thermal photon fields \citep{bonometto1973,krawczynski2012}.
Under hadronic models, a significant fraction of the high-energy emission is produced by relativistic protons, either directly through synchrotron radiation or indirectly through hadronic interactions that generate secondary particles, whose synchrotron emission contributes to the observed high-energy radiation \citep{mannheim1992, mucke2001}.

In this context, polarisation offers an independent diagnostic of the magnetic field structure and particle acceleration processes, providing complementary constraints that cannot be obtained from spectroscopy alone.
The Imaging X-ray Polarimetry Explorer (IXPE; \citealt{weisskopf2022imaging,soffitta2023}) has been observing a sample of HSP blazars close to their synchrotron peak, strongly contributing to our studies of these sources.
Only a few studies have combined SED modelling with multiwavelength polarisation constraints for these sources, showing that one-zone and multi-zone emission scenarios, stratified shocks, and anisotropic magnetic fields can reproduce their spectral and polarimetric properties \citep{magic2024_421,magic2024_501,liodakis2025,lisalda2026,tavecchio2026}. However, these studies have focused on individual sources and campaigns, motivating a systematic investigation across multiple HSP blazars and epochs.
We took the sample of HSP observed by IXPE to perform a study of their SED in different epochs, taking advantage of two different leptonic emission models: a one-zone model (JetSeT; \citealt{tramacere2020,tramacere2011, tramacere2009}) and a two-zone model \citep{magic2024_501}.
We compare the obtained parameters and investigate potential correlations between the polarimetric properties of these sources in their low-energy bump and their spectral parameters.

In Section \ref{sect_jetmodels} we describe the two models, and the sources and observations we used are described in Section \ref{sect_sample}.
In Section \ref{sect_data_analysis} we describe the data analysis we followed.
In Section \ref{sect_results} we present the results of our study, and we summarise them and draw our conclusions in Section \ref{sect_conclusions}.

\section{Jet models}
\label{sect_jetmodels}

\subsection{JetSeT}
\label{sect_jetset}

JetSeT \citep{tramacere2009, tramacere2011, tramacere2020} is an SED modelling framework developed for blazar jets.
We used its one-zone leptonic implementation, in which the emitting region is treated as a single homogeneous zone within a turbulent magnetic field framework.
In this picture, the one-zone emitting region can be interpreted as an effective representation of the average emission produced by multiple unresolved turbulent cells.

In addition to reproducing the broadband SED, JetSeT predicts the frequency-dependent polarisation properties of the synchrotron component, from millimetre to X-ray wavelengths.
In particular, it provides the expected ordered PD, corresponding to the polarisation that would be observed if all emitting cells shared the same magnetic field orientation.
The ordered PD therefore represents the intrinsic polarisation of a perfectly ordered magnetic field and not the total polarisation of the source.

\subsection{SSC two-zone model}
\label{sect_twozone}

We used the code of \cite{magic2024_501, magic2025_421} to model the SED within a two-zone scenario. This model assumes an energy-stratified leptonic emission region consisting of two overlapping blobs, referred to as the compact and extended regions.
Each region is characterised by its own radius, magnetic field, and population of relativistic electrons.
The compact zone mainly contributes to the X-ray and the $>100$ GeV emission, whereas the extended zone dominates in the optical-UV and MeV-GeV radiation.
The extended zone spans a larger volume than the compact zone, extending downstream from the particle acceleration site, possibly associated with a shock front.
The fact that the X-ray emission originates closer to the shock front than the optical emission is consistent with the energy-stratification scenario proposed in \cite{angelakis2016robopol}, and it is qualitatively motivated by recent observations of strong chromaticity in the polarisation degree from radio to X-ray \citep{cape2025}.
Since the two regions overlap, we also accounted for the emission arising from their interaction.
Specifically, the synchrotron photons produced in each region provide an additional target photon field for inverse Compton scattering by the relativistic electrons in the other region.
Although this two-zone model is a simplified representation of the jet, as a continuous distribution of emitting regions is expected to contribute to the observed SED, it has successfully reproduced spectral properties of Mrk~501 up to TeV energies during three IXPE observations in 2022, as well as the flaring activity of Mrk~421 in December 2023 \citep{magic2024_501, magic2025_421}, up to TeV energies.
We therefore considered this approximation adequate for the purposes of this work.

\section{Sample}
\label{sect_sample}

Our sample consisted of all HSP sources that were previously observed by IXPE, and the observations we took into account are the same as those analysed in \cite{cape2025}.
Ordered by increasing right ascension, the sources are 1ES~0229$+$200 \citep{ehlert2023x}, Mrk~421 \citep{di2022x, di2023discovery, kim2024magnetic, maksym2024two, magic2025_421}, PG~1553$+$113 \citep{middei2023ixpe, middei2026}, Mrk~501 \citep{liodakis2022polarized, chen2024x, magic2024_501, cape2025, lisalda2026}, 1ES~1959$+$650 \citep{errando2024detection, pacciani2025}, and PKS~2155$-$304 \citep{kouch2024ixpe}.
The observations we took into account are reported in Table \ref{tab:observations}.
Although the sample comprises only six blazars, multiple observations of the same source were treated as independent measurements under the assumption that different epochs probe distinct physical states of the jet.
Given the strong variability exhibited by blazars across all wavelengths, this approach allowed us to investigate correlations across a broader range of source conditions, although it might increase the effective sample size relative to the number of individual objects.

The multiwavelength data were taken from the Markarian Multiwavelength Data Center (\href{https://mmdc.am}{MMDC}) and the Space Science Data Center (\href{https://tools.ssdc.asi.it/SED/sed.jsp?ra=05+09+28.4&dec=%2B05+42+01.0}{SSDC}) SED builder tool.
We corrected the optical data for the contribution of the host galaxy and included only strictly simultaneous observations.
For the radio band, we allowed a $\pm$10-day window around the IXPE observation interval to ensure adequate SED coverage given the relatively low variability of the radio emission.
For the first three observations of Mrk~501 and the first two observations of Mrk~421, we included simultaneous MAGIC data from \cite{magic2024_501} and \cite{magic2024_421}, which reported on joint IXPE--TeV observations.
For 1ES~0229$+$200, we also included archival high-energy data from \cite{magic2020}. Although these observations are not simultaneous with the IXPE campaign, the source exhibits low variability at TeV energies, allowing us to use them to better constrain the shape of the high-energy peak, for which only upper limits and no contemporaneous detections were available.
For the one-zone model, we included upper limits because the JetSeT tool takes them into account during fitting.

For the IXPE polarisation data, we relied on the values reported in the literature (see the papers listed above).
For the multiwavelength polarisation analysis, we used the same dataset as in \cite{cape2025}, considering only strictly simultaneous observations from multiple observatories. An exception was made for the radio data, for which observations acquired within a $\pm10$-day window of the observing interval were included.

\begin{table}[h!]
\centering
\caption{HSP sources and IXPE observations.}
\label{tab:observations}
\begin{tabular}{ccc}
\hline
Source & obsID & Dates \\
\hline
\multirow{4}{*}{1ES~1959$+$650}
 & 01006201 & May 3-4, 2022 \\
 & 01006001 & Jun 9-12, 2022 \\
 & 02004801 & Oct 28-31, 2022\\
 & 02250801 & Aug 14-19, 2023\\
\hline
\multirow{4}{*}{Mrk~421}
 & 01003701 & May 4-6, 2022\\
 & 01003801 & Jun 4-6, 2022\\
 & 02004401 & Dec 6-8, 2022\\
 & 02008199 & Dec 6-22, 2023\\
 \hline
\multirow{10}{*}{Mrk~501}
 & 01004501 & Mar 8-10, 2022\\
 & 01004601 & Mar 27-29, 2022\\
 & 01004701 & Jul 9-12, 2022\\
 & 02004601 & Feb 12-14, 2023\\
 & 02004501 & Mar 19-21, 2023\\
 & 02004701 & Apr 16-18, 2023\\
 & 03008801 & Apr 8-10, 2024\\
 & 03008901 & Jul 7-9, 2024\\
 & 03009001 & Aug 10-12, 2024\\
 & 03009101 & Jan 3-5, 2025\\
 \hline
 PKS~2155$-$304 & 02005601 & Oct 27-Nov 7, 2023\\
 \hline
 PG~1553$+$113 & 02004999 & Feb 1-9, 2023\\
 \hline
 1ES~0229$+$200 & 01006499 & Jan 15-Feb 1, 2023 \\
 \hline
\end{tabular}
\end{table}

\section{Modelling the spectral energy distribution}
\label{sect_data_analysis}

\begin{figure}[h!]
    \centering
    \includegraphics[width=0.98\linewidth]{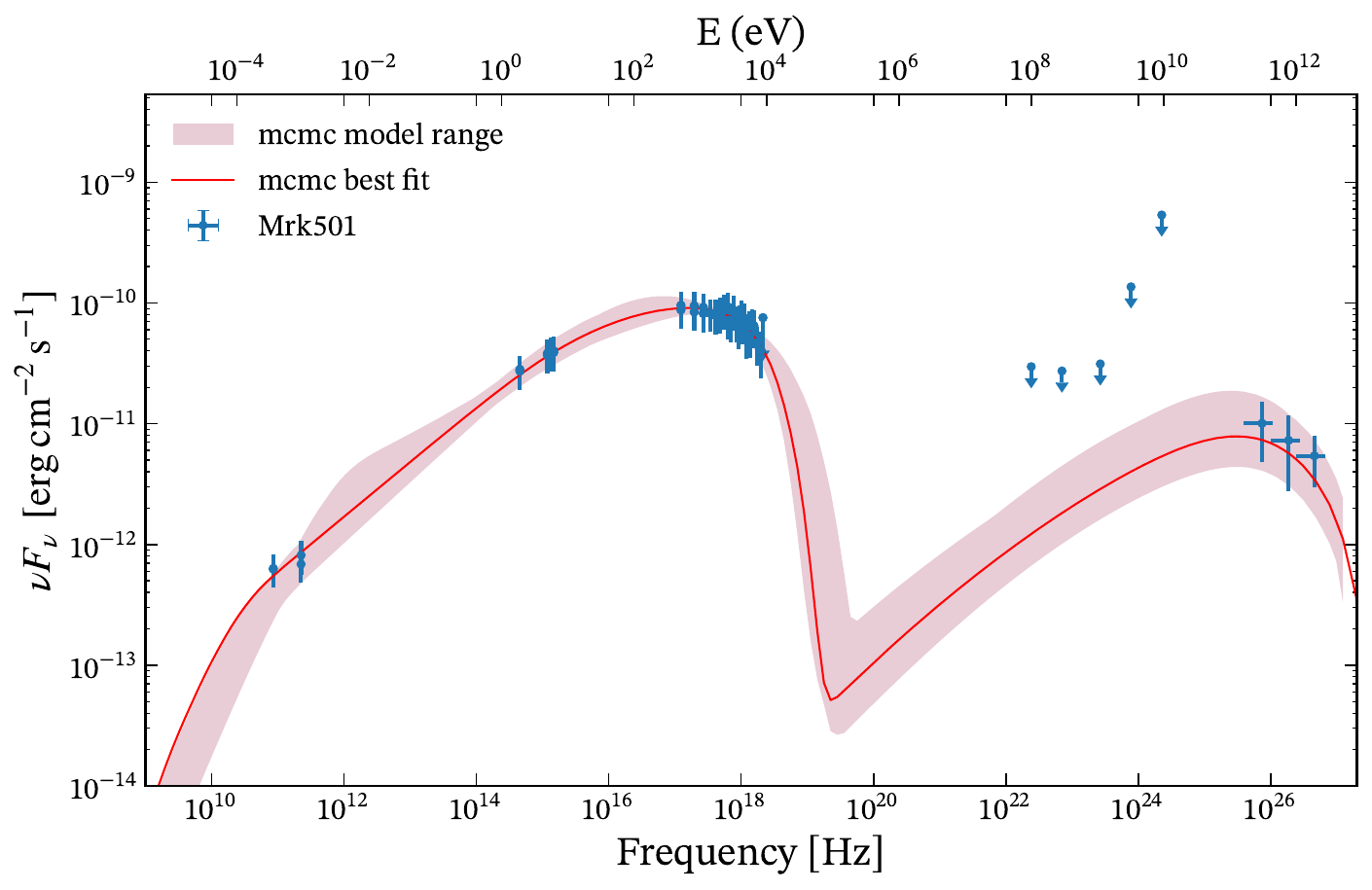}
    \caption{Fitted SED of Mrk~501 (obsID: 01004701) obtained with the one-zone model (JetSeT).
    The red line represents the best fit, the red shaded area the $1\sigma$ envelope, the blue dots the data points, and the downward blue arrows the upper limits.
    }
    \label{fig:fit_onezone}
\end{figure}

We started our data analysis by performing the Markov Chain Monte Carlo (MCMC) SED fitting through JetSeT.
Prior to the fit, we binned the data by frequency and, where necessary, we added systematic uncertainties to obtain more uniform error bars across the SED.
For the electron energy distribution (EED) within the emission zone, we used a log-parabola particle distribution.
An example of the fit obtained for Mrk~501 (obsID: 01004701) can be found in Fig. \ref{fig:fit_onezone}.

Based on the assumption that the emission arises from a number of spatially separated emitting cells with a uniform magnetic field, JetSeT also predicts the synchrotron polarisation degree expected within the framework of leptonic models.
This ordered polarisation degree, $\Pi_{\nu}^{\mathrm{ord}}$, represents the polarisation that would be observed if all emitting cells shared the same magnetic field orientation.
Under the assumption that the observed emission originates from $N_{\nu}$ emitting cells, where the only difference in the cell properties is the random orientation of the magnetic field, the observed polarisation degree is reduced according to:
\begin{equation}
    \langle \Pi_{\nu} \rangle \simeq
    \frac{\Pi_{\nu}^{\mathrm{ord}}}{\sqrt{N_{\nu}}},
\label{eq:n_cells}
\end{equation}
where $\langle \Pi_{\nu} \rangle$ is the observed PD, and $N_{\nu}$ is the  number of emitting cells contributing at frequency $\nu$ \citep{marscher2022}.

\cite{tramacere2025} showed that when the cell-to-cell randomisation is not limited to the magnetic field orientation but also includes variations in the EED parameters, Equation~\ref{eq:n_cells} yields an effective number of emitting cells, $N_{\nu}^{\rm eff}$. 
The quantity $N_{\nu}^{\rm eff}$ provides a flux-weighted estimate of the cells contributing to the total fractional polarisation and satisfies $N_{\nu}^{\rm eff}<N_{\nu}$. \cite{tramacere2025} also showed that dispersion in the EED parameters can modulate $N_{\nu}^{\rm eff}$ in a way that reproduces the multiwavelength polarisation chromaticity observed by IXPE in high-synchrotron-peaked BL Lac objects and the limiting envelope seen in the Robopol optical-polarisation data \citep{angelakis2016robopol}.
The comparison of the observed PD with the value predicted by JetSeT therefore provides an estimate of the effective number of emitting cells at each frequency.

For the two-zone model, we considered two emitting zones shaped as homogeneous spherical blobs, each with their own radius $R$ and magnetic field $B$, and the population of relativistic electrons within the emission regions was assumed to be a power law for the extended zone, and a broken power law for the compact zone.

For Mrk~501, we started from the parameters reported in \cite{magic2024_501}, where the same model is applied to three of our observations, and found the best parameters for the remaining observation intervals.
For 1ES~0229+200, the Doppler factor $\delta$ was taken from \cite{zech2021}, and for PG~1553+113, the Doppler factor was taken from \cite{aleksi2012}.
For these sources, we assumed a Doppler factor equal to the bulk Lorentz factor $\Gamma$  and computed the viewing angle as $\theta = 1/\delta$ rad.
For Mrk~421, we took the bulk Lorentz factor and viewing angle from \cite{magic2025_421}, for 1ES~1959+650 from \cite{aliu2013}, and for PKS~2155-304 from \cite{petropoulou2014}.
The Doppler factor for these sources was calculated as
$\delta = 1/{\Gamma \left(1 - \beta \cos\theta \right)}$, where $\beta = \sqrt{1 - \Gamma^{-2}}$ is the dimensionless velocity.
The values of these parameters are listed in the tables on the \href{https://zenodo.org/records/22030471}{Zenodo} platform.

\begin{figure}[h!]
    \centering
    \includegraphics[width=0.98\linewidth]{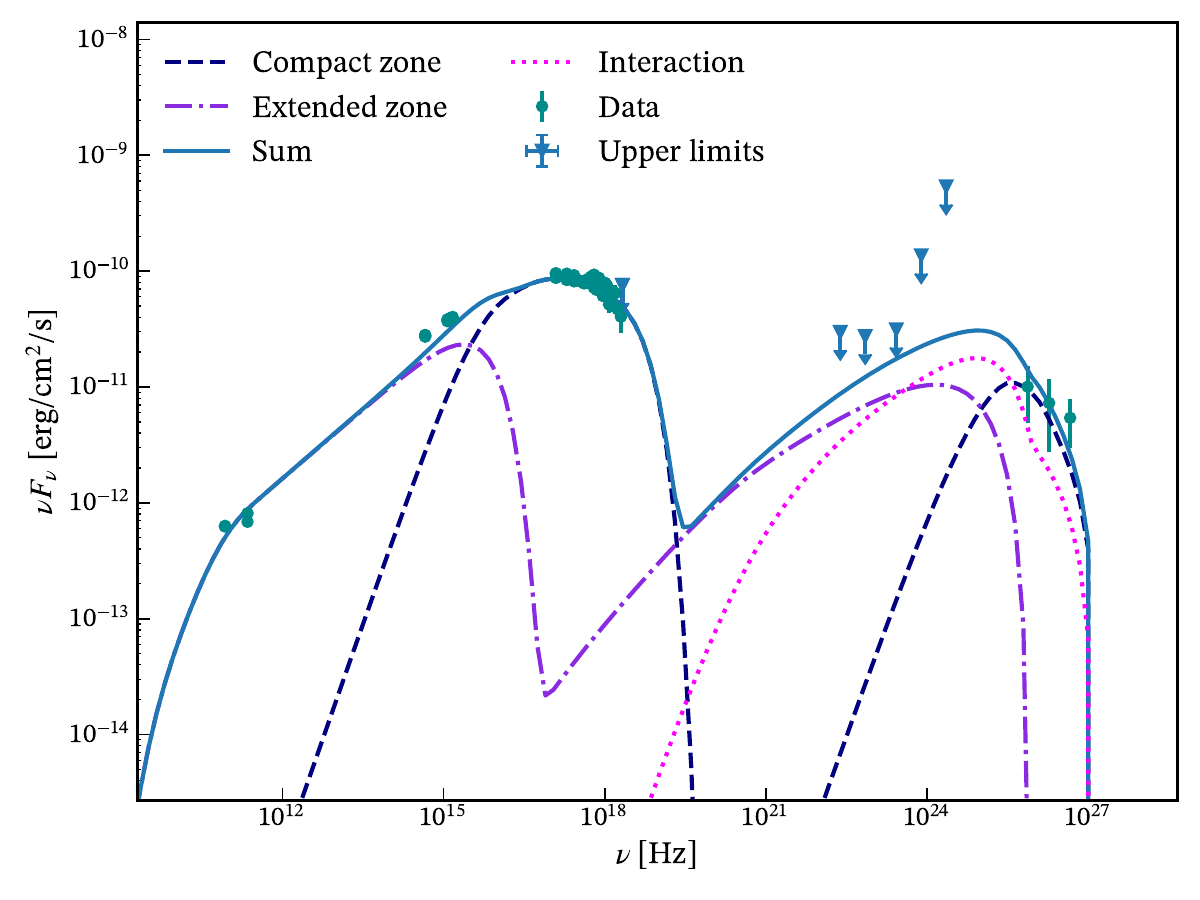}
    \caption{SED of Mrk~501 (obsID: 01004701) obtained with the two-zone model.
    The dash-dotted purple line represents the extended zone, the dashed black line shows the compact zone, and the dotted pink line shows the interaction between the two.
    The solid blue line represents the total SED, the cyan dots represent the observational data, and the downward blue arrows represent upper limits.}
    \label{fig:fit_twozone}
\end{figure}

In the extended zone, the power-law electron distribution was defined between a minimum and a maximum Lorentz factor ($\gamma_{min}$, $\gamma_{max}$), and the slope was assumed to be $p=2.2$ for every observation.
In the compact zone, the distribution of electrons is defined between a minimum and a maximum Lorentz factor ($\gamma_{min}$, $\gamma_{max}$), with the break located in $\gamma_{break}$; the slopes below and above the break are $n_1$ and $n_2$, respectively.
Both regions were defined with an electron energy density $U'_e$.
The radius $R_2$ of the compact zone was adopted following \cite{magic2024_501}, such that the ratio of the radii satisfies
\begin{equation}
\frac{R_{1}}{R_{2}} \approx l^{1/3},
\end{equation}
where $R_1$ is the radius of the extended zone, and $l$ is inferred from the median PD.
The expected average PD from an emission zone consisting of $N$ separate cells can indeed be approximated as $PD\approx 75\%/\sqrt{N}$, from which $N$ can be inferred using the median X-ray PD \citep{marscher2014turb, tavecchio2021, zhang2024}.
Assuming the extended zone, responsible for the bulk of the optical emission, to be a factor of $l$ larger than the compact region, such that in the optical range $PD_{\rm opt} \approx 75\%/\sqrt{lN}$, and assuming that turbulent cells have equal volumes, the radius of the extended zone is then higher by a factor of $l^{1/3}$ than the radius of the compact zone.
The adopted values for the two radii reproduce this relation within 15\% for all considered observations, with the exception of the fifth observation of Mrk~501, for which the X-ray PD was slightly lower than the optical PD, and the discrepancy is about 25\%.
An example of the SED of Mrk~501 (obsID: 01004701) obtained with this model is reported in Fig. \ref{fig:fit_twozone}, which highlights the contributions from the two zones.

After obtaining the best parameters from the two models (reported on the \href{https://zenodo.org/records/22030471}{Zenodo} platform), we investigated possible correlations between the two through the Kolmogorov--Smirnov (K--S) and the Wilcoxon statistical tests.
The K--S test assesses whether two independent samples are drawn from the same parent distribution, while the Wilcoxon test assesses whether two paired samples differ systematically by testing whether the median of their pairwise difference is zero.
We also investigated possible correlations between the model parameters and the median PD in the different bands, and between the parameters and the median PD ratio calculated between the different bands.
These were assessed through the Spearman rank correlation test, which evaluates the monotonic relation between two variables.

\section{Results}
\label{sect_results}

We present here the results obtained for the two models, compare their predictions, and discuss the number of emitting cells inferred from the JetSeT polarisation modelling.
For the separate analysis of the models, we used the Spearman rank correlation test to investigate the relation between the median PD in the X-ray, optical (R), and radio (225 GHz) bands and the model-predicted emitting-region radius $R$, magnetic field strength $B$, and synchrotron peak frequency $\nu_{\rm synch}$.
We also investigated correlations between these parameters and the median X-ray-to-optical and X-ray-to-radio ratios.
We did not include lower-frequency radio observations, as the emission at 225 GHz is expected to be optically thin and is significantly less affected by Faraday rotation and depolarisation effects, making it a more reliable tracer of the intrinsic synchrotron polarisation \citep{agudo2010,trippe2012}.
For the one-zone model, we also searched for possible correlations between the spectral slope $r$ and the X-ray PD, as well as between $r$ and $\nu_{\rm synch}$, using the Spearman test.
For the comparison between the two models, we used the K--S and the Wilcoxon tests applied to the two distributions of $R$, $B$, and $\nu_{\rm synch}$, considering the extended zone and the compact zone of the two-zone model separately.

\subsection{One-zone model}
\label{subs_one_zone}

The best fits obtained with JetSeT are reported in Appendix \ref{appendix_jetset_fits}; the parameters, as well as the results of the Spearman rank correlation test applied to the median PD and the model parameters for the three bands considered, are reported on the \href{https://zenodo.org/records/22030471}{Zenodo} platform.
We report the plots of the parameters versus median PD in Appendix \ref{appendix_medianpd} (see Fig. \ref{fig:R_onezone}, Fig. \ref{fig:B_onezone}, and Fig. \ref{fig:nu_onezone} for $R$, $B$ and $\nu_{\rm synch}$, respectively).

The resulting p-values are all $\geq 0.05$, indicating that there is insufficient evidence for a statistically significant monotonic correlation between the model parameters $R$ and $\nu_{\rm synch}$ and the median polarisation degree in the three bands considered.
There is an exception for the parameter $B$, which shows a statistically significant positive correlation with the radio PD ($\rho=0.778$, $p=0.014$), but not with the PD in the other bands.

Following \cite{tramacere2011}, we investigated correlations between the synchrotron peak frequency and the spectral curvature $r$ estimated with the one-zone model.
We applied the Spearman correlation test and found a statistically significant negative correlation between the two quantities ($\rho = -0.542$, $p=0.011$).
This indicates that observations with higher synchrotron peak frequencies tend to exhibit lower values of the curvature parameter (see Appendix \ref{appendix_spectralcurv}). 

Following \cite{saade2025}, who suggested a possible correlation between $r$ and the X-ray PD, we used the Spearman test to investigate this relation, but found no statistically significant correlation ($\rho=0.173$, $p=0.454$).

We also investigated the relation between the median PD ratio (X-ray over radio and X-ray over optical) and the parameters $R$, $B$, and $\nu_{\rm synch}$, but found no statistically significant correlations.
We report the plots representing these relations and the values of the Spearman $\rho$ and $p$ in Appendix \ref{appendix_medianpdratio}.

\subsection{Two-zone model}
\label{subs_two_zone}

We took the physical parameters of the extended zone and of the compact zone into account separately.
The compact zone primarily accounts for the high-energy emission (X-ray and $>100 \ \rm GeV$) and was therefore used to investigate correlations with the X-ray polarisation properties.
The extended zone, on the other hand, accounts for the bulk of the emission in the radio, optical/UV, and MeV--GeV energy range, and was therefore used to investigate correlations with the optical (R) and radio (225 GHz) bands.
The best fits obtained with the two-zone model are reported in Appendix \ref{appendix:twozone_fits}; the best parameters obtained for the two zones, as well as the results of the statistical tests assessing the relations between the model parameters and the median PD and PD ratios, are reported on the \href{https://zenodo.org/records/22030471}{Zenodo} platform.

For the X-ray band, we used the parameters of the compact region, and for the optical and radio bands, we used those of the extended region.
The synchrotron peak frequency does not depend on the region considered, as inferred from the total SED.
We report the plots of the parameter versus median PD in Appendix \ref{appendix_medianpd} (see Fig. \ref{fig:R_twozone}, Fig. \ref{fig:B_twozone}, and Fig. \ref{fig:nu_twozone} for $R$, $B$ and $\nu_{\rm synch}$, respectively).

As for the one-zone model case, the p-values resulting from the tests are $\geq 0.05$, indicating no significant correlations between $R$, $B$, or $\nu_{\rm synch}$ and the median polarisation degree in the radio, optical, and X-ray bands.
The only significant correlation is found between the radius of the compact emission region and the X-ray PD ($\rho=-0.639$, $p=0.002$), with smaller emitting regions corresponding to higher X-ray polarisation.
This behaviour is expected because the X-ray emission predominantly originates in the compact region.
A possible interpretation is that a more confined emitting volume preserves a higher degree of magnetic field coherence, whereas a larger region averages over a greater number of emitting cells with different magnetic field orientations, resulting in a lower net polarisation.

For the ratio of the X-ray and optical polarisation degree, we found no significant correlations with the physical properties of the extended emission region or with the synchrotron peak frequency.
In contrast, the compact emission region exhibits significant correlations with the magnetic field strength ($\rho=0.494$, $p=0.023$) and the emitting-region radius ($\rho=-0.602$, $p=0.004$).
Specifically, higher magnetic field strengths are associated with higher X-ray-to-optical polarisation ratios, while larger compact emitting regions correspond to lower values of this ratio.
This behaviour is consistent with the construction of the two-zone model, in which the compact region is responsible for the X-ray synchrotron emission, while the extended region dominates the lower-energy emission.
A possible interpretation is that a more confined emitting volume preserves a higher degree of magnetic field coherence, whereas a larger region averages over a greater number of emitting cells with different magnetic field orientations, resulting in a lower net polarisation.

No statistically significant correlations were found between the X-ray-to-radio polarisation degree ratio and any of the investigated physical parameters.
This holds for the extended and compact emission regions and for the synchrotron peak frequency.
The plots representing the relations of the parameters with the PD ratios are reported in Appendix \ref{appendix_medianpdratio}, together with the values of $\rho$ and $p$ calculated with the Spearman statistical test.

\subsection{Model comparison}
\label{subs_comparison}

In order to investigate whether the two models yield similar physical properties for the emitting region, we compared the parameters of the one-zone model with those of the two-zone model using the K--S and Wilcoxon statistical tests.
We compared the one-zone parameters with the parameters of the two-zone extended and compact regions separately.

\begin{figure}
    \centering

    \begin{subfigure}{0.85\linewidth}
        \centering
        \includegraphics[width=\linewidth]{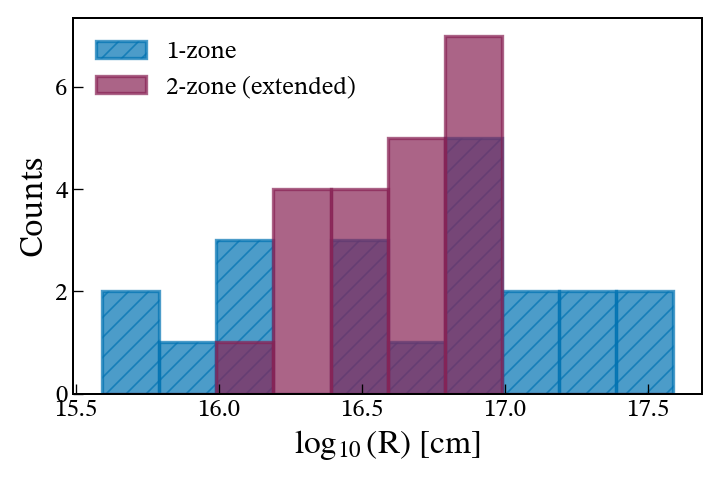}
        \caption{}
        \label{fig:hist_R_extended}
    \end{subfigure}

    \vspace{0.5cm}

    \begin{subfigure}{0.85\linewidth}
        \centering
        \includegraphics[width=\linewidth]{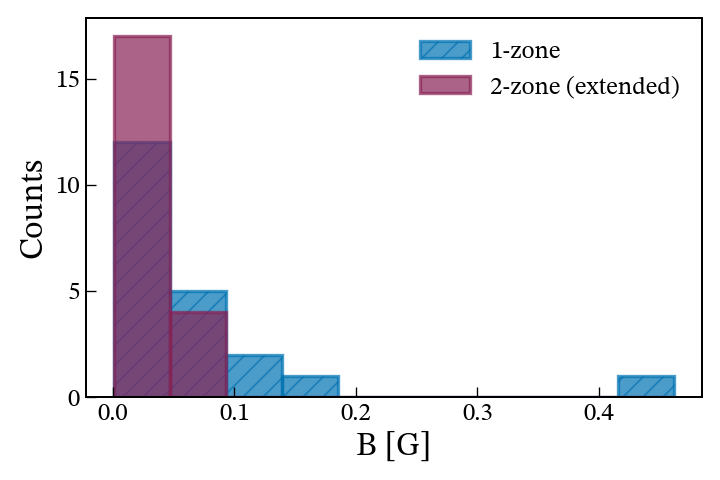}
        \caption{}
        \label{fig:hist_B_extended}
    \end{subfigure}

    \caption{Distributions of the physical parameters inferred from the one-zone (blue) and two-zone (red) models. (a) Emitting region radius $R$. (b) Magnetic field strength $B$. For the two-zone model, the distributions correspond to the extended emission region.}
    \label{fig:histograms_extended}
\end{figure}

\begin{figure}
    \centering

    \begin{subfigure}{0.85\linewidth}
        \centering
        \includegraphics[width=\linewidth]{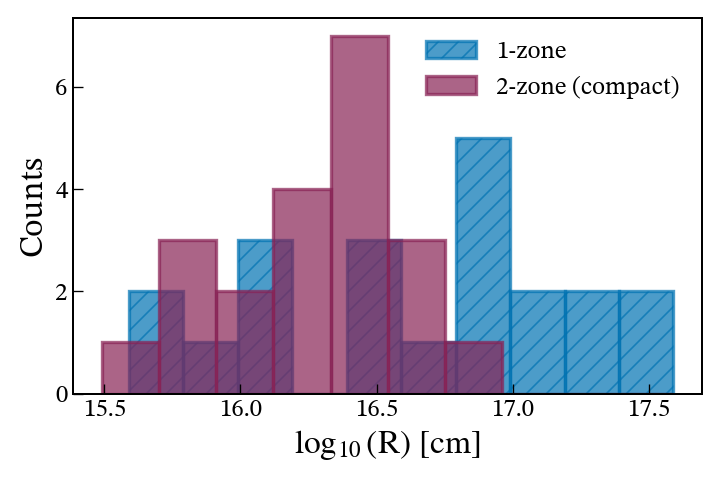}
        \caption{}
        \label{fig:hist_R_compact}
    \end{subfigure}

    \vspace{0.5cm}

    \begin{subfigure}{0.85\linewidth}
        \centering
        \includegraphics[width=\linewidth]{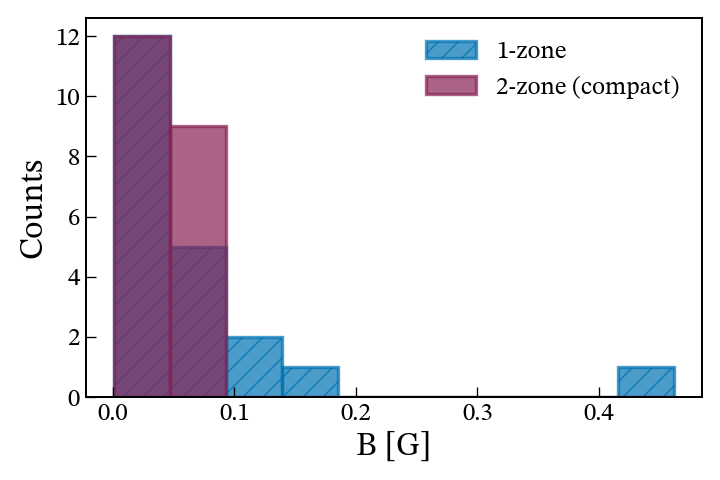}
        \caption{}
        \label{fig:hist_B_compact}
    \end{subfigure}

    \caption{Distributions of the physical parameters inferred from the one-zone (blue) and two-zone (red) models. (a) Emitting region radius $R$. (b) Magnetic field strength $B$. For the two-zone model, the distributions correspond to the compact emission region.}
    \label{fig:histograms_compact}
\end{figure}

\begin{figure}[h!]
    \centering
    \includegraphics[width=0.9\linewidth]{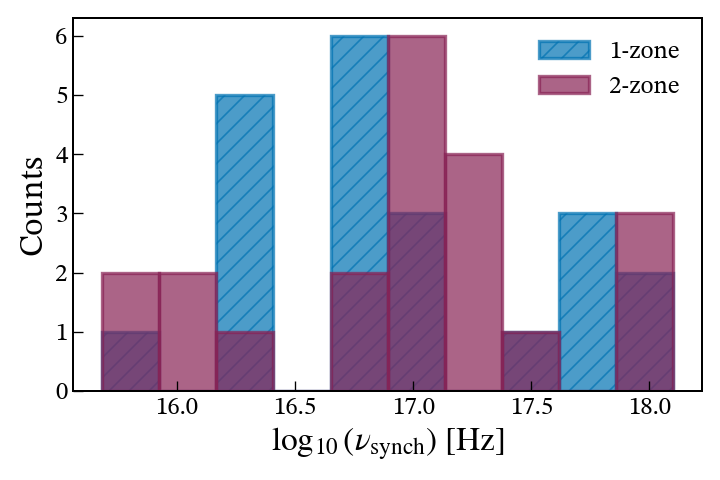}
    \caption{Distributions of the synchrotron peak frequencies inferred from the one- (blue) and two-zone model (red).}
    \label{fig:hist_nu}
\end{figure}

The results of the statistical tests applied to the two models are reported on the \href{https://zenodo.org/records/22030471}{Zenodo} platform.
Figures \ref{fig:histograms_extended} and \ref{fig:histograms_compact} represent the distributions of the parameters $R$ and $B$ of all the considered observations for the two models (extended region and compact region of the two-zone model, respectively).
Figure \ref{fig:hist_nu} represents the distribution of the synchrotron peak frequencies.

The comparison between the one-zone model and the extended emission region of the two-zone model shows broad agreement in the inferred physical parameters.
Neither the synchrotron peak frequency nor the magnetic field strength exhibit significant differences in either the paired values or their overall distributions.
For the emission region radii, the Wilcoxon signed-rank test detected no significant paired difference between the values inferred by the two models.
In contrast, the K--S test indicated that their overall distributions differ significantly.
This suggests that the two models produce different distributions of emitting-region sizes, even though neither model systematically yields larger or smaller radii for individual observations.

The comparison between the one-zone model and the compact emission region yields a different picture.
The emitting-region radius differs significantly in the paired estimates and in its overall distribution, indicating that the compact region is systematically smaller than the single emitting region inferred in the one-zone scenario.
This result should be interpreted with caution, however, because the two-zone model is intrinsically constructed such that the compact region represents the smaller of the two emitting zones.
Therefore, the observed difference is expected, and reflects the underlying assumptions of the model.
In contrast, the magnetic field strengths exhibit no significant paired offset according to the Wilcoxon signed-rank test.
However, the K--S test indicates that their overall distributions differ significantly.
This suggests that although the two models infer comparable magnetic field strengths for individual observations, the distribution of the inferred values across the sample is not the same.

\subsection{Effective number of emitting cells under one-zone model polarisation predictions}
\label{subs_polarisation}

\begin{figure}[h!]
    \centering
    \includegraphics[width=0.98\linewidth]{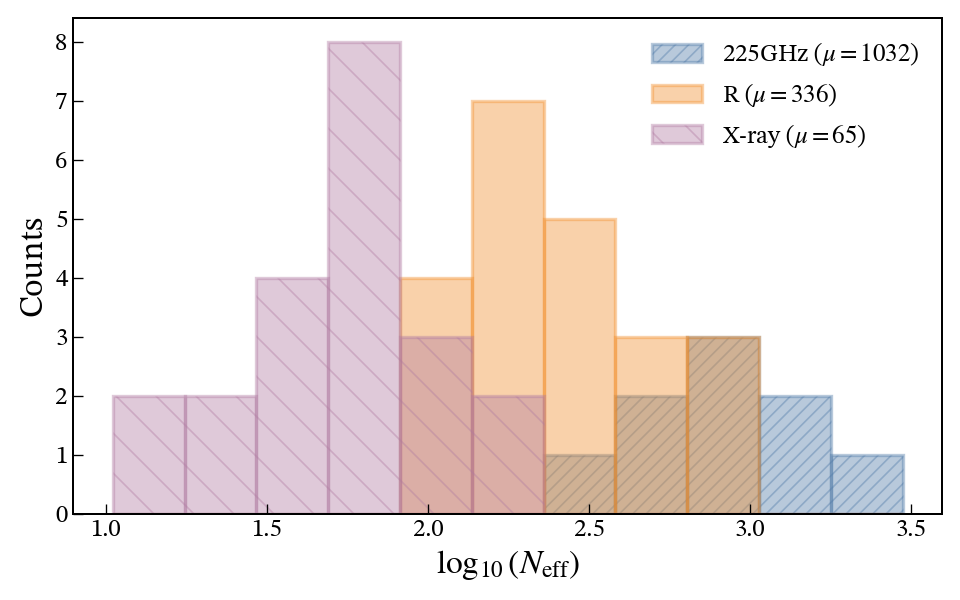}
    \caption{Histograms of $\rm log_{10}$$(N_{\rm eff})$ derived from the ordered and observed polarisation degrees for the 225 GHz (blue), optical (green), and X-ray bands (pink). The legend reports the average of $N_{\rm eff}$ for each band.}
    \label{fig:hist_n}
\end{figure}

By comparing the polarisation predicted by JetSeT with the median observed PD, we estimated the effective number of emitting cells in each band ($N_{\nu}^{\rm eff}$).
This is plotted in Fig. \ref{fig:hist_n}.
As discussed in Section \ref{sect_data_analysis}, under the assumption of a fully turbulent model (randomisation not limited to the magnetic field orientation), $N_{\nu}^{\rm eff}$ provides the flux-weighted estimate of cells that would reproduce the observed degree of polarisation and should therefore be interpreted as an equivalent measure of the magnetic field disorder, plus dispersion in cells sizes and EED parameters, rather than the actual physical number of emitting cells in the jet.
Under the assumption that randomisation is limited to the orientation of the magnetic field, that is, identical cells with randomly oriented magnetic fields, we have $N_{\nu}^{\rm eff}=N_{\nu}$, and this number should therefore be interpreted as an equivalent measure of the magnetic field disorder and as an estimate of the actual number of cells.

Our results yield an average $N_{\nu}^{\rm eff}$ ($\mu$) and standard deviation ($\sigma$) of $\mu=64.7$, $\sigma=43.9$ for the X-ray band, $\mu=336.0$, $\sigma=274.9$ for the optical band, and $\mu=1032.3$, $\sigma=755.7$ for the radio band.
The average effective number of emitting cells increases with decreasing frequency.
The trend is consistent with energy-stratification models, in which the highest-energy emission originates from compact regions located closer to the jet base, where only a limited number of turbulent cells contributes to the observed radiation.
At lower frequencies, the emission is produced farther downstream in more extended regions, where a larger number of emitting cells contribute simultaneously, leading to a higher effective number of cells.
The trend and the estimated values of $N_{\nu}^{\rm eff}$ are also consistent with the predictions of the fully turbulent model \citep{tramacere2025}, which naturally reproduces the observed decrease in the number of cells as the EED cutoff energy increases.
More in detail, the $N_{\nu}^{\rm eff}$ values inferred from the ratio of the maximum to the observed polarisation degree are consistent to within $\sim20\%$ with those obtained from the Monte Carlo simulations of \citet{tramacere2025}, reproducing the observed $N_{\nu}^{\rm eff}$ trend across the IXPE HSP BL Lac sample.

\section{Summary and discussion}
\label{sect_conclusions}

We analysed a sample of six HSP blazars observed by IXPE, comprising a total of 21 observations.
For each observation, we collected simultaneous multiwavelength data and modelled the spectral energy distribution using a one- and a two-zone leptonic scenario.
The primary aim of this work was to assess how the inferred physical parameters depend on the adopted emission model and to investigate possible correlations between these parameters and the observed polarisation properties of the sources.
For the two-zone model, these correlations were investigated separately for the extended and compact emission regions.

We started by performing the correlation analysis separately for the one-zone and two-zone models.
For each model, we used the Spearman rank correlation test to investigate correlations between the three physical parameters we considered (the radius $R$ of the emitting region, the uniform magnetic field $B$ within the emission zone, and the frequency of the synchrotron peak $\nu_{\rm synch}$) and the median PD of the sources during the IXPE observations in the radio (225 GHz), optical (R), and X-ray bands.
No statistically significant correlations were found between any of the investigated parameters and the observed median PD for either model, with the exception of the magnetic field strength inferred from the one-zone model, which shows a positive correlation with the radio PD (Fig. \ref{fig:B_onezone}, third panel), and the radius of the compact region in the two-zone model, which shows an anti-correlation with the X-ray PD (Fig. \ref{fig:R_twozone}, first panel).
The former might indicate that observations with stronger magnetic fields are associated with a higher degree of magnetic field ordering, leading to enhanced synchrotron polarisation.
The latter is qualitatively expected because the X-ray emission predominantly originates in the compact region: a smaller emitting volume is likely to preserve a greater degree of magnetic field coherence, whereas a larger region averages over a greater number of emitting cells and magnetic field orientations, resulting in a lower net polarisation.
We found a statistically significant negative correlation between the synchrotron peak frequency and the spectral curvature parameter inferred from the one-zone model (Fig. \ref{fig:nu_vs_r}).
This result is consistent with \citet{tramacere2011}, who reported a decrease in spectral curvature with increasing synchrotron peak frequency, and it might be a signature of a stochastic component in the acceleration process \citep{tramacere2007,tramacere2009} or an energy-dependent acceleration mechanism \citep{massaro2004log}.

We also searched for a correlation between the spectral curvature estimated by JetSeT and the observed X-ray PD.
This analysis could only be performed for the one-zone model because the electron energy distribution in the two-zone model was assumed to be a pure power law, yielding a fixed spectral slope.
In contrast to the potential indication of a correlation reported by \cite{saade2025}, no statistically significant correlation was found in our sample.

We compared the physical parameters inferred from the one-zone (JetSeT) model and the two emitting regions of the two-zone model using the K--S and Wilcoxon signed-rank tests. 
We note that agreement between parameter distributions, or the absence of a significant systematic offset, does not necessarily imply that parameters inferred for a given observation in one model predict those inferred for the same observation in the other model.
Although both models successfully reproduce the observed SEDs, the inferred parameters show different levels of agreement.
For the extended region, the magnetic field strengths are statistically consistent with those inferred by the one-zone model, as neither the K--S nor the Wilcoxon signed-rank test rejected the null hypothesis of equal distributions or paired values.
For the compact region, the Wilcoxon signed-rank test did not reveal a significant paired difference in the magnetic field strength with respect to the one-zone model, whereas the K--S test indicated that the overall distributions differ significantly.
A similar result was obtained for the radius of the extended region: no significant paired offset was detected, but the overall distributions differ significantly.
As expected from the construction of the two-zone model, the radius of the compact region is systematically smaller than that inferred in the one-zone scenario, reflecting the assumption that the high-energy emission originates from a more localised region rather than providing an independent observational constraint.

Investigating the synchrotron peak frequencies inferred by the two models, the K--S and the Wilcoxon tests indicate statistical consistency between the one-zone and two-zone estimates.
Given the relatively small number of observations, however, these results should be interpreted with caution, as the statistical significance remains limited.
These findings suggest that broadband SED fitting alone cannot uniquely constrain the underlying jet properties, as physically distinct scenarios can reproduce the observed spectra equally well.
This highlights the importance of polarisation measurements as an additional observable for breaking these degeneracies and distinguishing between competing emission models.

We searched for correlations between the PD ratio in different bands (X-ray over optical and X-ray over radio) and the three parameters.
For the one-zone model, no correlations were found.
For the two-zone model, significant correlations were found between the X-ray-to-optical PD ratio and both the radius and the magnetic field strength of the compact region (Fig. \ref{fig:twozone_compact_ratio_opt}).
These results highlight the connection between the physical properties of the compact emitting region and the relative degree of X-ray polarisation, with larger emitting-region radii corresponding to lower X-ray-to-optical PD ratios and stronger magnetic fields corresponding to higher X-ray-to-optical PD ratios.

Finally, we  used JetSeT to estimate the ordered PD at different frequencies, and compared it to the observed PD through the relation reported in Eq. \ref{eq:n_cells}, obtaining the effective number of emitting cells.
We found an increasing average number of cells with decreasing frequency, in agreement with the energy-stratified scenario \citep{angelakis2016robopol,liodakis2022polarized}: the high-energy emission is limited to compact regions closer to the jet base, while the lower-energy emission originates farther downstream in the jet, in more extended regions.
This is also in agreement with the previous analysis of the same IXPE pointings from a polarimetric point of view \citep{cape2025}, which also indicated an energy-stratified acceleration scenario.

Overall, our results highlight that broadband SED modelling alone is insufficient to uniquely determine the physical properties of blazar jets.
While the one-zone and two-zone models both successfully reproduced the observed emission, they implied different physical conditions within the emitting region, demonstrating the intrinsic degeneracy of SED modelling.
At the same time, the frequency dependence of the effective number of emitting cells provided independent evidence in favour of an energy-stratified emission scenario or of a fully turbulent scenario with a randomisation in the EED parameters \citep{tramacere2025}.
These results emphasise the importance of combining spectroscopy and polarimetry to better constrain the physical properties of relativistic jets.
Future X-ray polarimetric observations, together with simultaneous multiwavelength campaigns and multi-zone models, will provide increasingly stringent constraints on the physical processes dominating blazar emission.

\section*{Data availability}

Supplementary tables with the best parameters obtained with both the one-zone and the two-zone models and the results of all the statistical tests applied in this work have been uploaded on the \href{https://zenodo.org/records/22030471}{Zenodo} platform.

\begin{acknowledgements}

      This research has made use of data from the RoboPol programme, a collaboration between Caltech, the University of Crete, IA-FORTH, IUCAA, the MPIfR, and the Nicolaus Copernicus University, which was conducted at Skinakas Observatory in Crete, Greece.

      Part of this work is based on archival data, software or online services provided by the Space Science Data Center - ASI.
      
      Some of the data used in this work were obtained from the MMDC.

      S. C. and I. L. were funded by the European Union ERC-2022-STG - BOOTES - 101076343. Views and opinions expressed are however those of the author(s) only and do not necessarily reflect those of the European Union or the European Research Council Executive Agency. Neither the European Union nor the granting authority can be held responsible for them.
      
      A. A.-E. acknowledges the support of the Swiss National Science Foundation (SNSF) through the Ambizione grant number 232998.

      This research was partially funded by the Deutsche Forschungsgemeinschaft (DFG, German Research Foundation) as part of the DFG Research Unit FOR5195 – project number 443220636.

      R. M. acknowledges financial support from the INAF Scientific Directorate. 

      The Imaging X-ray Polarimetry Explorer (IXPE) is a joint US and Italian mission.  The US contribution is supported by the National Aeronautics and Space Administration (NASA) and led and managed by its Marshall Space Flight Center (MSFC), with industry partner Ball Aerospace (now, BAE Systems).  The Italian contribution is supported by the Italian Space Agency (Agenzia Spaziale Italiana, ASI) through contract ASI-OHBI-2022-13-I.0, agreements ASI-INAF-2022-19-HH.0 and ASI-INFN-2017.13-H0, and its Space Science Data Center (SSDC) with agreements ASI-INAF-2022-14-HH.0 and ASI-INFN 2021-43-HH.0, and by the Istituto Nazionale di Astrofisica (INAF) and the Istituto Nazionale di Fisica Nucleare (INFN) in Italy.  This research used data products provided by the {\it IXPE} Team (MSFC, SSDC, INAF, and INFN) and distributed with additional software tools by the High-Energy Astrophysics Science Archive Research Center (HEASARC), at NASA Goddard Space Flight Center (GSFC).

\end{acknowledgements}

\newpage

\bibliographystyle{aa} 
\bibliography{bibliography.bib}

\newpage

\begin{appendix}
\onecolumn

\clearpage
\section{JetSeT fits}
\label{appendix_jetset_fits}

We report here the best fits obtained by JetSeT (source and obsID reported in the caption).
The red line represents the fit, the red shaded area the 1$\sigma$ envelope, and the blue dots represent the data (with arrows indicating upper limits).

\begin{figure*}[h!]
\centering

\begin{minipage}{0.32\textwidth}
    \centering
    \includegraphics[width=\linewidth]{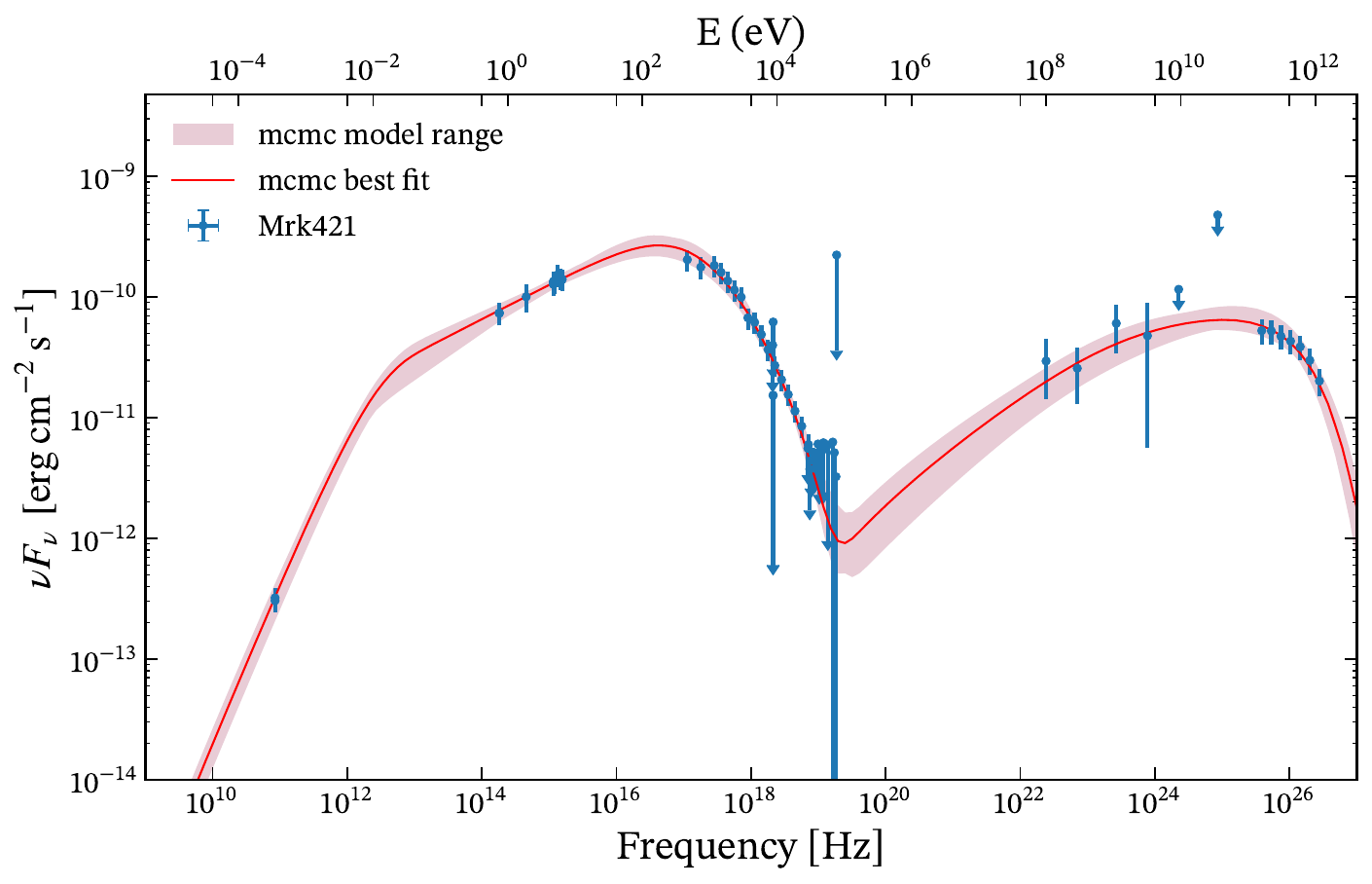}
    \small Mrk~421 -- 01003701
\end{minipage}
\hfill
\begin{minipage}{0.32\textwidth}
    \centering
    \includegraphics[width=\linewidth]{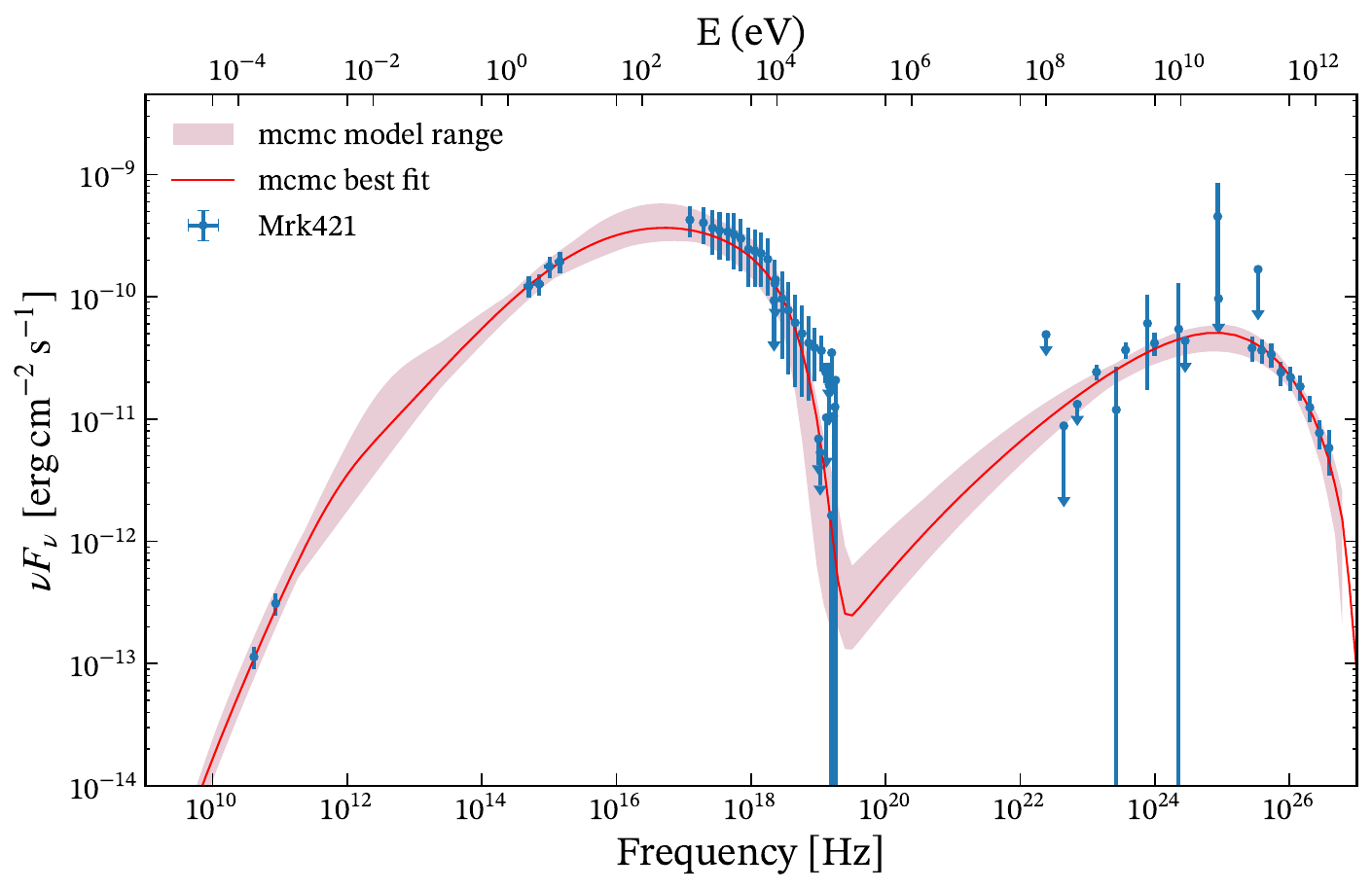}
    \small Mrk~421 -- 01003801
\end{minipage}
\hfill
\begin{minipage}{0.32\textwidth}
    \centering
    \includegraphics[width=\linewidth]{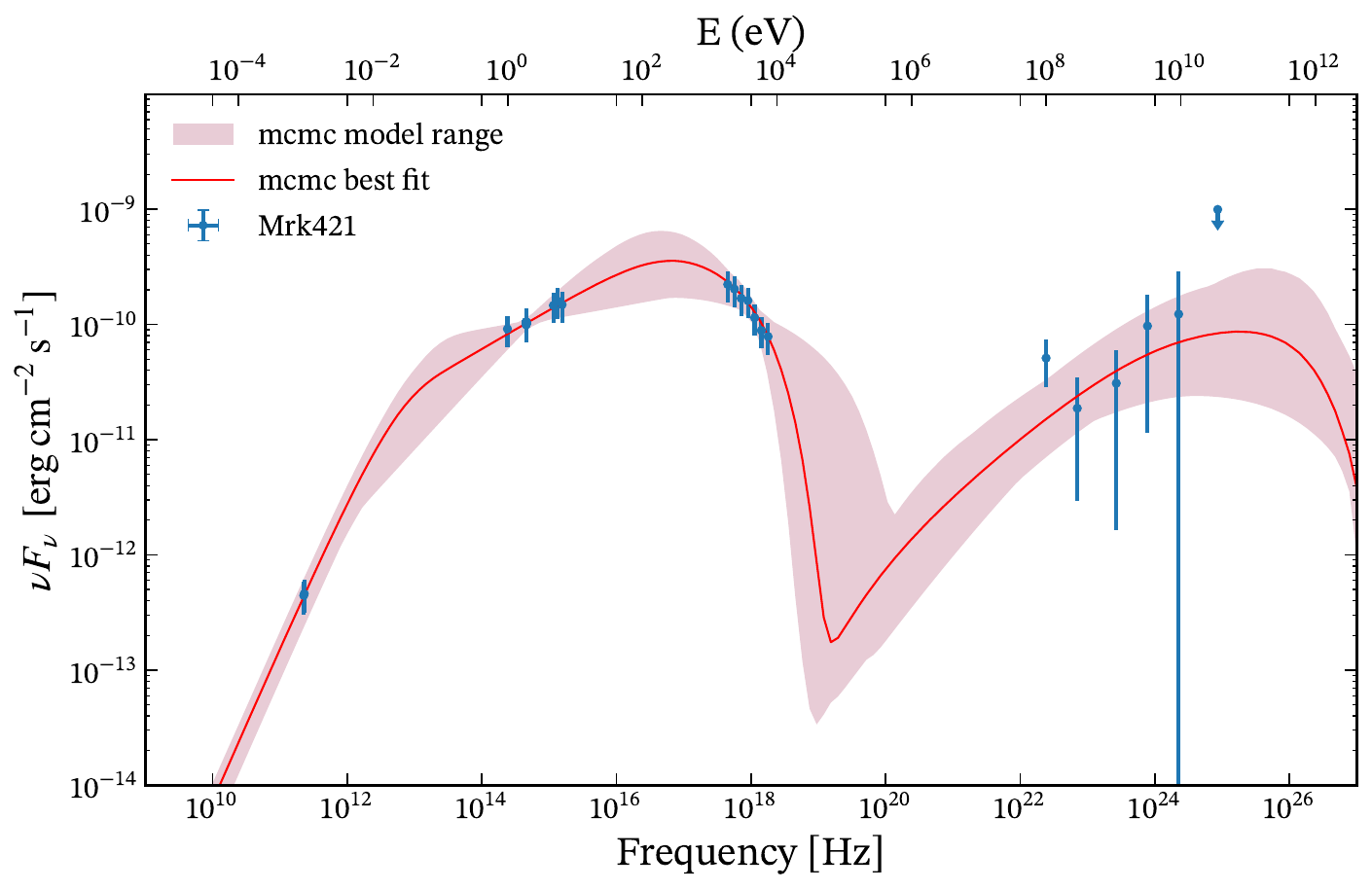}
    \small Mrk~421 -- 02004401
\end{minipage}

\vspace{0.4cm}

\begin{minipage}{0.32\textwidth}
    \centering
    \includegraphics[width=\linewidth]{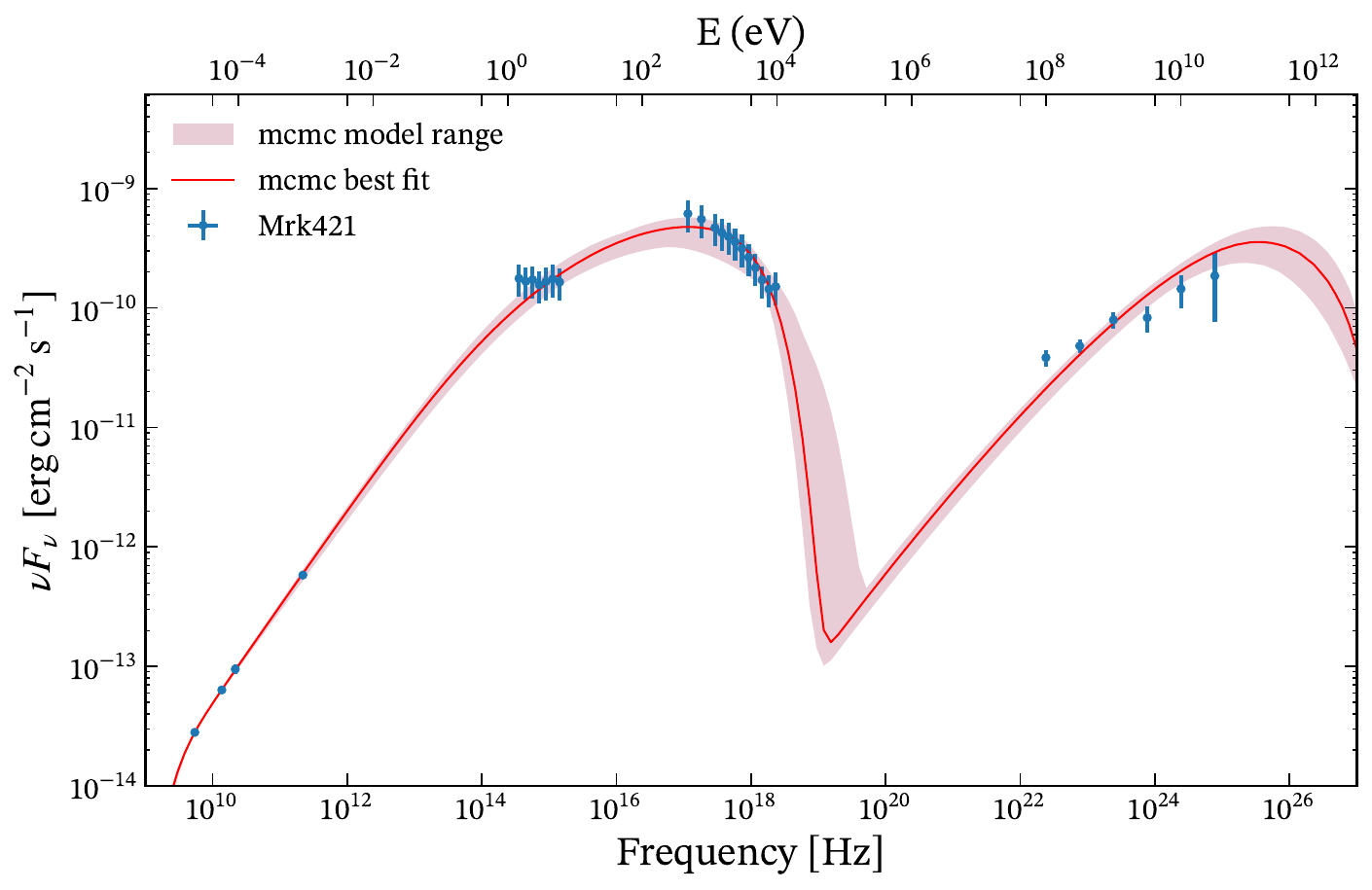}
    \small Mrk~421 -- 02008199
\end{minipage}
\hfill
\begin{minipage}{0.32\textwidth}
    \centering
    \includegraphics[width=\linewidth]{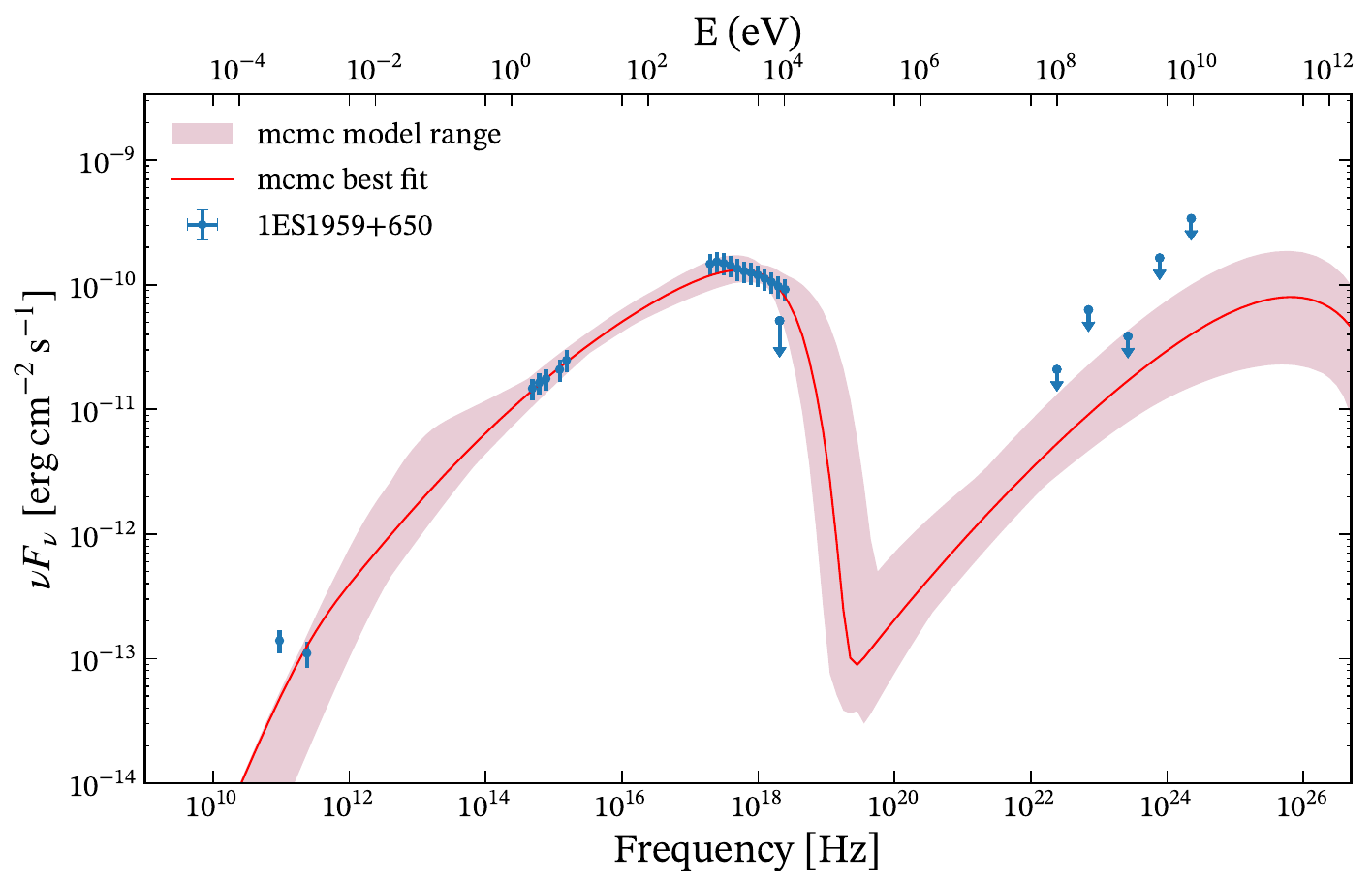}
    \small 1ES~1959$+$650 -- 01006201
\end{minipage}
\hfill
\begin{minipage}{0.32\textwidth}
    \centering
    \includegraphics[width=\linewidth]{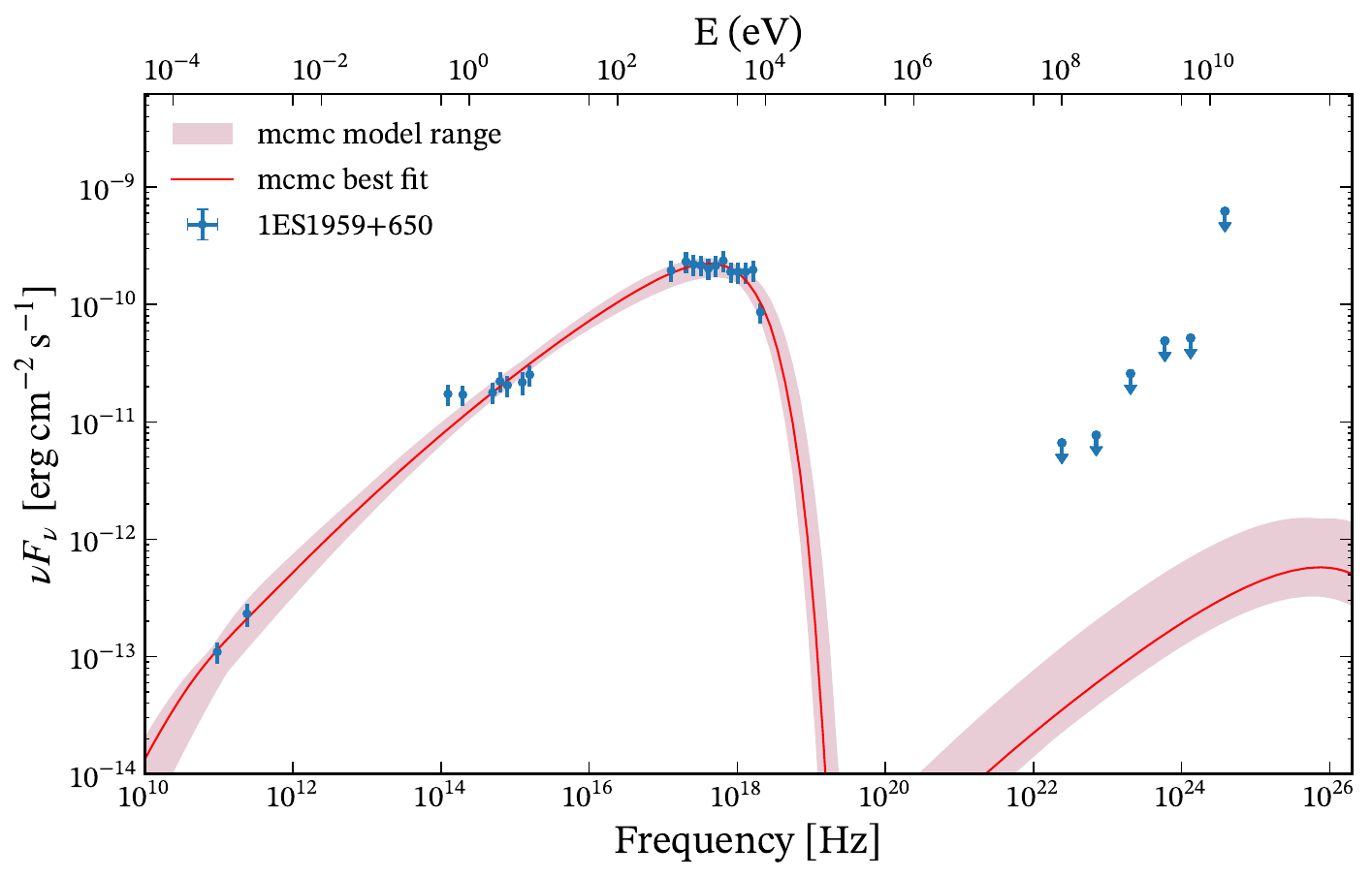}
    \small 1ES~1959$+$650 -- 01006001
\end{minipage}

\vspace{0.4cm}

\begin{minipage}{0.32\textwidth}
    \centering
    \includegraphics[width=\linewidth]{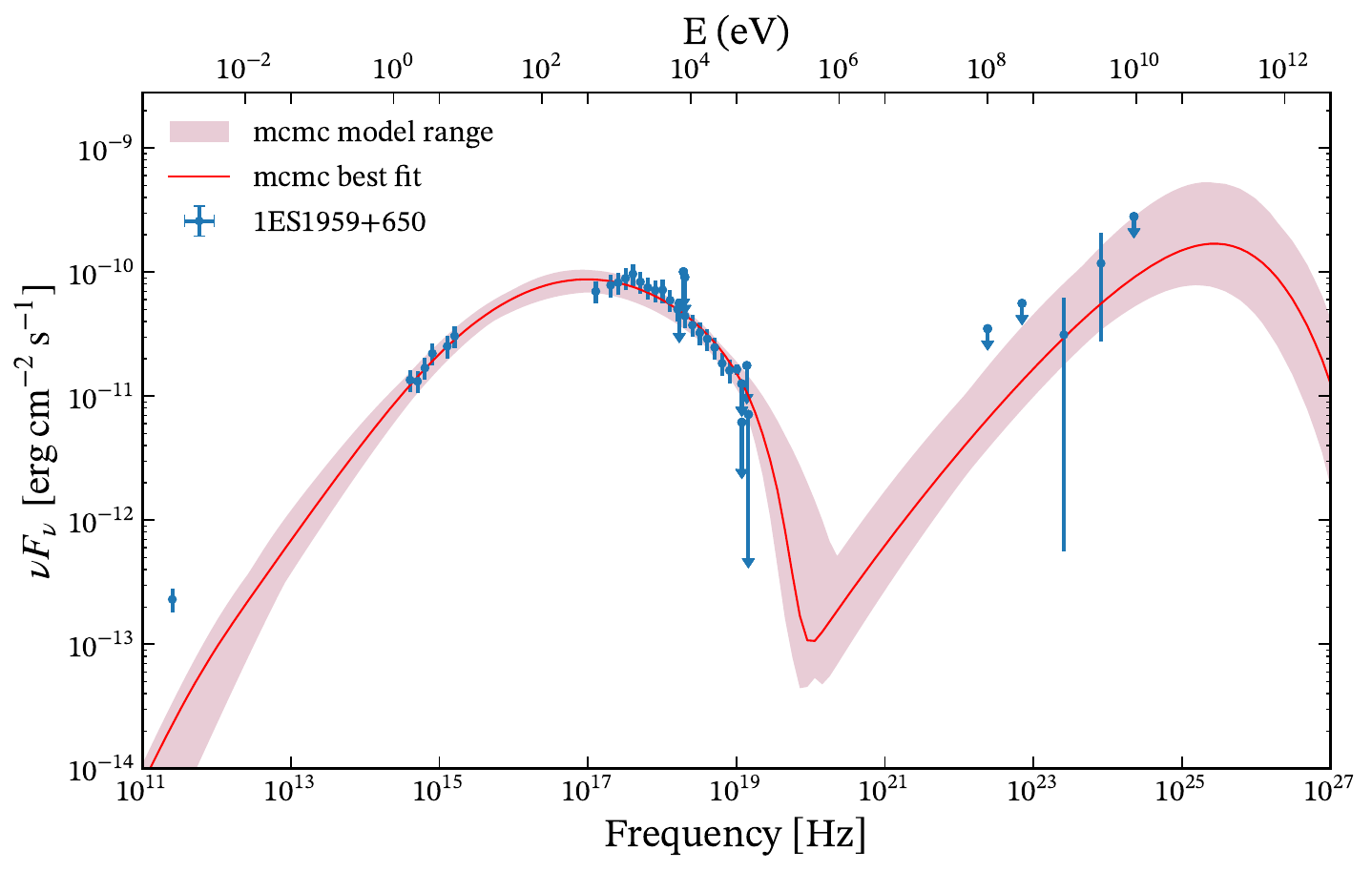}
    \small 1ES~1959$+$650 -- 02004801
\end{minipage}
\hfill
\begin{minipage}{0.32\textwidth}
    \centering
    \includegraphics[width=\linewidth]{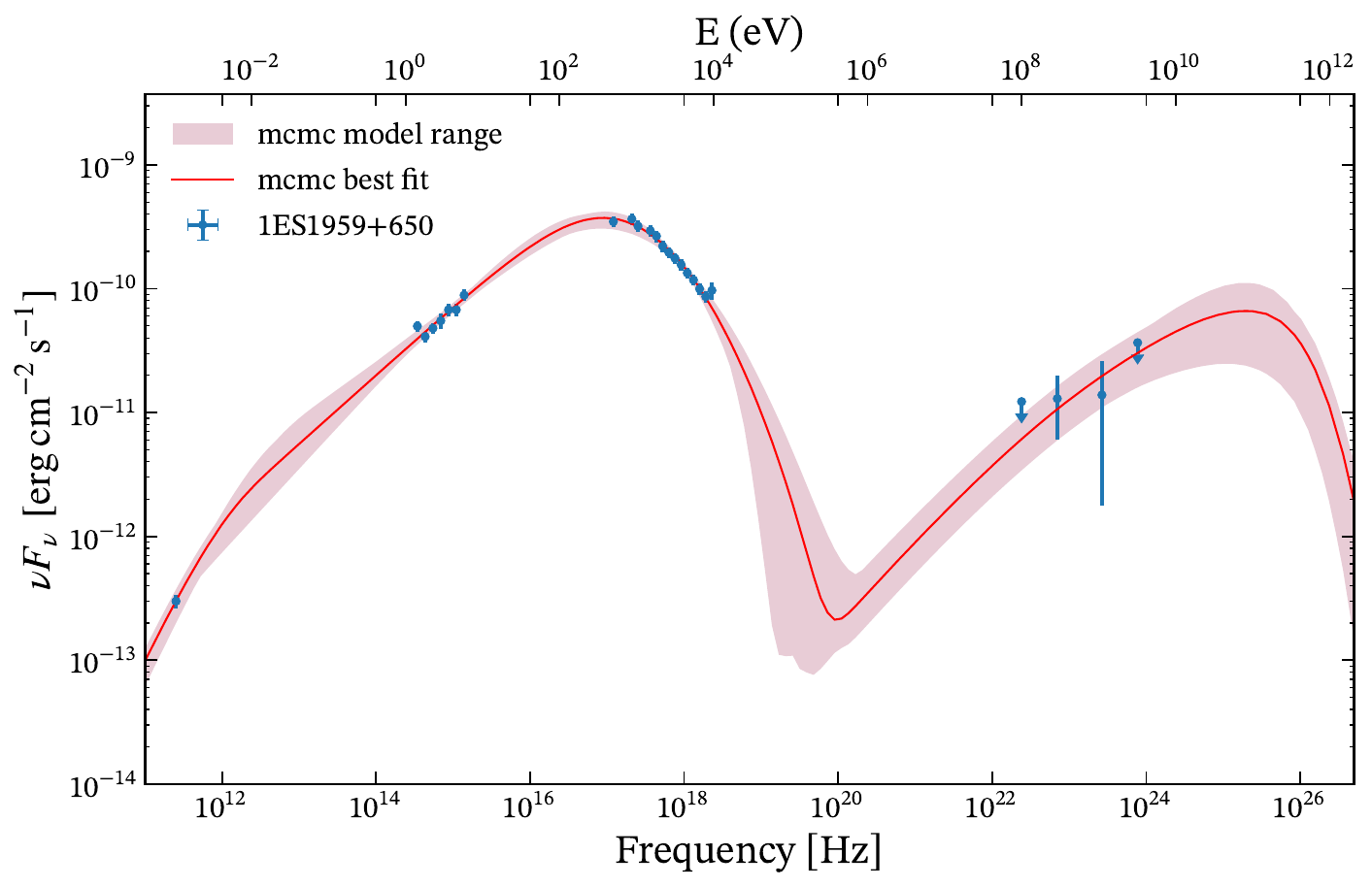}
    \small 1ES~1959$+$650 -- 02250801
\end{minipage}
\hfill
\begin{minipage}{0.32\textwidth}
    \centering
    \includegraphics[width=\linewidth]{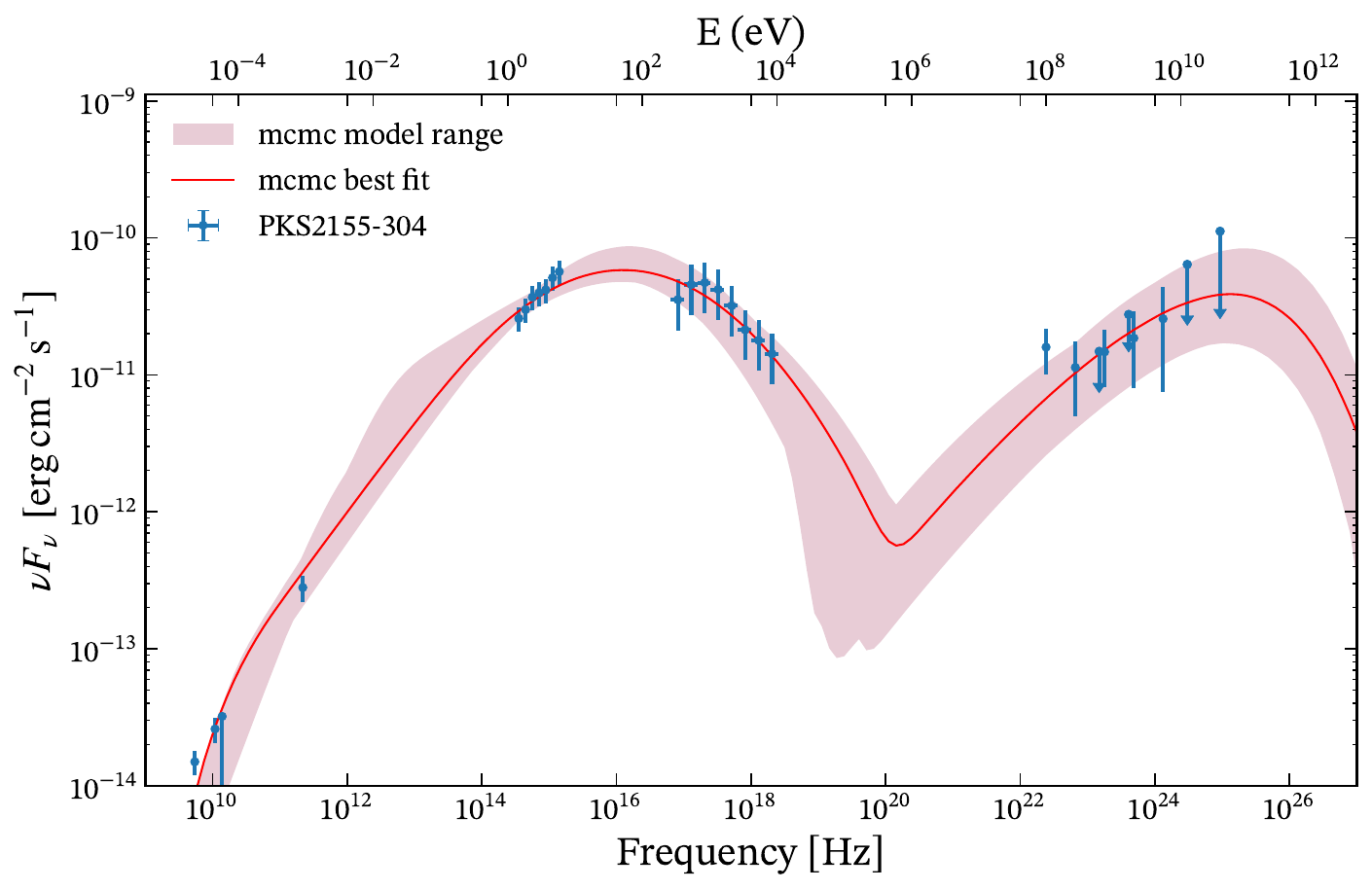}
    \small PKS~2155$-$304 -- 02005601
\end{minipage}

\vspace{0.4cm}

\begin{minipage}{0.32\textwidth}
    \centering
    \includegraphics[width=\linewidth]{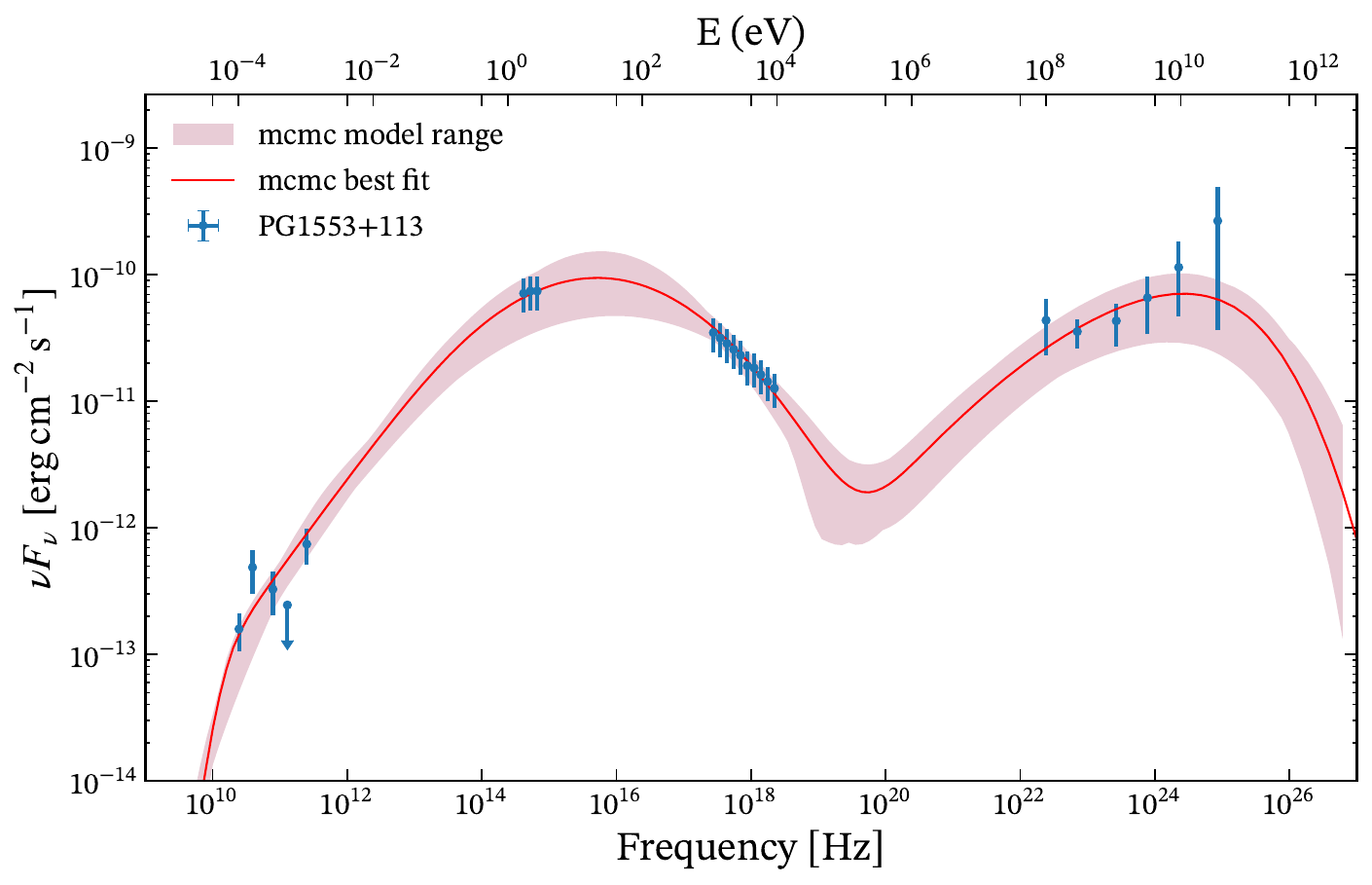}
    \small PG~1553$+$113 -- 02004999
\end{minipage}
\hfill
\begin{minipage}{0.32\textwidth}
    \centering
    \includegraphics[width=\linewidth]{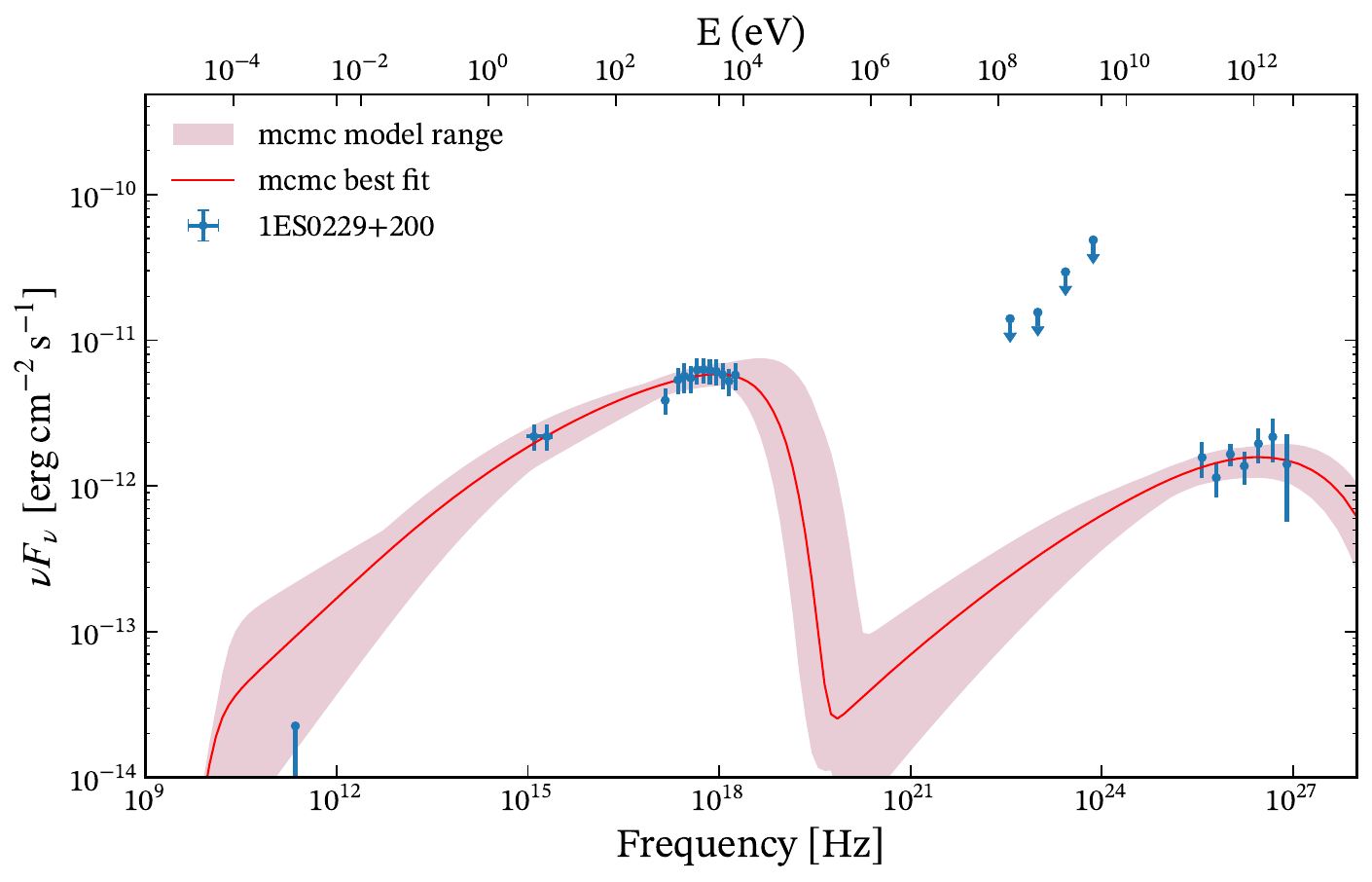}
    \small 1ES~0229$+$200 -- 01006499
\end{minipage}
\hfill
\begin{minipage}{0.32\textwidth}
    \centering
    \includegraphics[width=\linewidth]{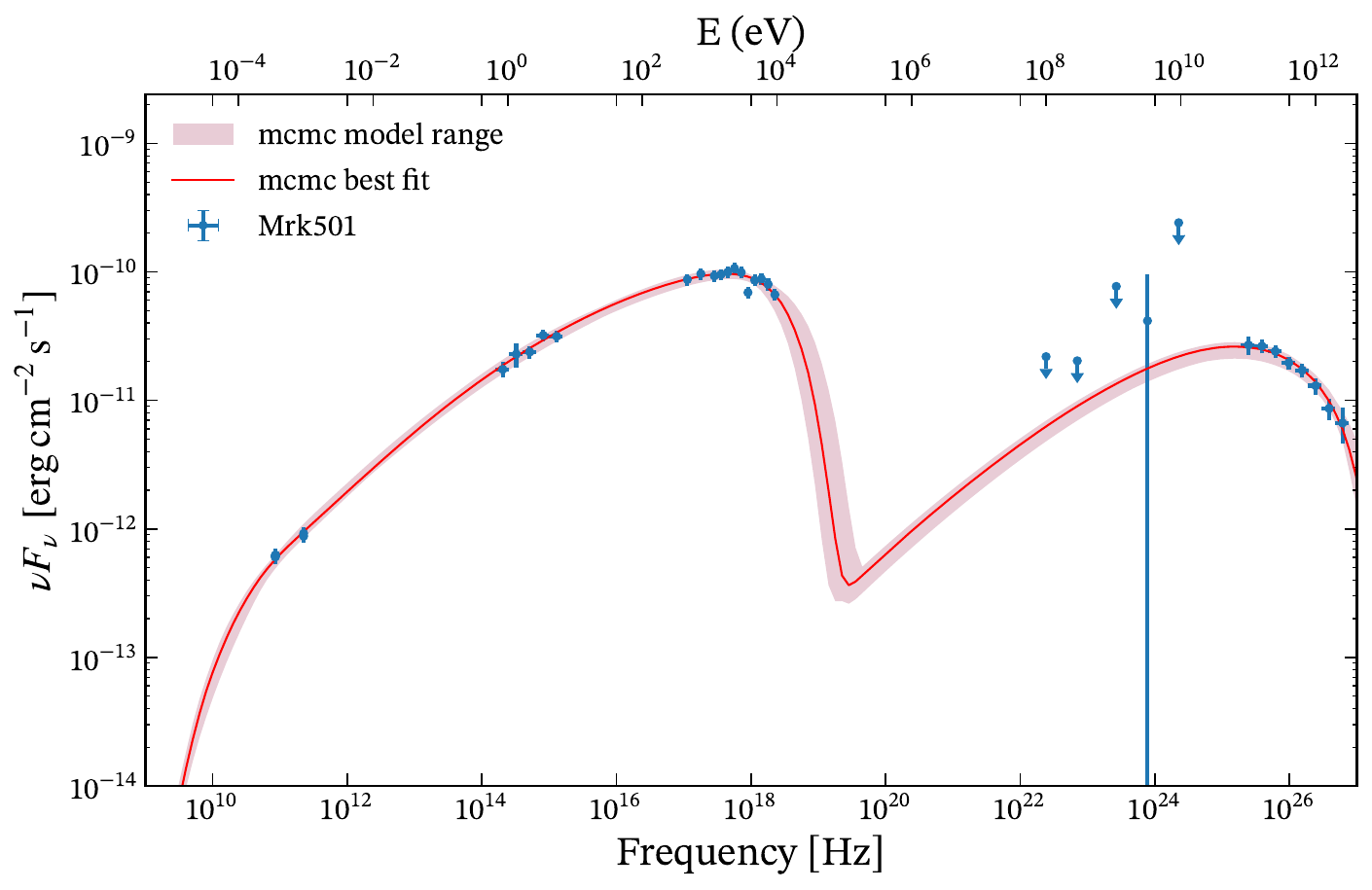}
    \small Mrk~501 -- 01004501
\end{minipage}

\vspace{0.4cm}

\begin{minipage}{0.32\textwidth}
    \centering
    \includegraphics[width=\linewidth]{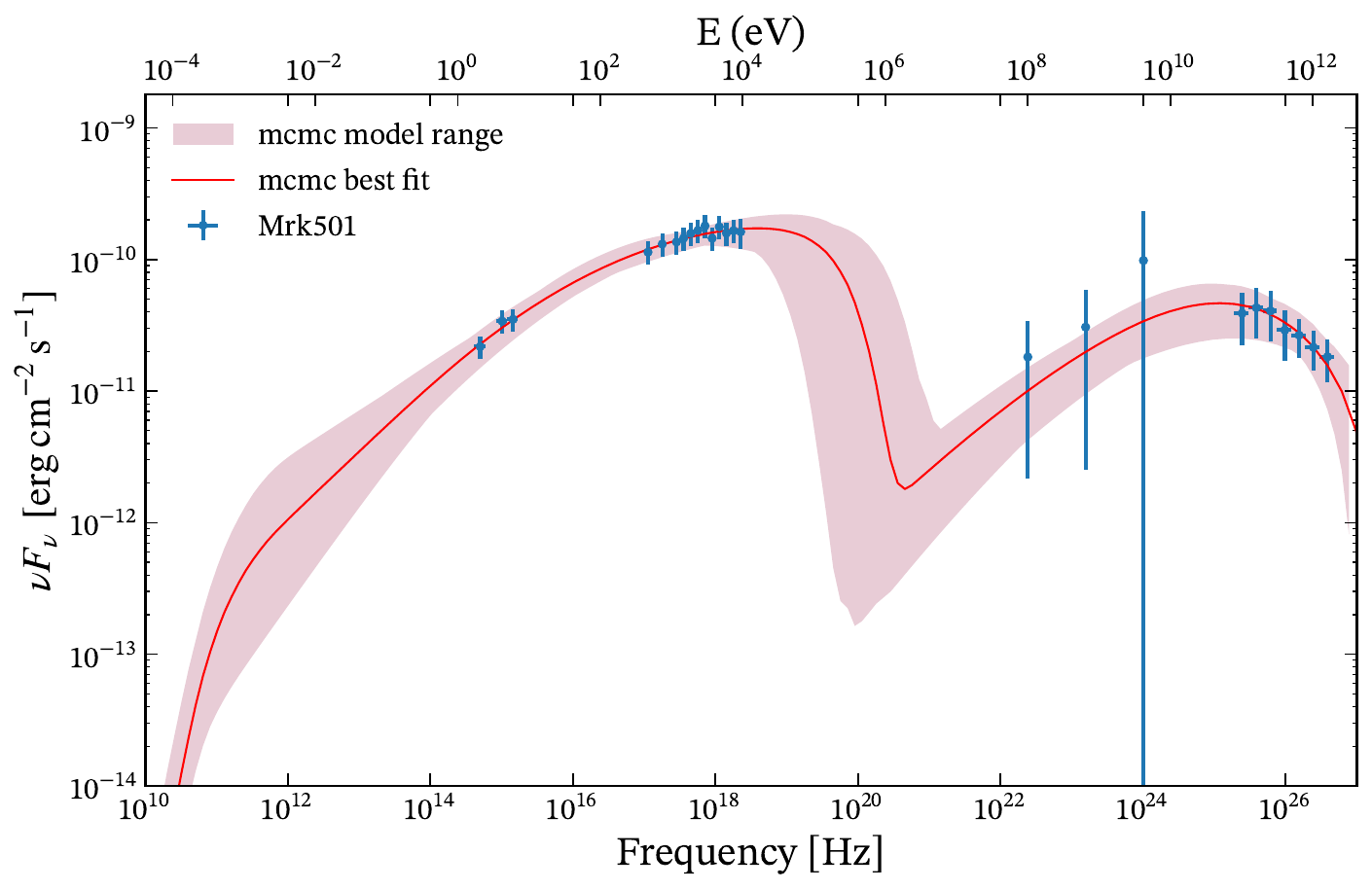}
    \small Mrk~501 -- 01004601
\end{minipage}
\hfill
\begin{minipage}{0.32\textwidth}
    \centering
    \includegraphics[width=\linewidth]{Images/model_501_3.pdf}
    \small Mrk~501 -- 01004701
\end{minipage}
\hfill
\begin{minipage}{0.32\textwidth}
    \centering
    \includegraphics[width=\linewidth]{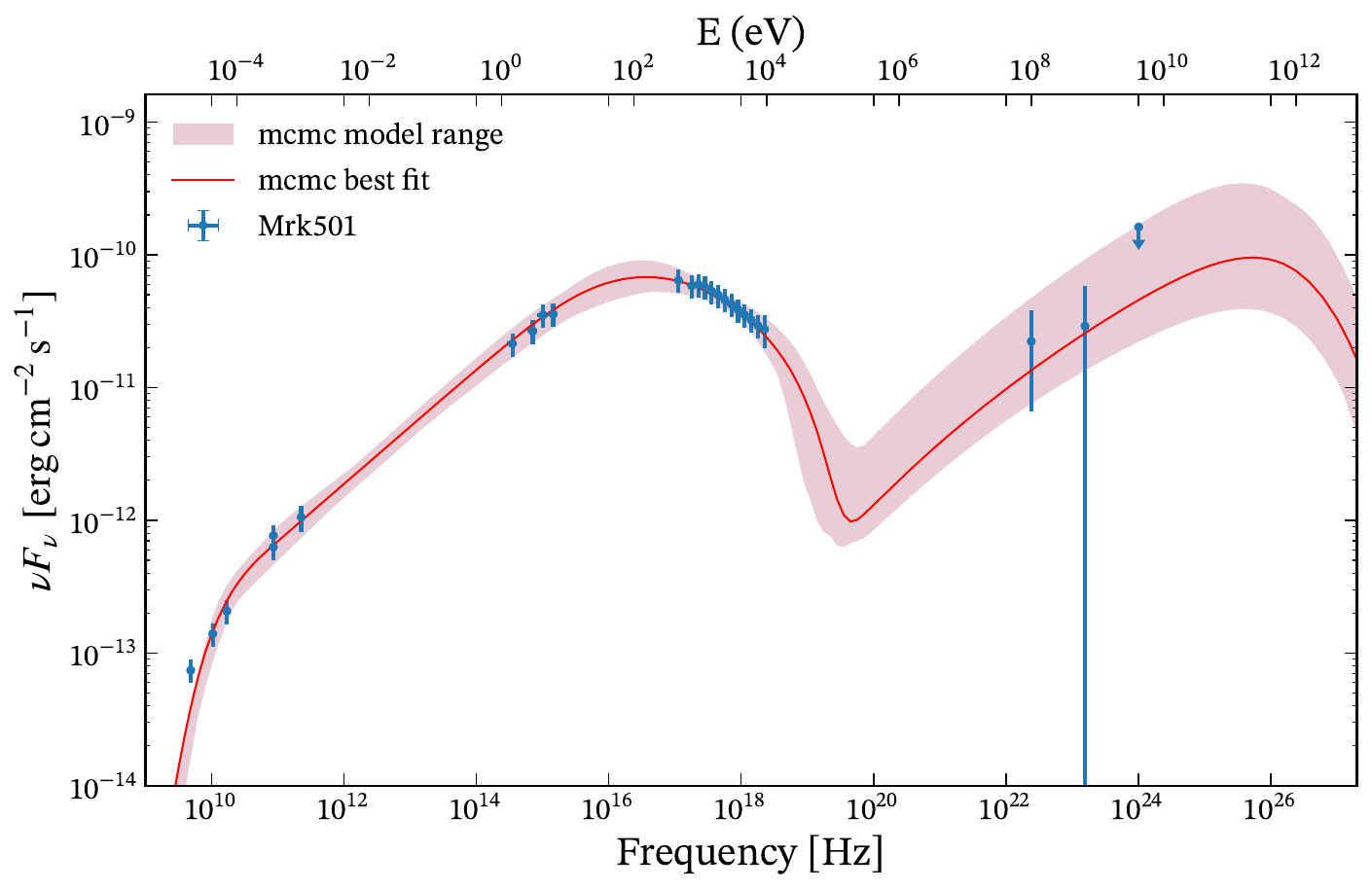}
    \small Mrk~501 -- 02004601
\end{minipage}

\end{figure*}

\begin{figure*}[h!]
\centering

\begin{minipage}{0.32\textwidth}
    \centering
    \includegraphics[width=\linewidth]{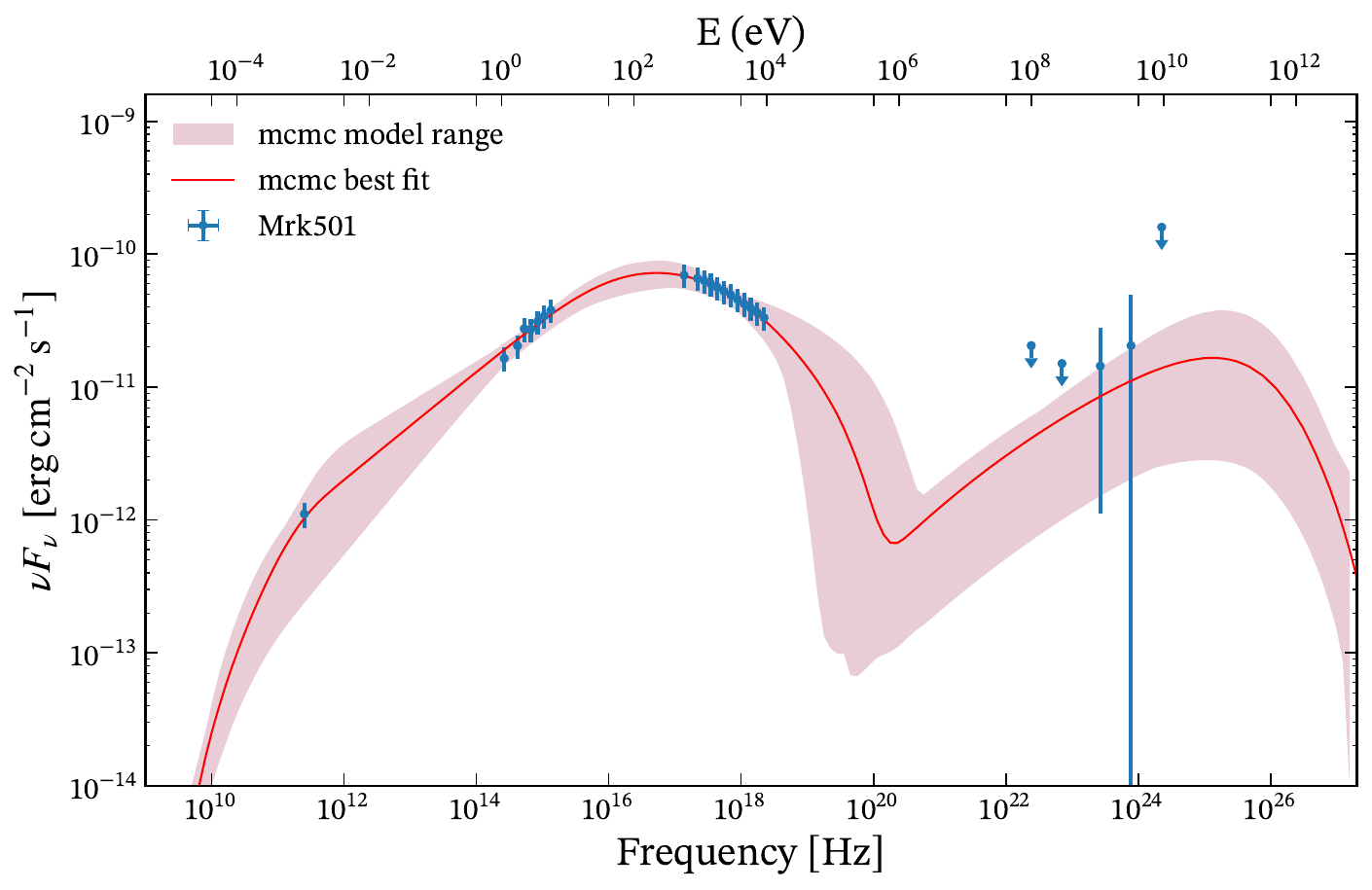}
    \small Mrk~501 -- 02004501
\end{minipage}
\hfill
\begin{minipage}{0.32\textwidth}
    \centering
    \includegraphics[width=\linewidth]{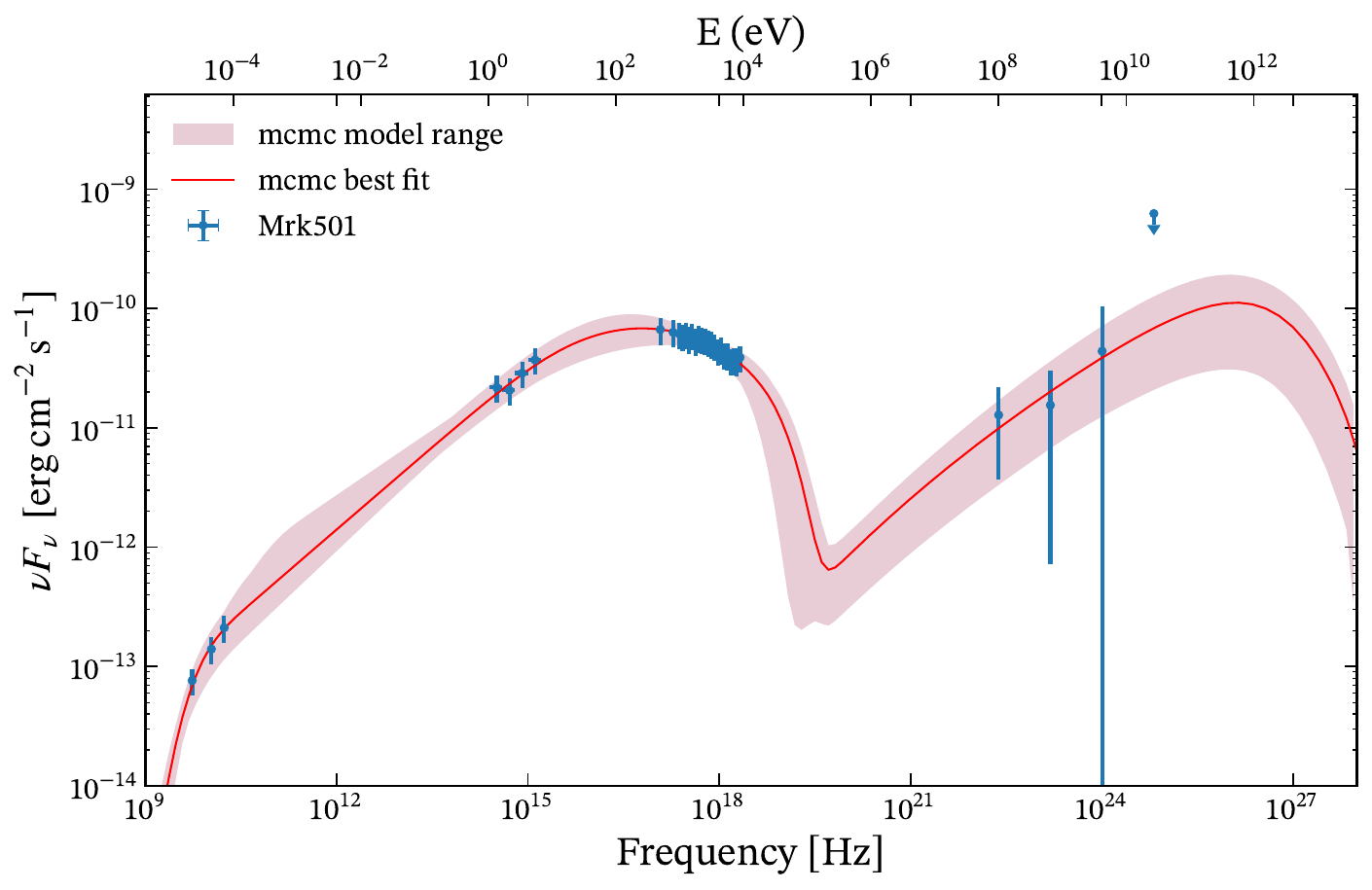}
    \small Mrk~501 -- 02004701
\end{minipage}
\hfill
\begin{minipage}{0.32\textwidth}
    \centering
    \includegraphics[width=\linewidth]{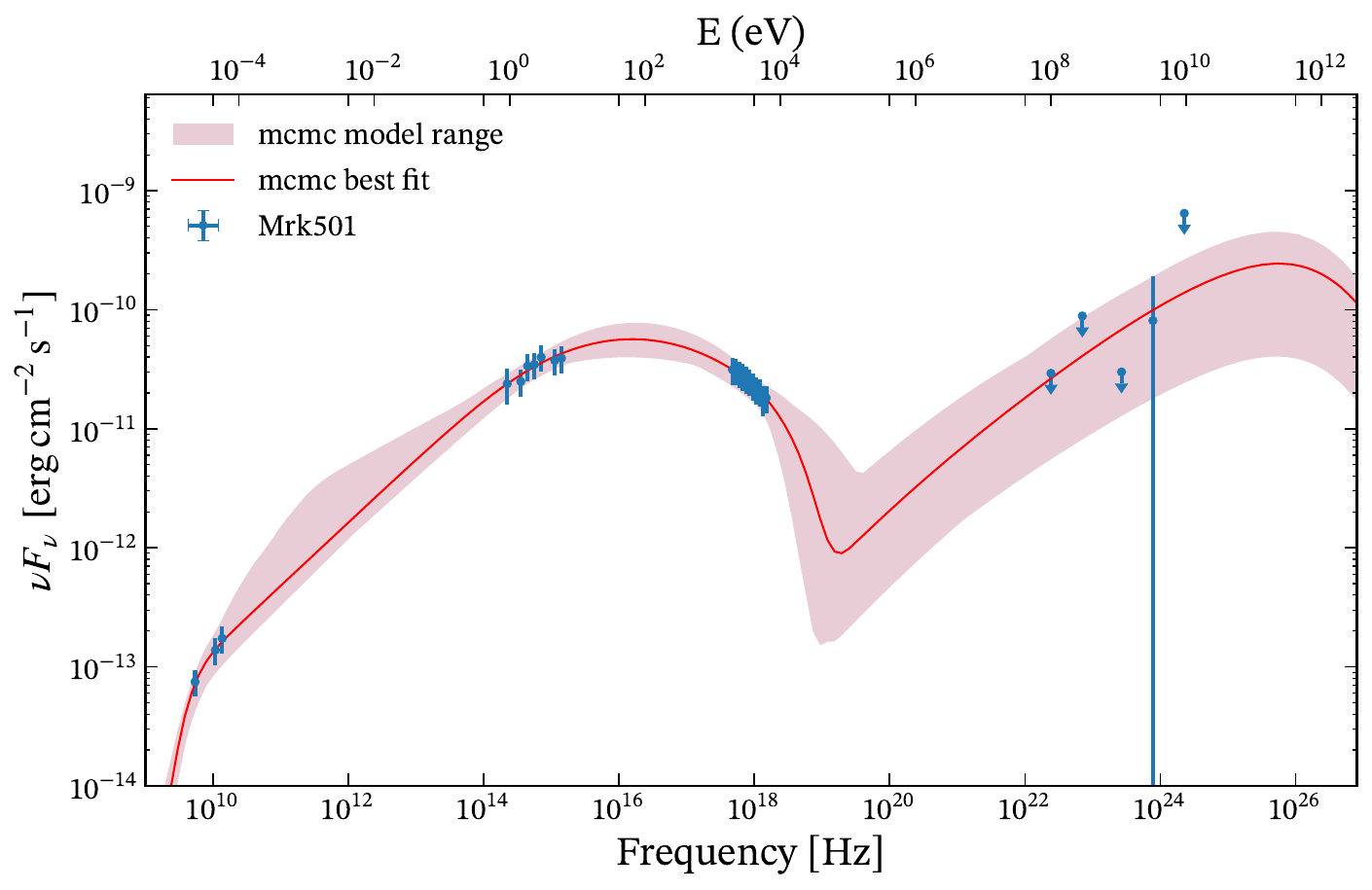}
    \small Mrk~501 -- 03008801
\end{minipage}

\vspace{0.4cm}

\begin{minipage}{0.32\textwidth}
    \centering
    \includegraphics[width=\linewidth]{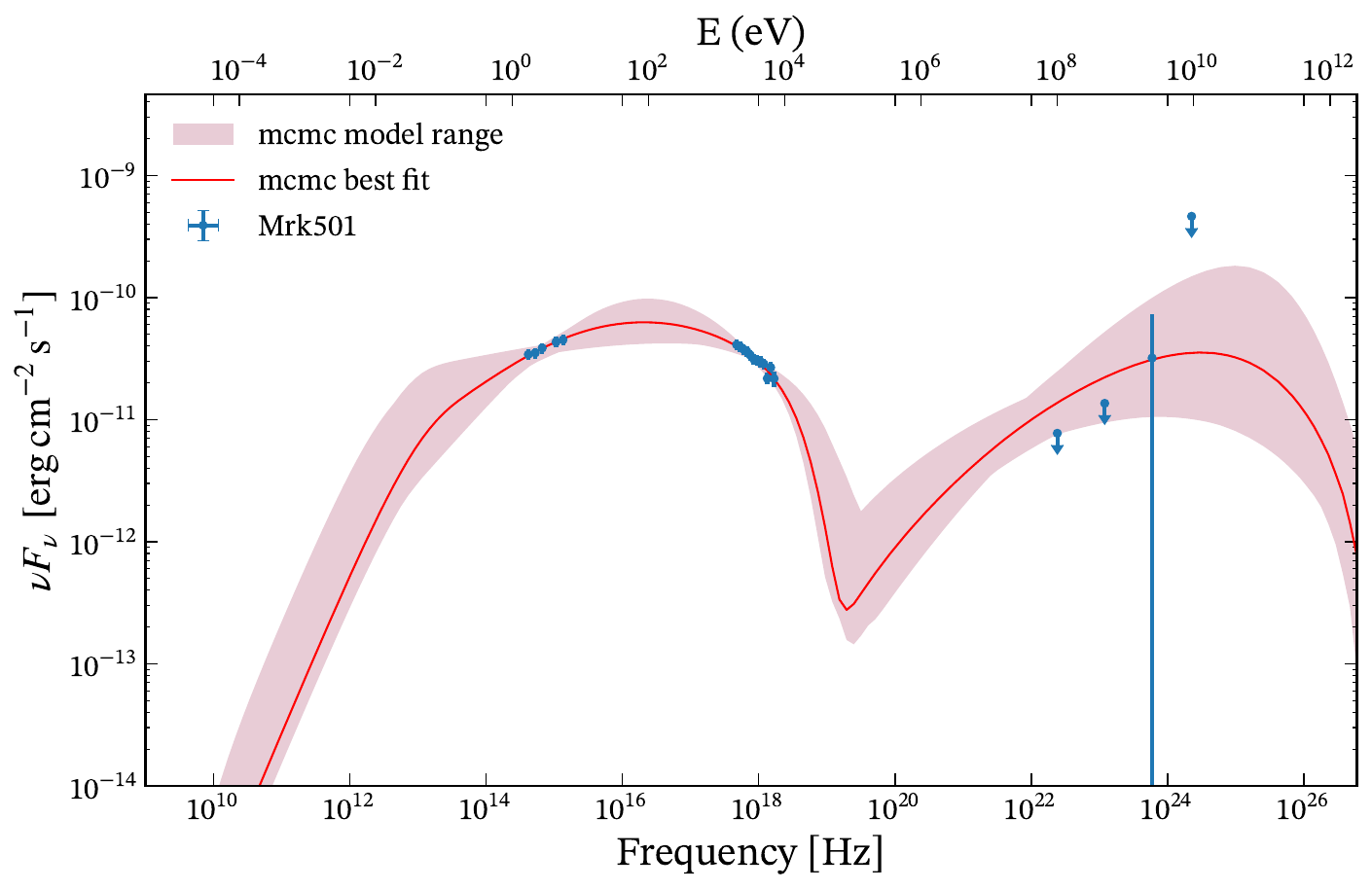}
    \small Mrk~501 -- 03008901
\end{minipage}
\hfill
\begin{minipage}{0.32\textwidth}
    \centering
    \includegraphics[width=\linewidth]{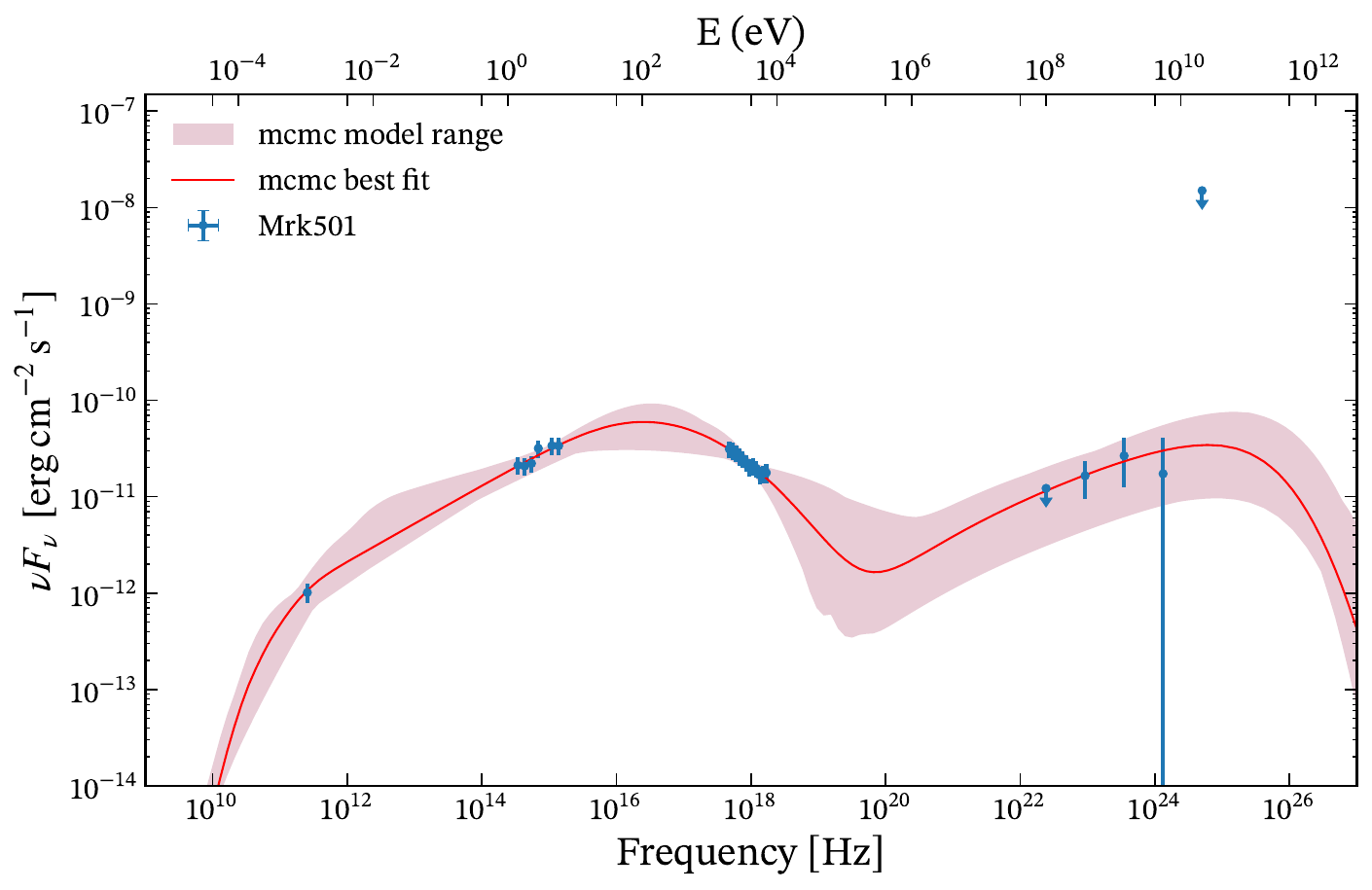}
    \small Mrk~501 -- 03009001
\end{minipage}
\hfill
\begin{minipage}{0.32\textwidth}
    \centering
    \includegraphics[width=\linewidth]{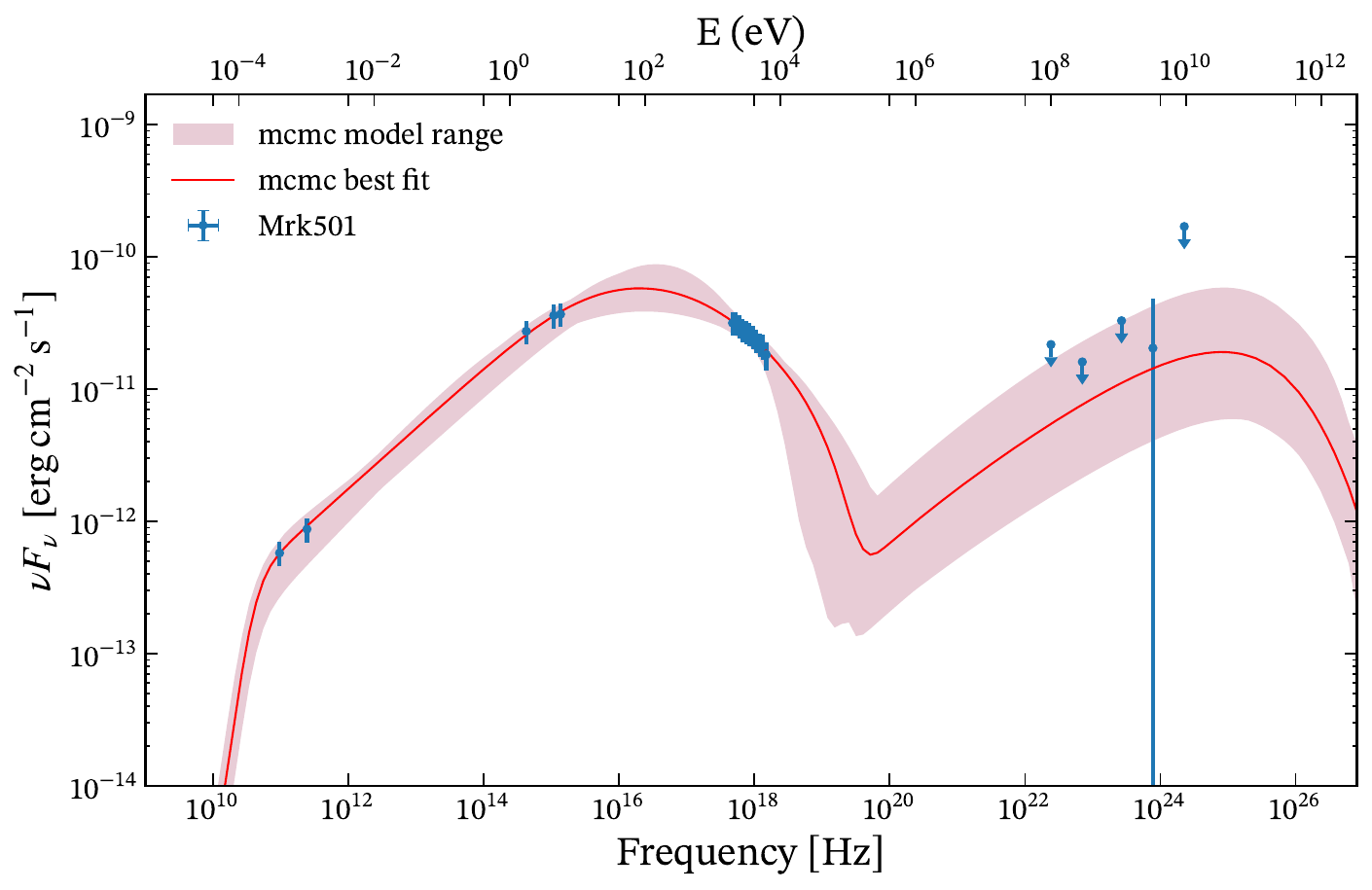}
    \small Mrk~501 -- 03009101
\end{minipage}
\caption{Best fits obtained by JetSeT. Source and obsID reported in the caption. The red line represents the fit, the red shaded area the 1$\sigma$ envelope, and the blue dots represent the data, with arrows indicating upper limits.}

\end{figure*}

\clearpage
\section{Median PD compared to model parameters}
\label{appendix_medianpd}

\begin{figure*}[h!]
    \centering

    \begin{subfigure}[t]{0.32\textwidth}
        \centering
        \includegraphics[width=\linewidth]{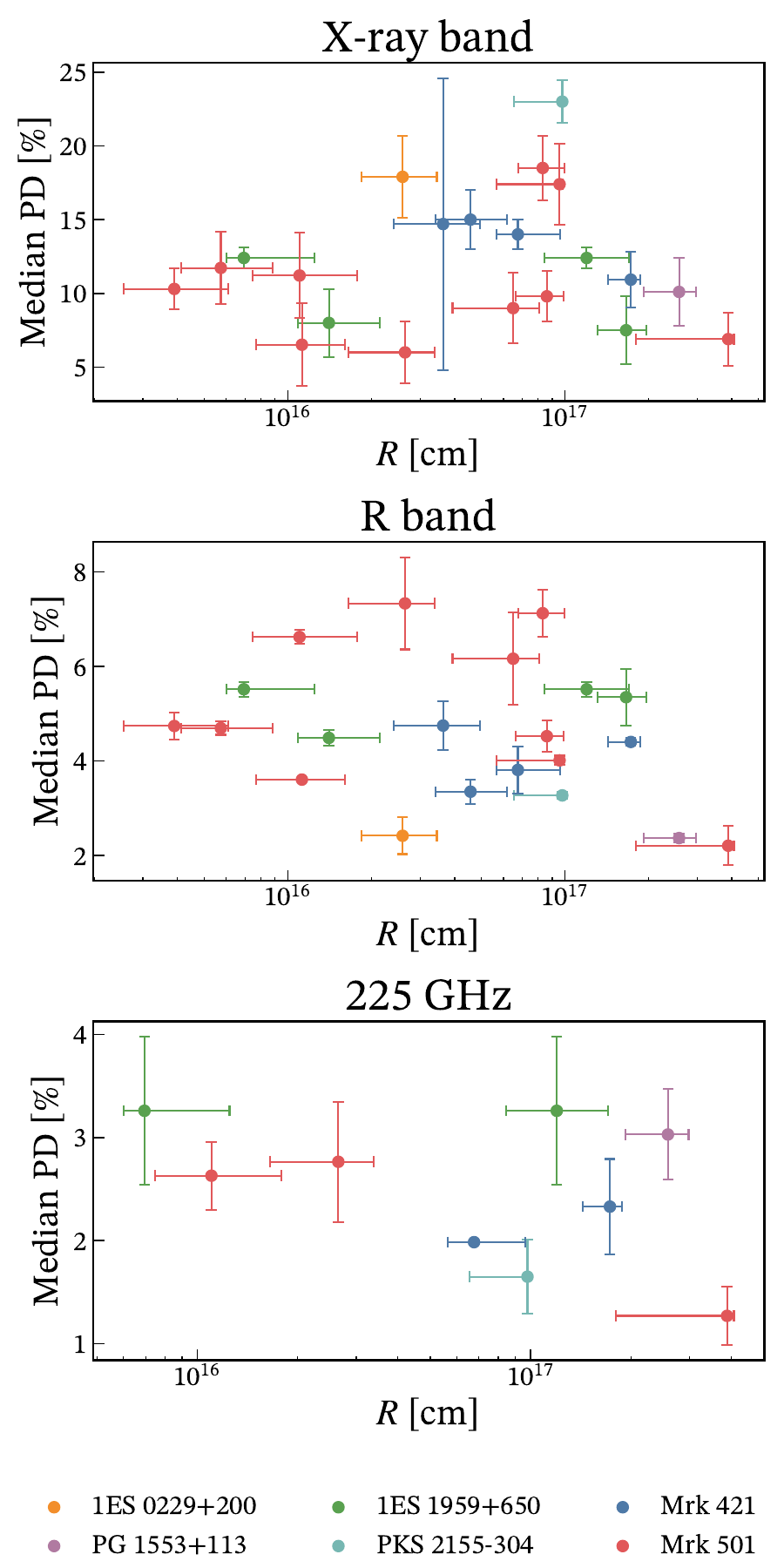}
        \caption{Radius $R$.}
        \label{fig:R_onezone}
    \end{subfigure}
    \hfill
    \begin{subfigure}[t]{0.32\textwidth}
        \centering
        \includegraphics[width=\linewidth]{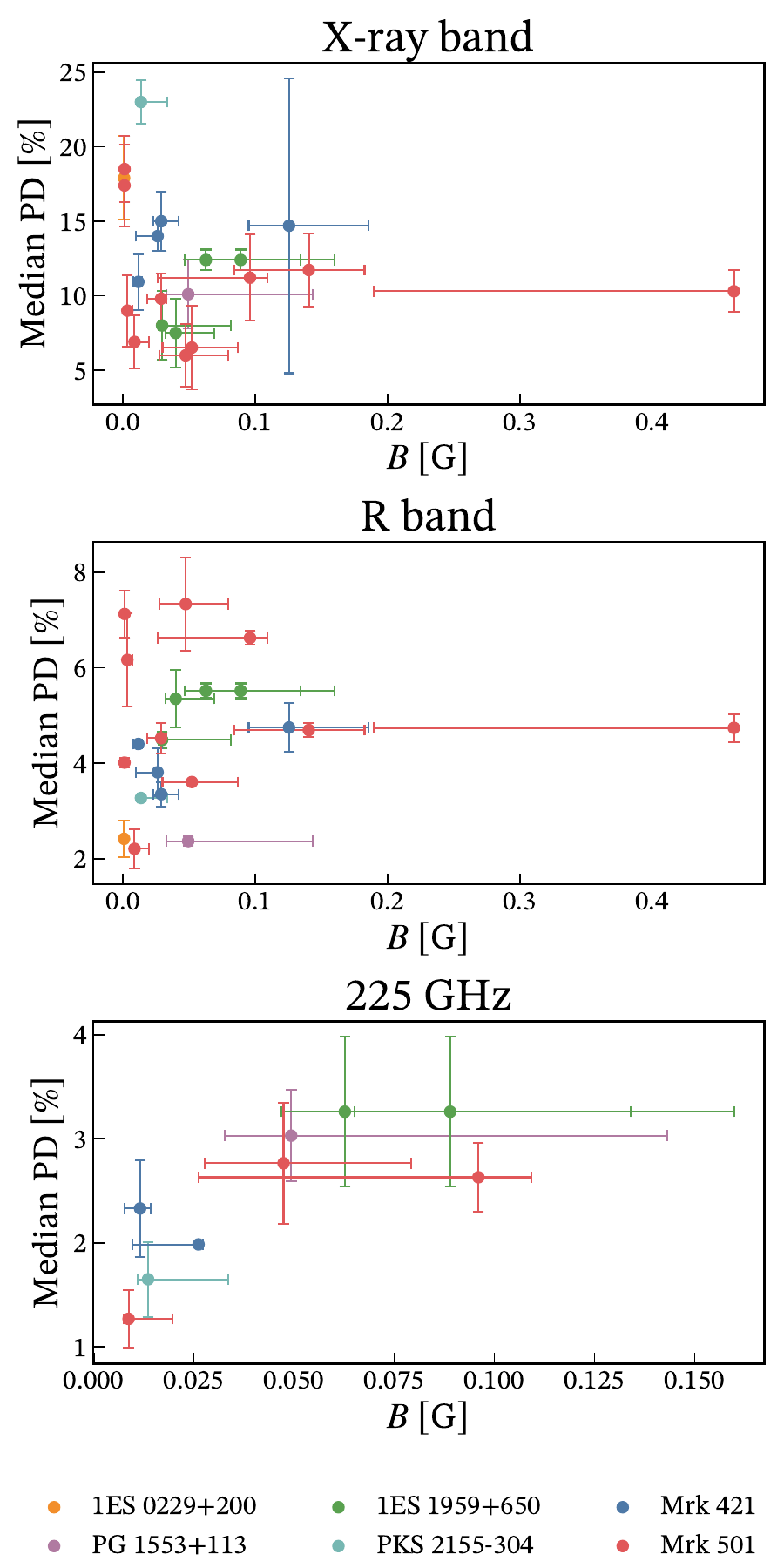}
        \caption{Magnetic field $B$.}
        \label{fig:B_onezone}
    \end{subfigure}
    \hfill
    \begin{subfigure}[t]{0.32\textwidth}
        \centering
        \includegraphics[width=\linewidth]{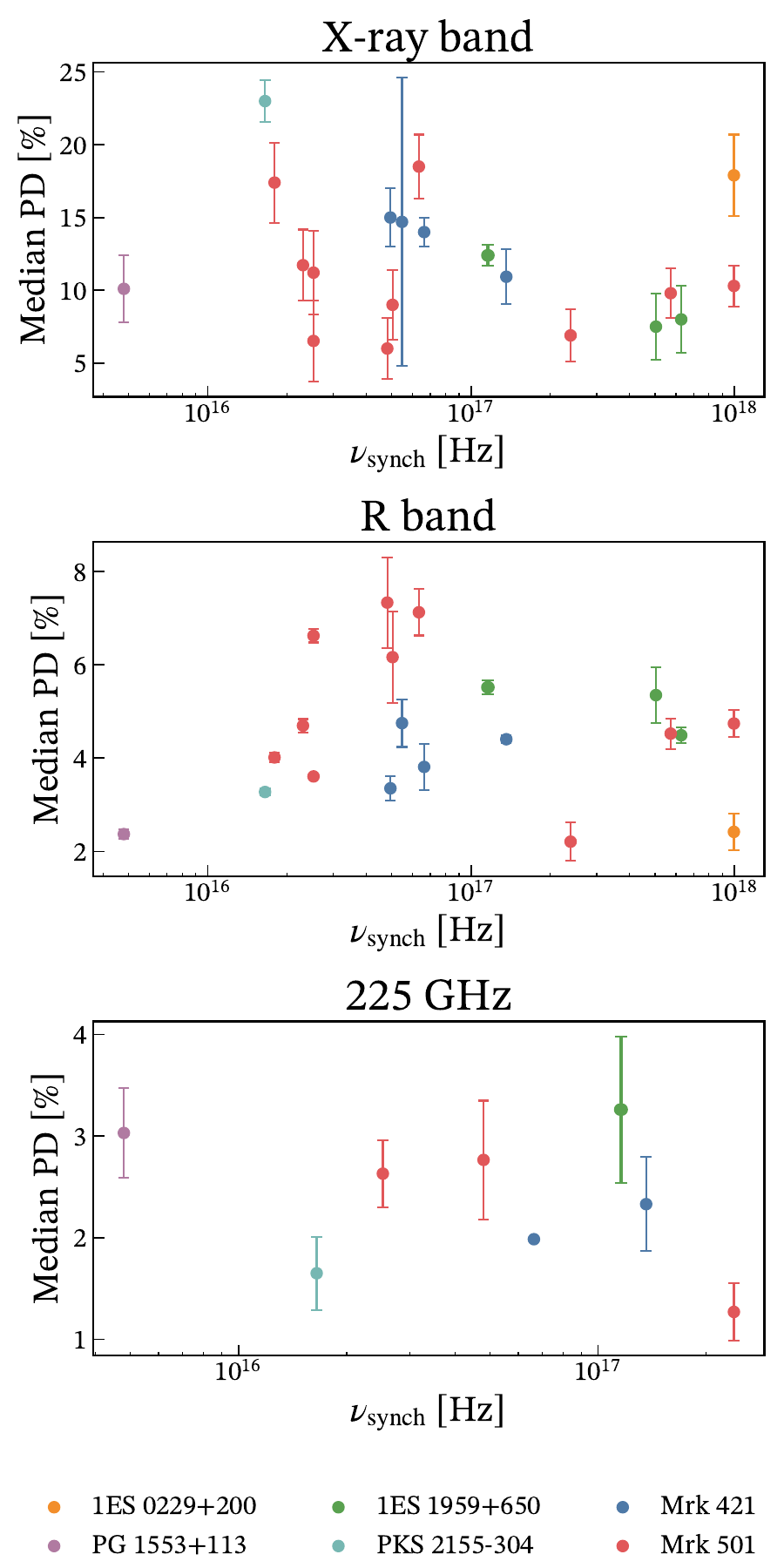}
        \caption{Synchrotron peak frequency $\nu_{\rm synch}$.}
        \label{fig:nu_onezone}
    \end{subfigure}

    \caption{Median polarisation degree in the X-ray, R, and 225 GHz bands (top to bottom within each panel) as a function of the physical parameters inferred from the one-zone model: (a) emitting region radius $R$, (b) magnetic field strength $B$, and (c) synchrotron peak frequency $\nu_{\rm synch}$.}
    \label{fig:onezone_parameters}
\end{figure*}

\begin{figure}[h!]
    \centering

    \begin{subfigure}[t]{0.32\textwidth}
        \centering
        \includegraphics[width=\linewidth]{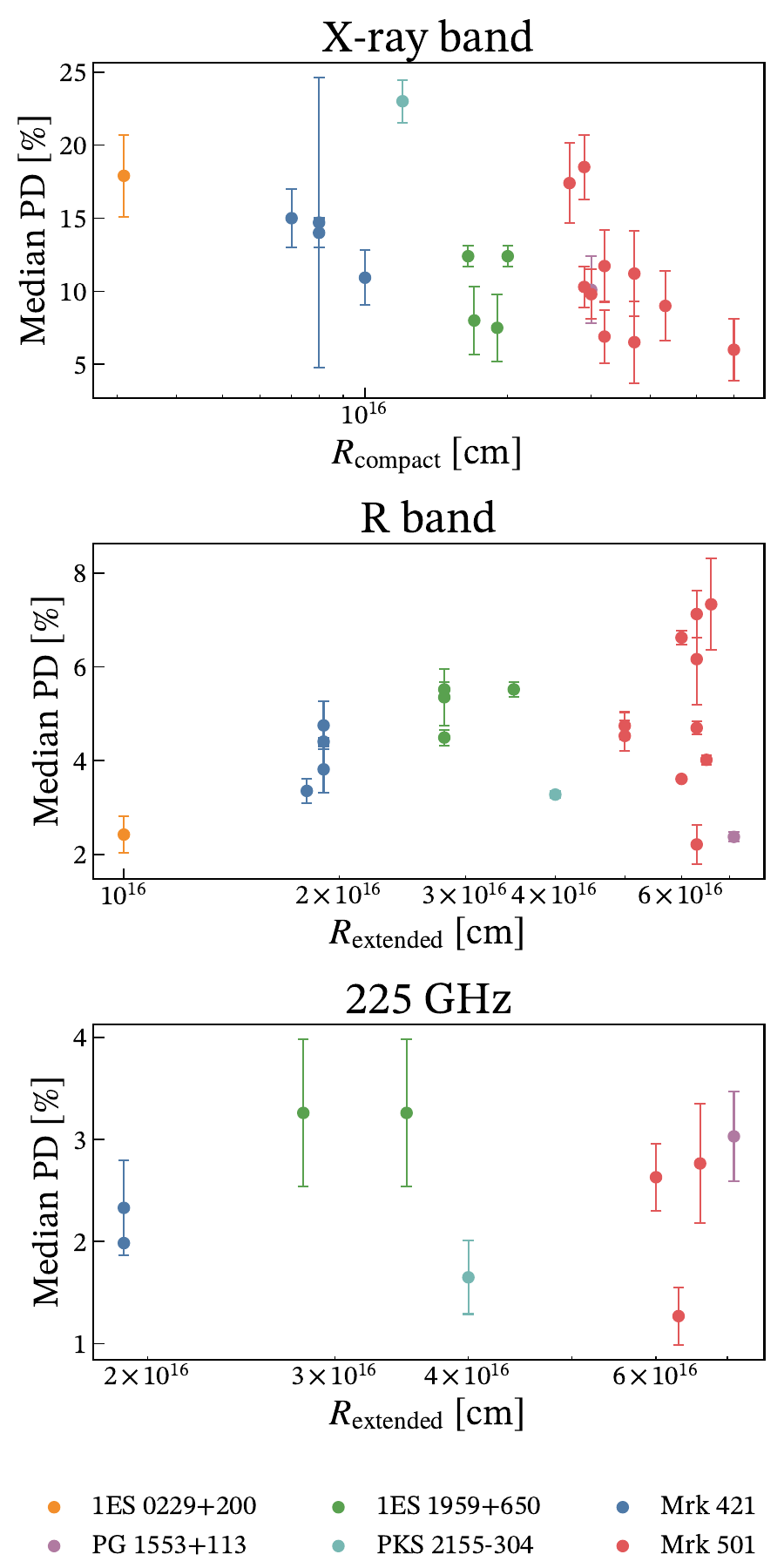}
        \caption{Radius $R$.}
        \label{fig:R_twozone}
    \end{subfigure}
    \hfill
    \begin{subfigure}[t]{0.32\textwidth}
        \centering
        \includegraphics[width=\linewidth]{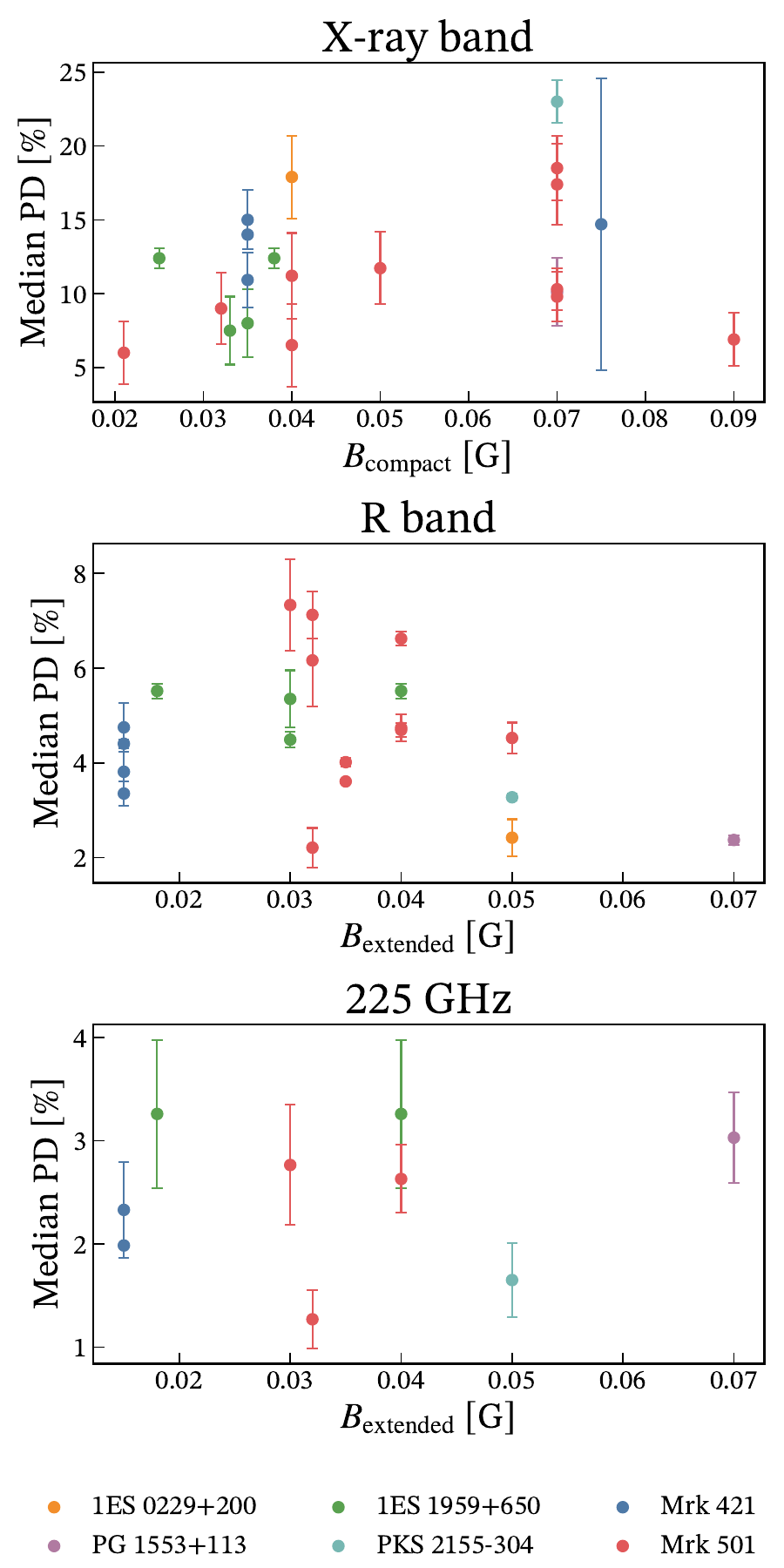}
        \caption{Magnetic field $B$.}
        \label{fig:B_twozone}
    \end{subfigure}
    \hfill
    \begin{subfigure}[t]{0.32\textwidth}
        \centering
        \includegraphics[width=\linewidth]{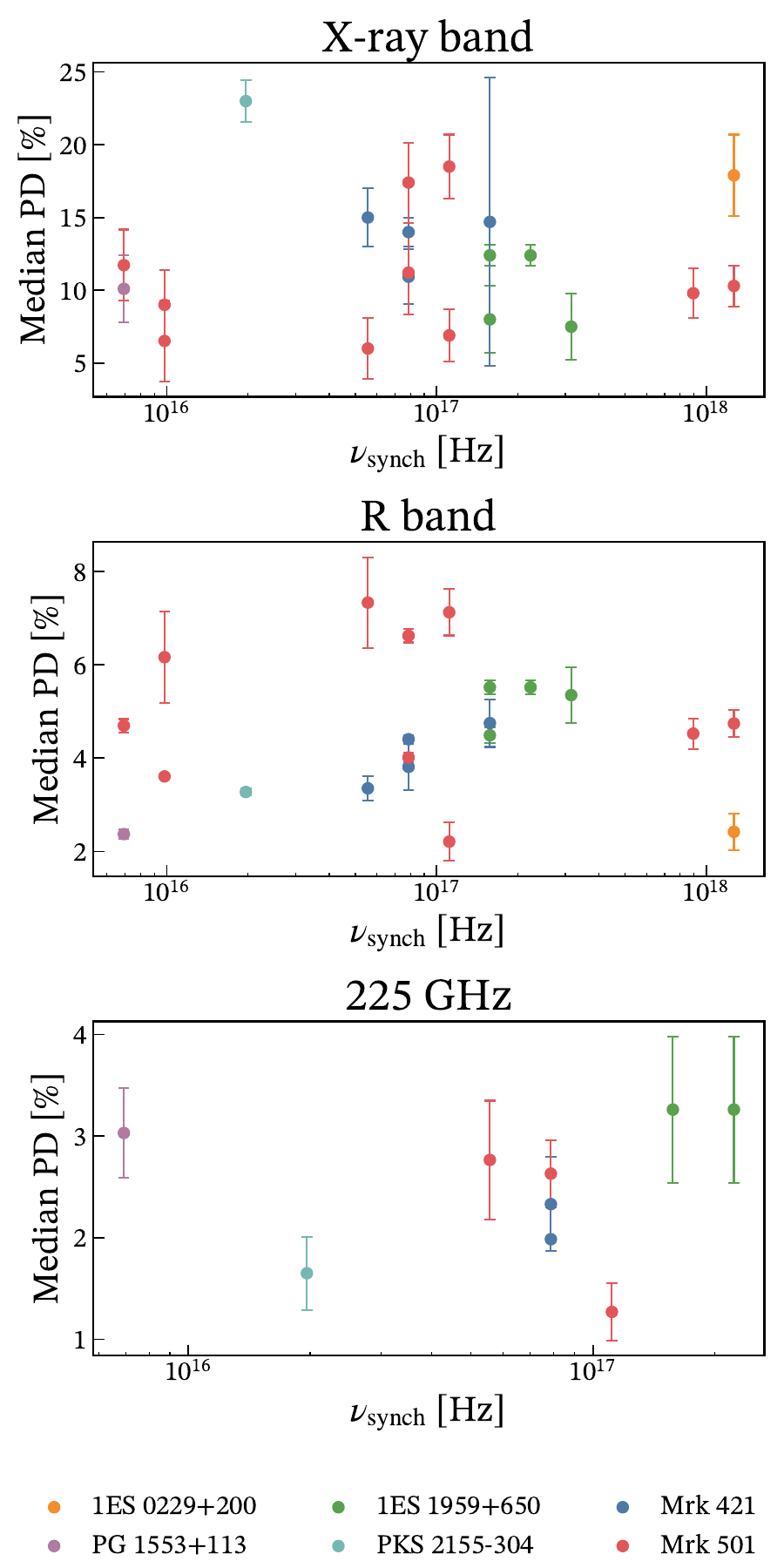}
        \caption{Synchrotron peak frequency $\nu_{\rm synch}$.}
        \label{fig:nu_twozone}
    \end{subfigure}

    \caption{Median polarisation degree in the X-ray, R, and 225 GHz bands (top to bottom within each panel) as a function of the physical parameters inferred from the two-zone model: (a) emitting region radius $R$, (b) magnetic field strength $B$, and (c) synchrotron peak frequency $\nu_{\rm synch}$. For the comparison with the X-ray PD, we used the $R$ and $B$ parameters of the compact zone; for the optical and radio PD, we used the ones of the extended zone. For $\nu_{\rm synch}$, there is no distinction between the compact and extended regions, as it is inferred from the total SED.}
    \label{fig:twozone_parameters}
\end{figure}

We report here the plots of the model parameters versus the median PD in the three considered bands: X-ray, R, and radio (225 GHz).
Figures \ref{fig:R_onezone}, \ref{fig:B_onezone}, and \ref{fig:nu_onezone} refer to the one-zone model; figures \ref{fig:R_twozone}, \ref{fig:B_twozone}, and \ref{fig:nu_twozone} refer to the two-zone model.
The error bars represent the 95\% confidence level on the estimate of the parameters.
The plot for $\nu_{\rm synch}$ does not present error bars on the $x$ axis, as the frequency was extracted from the fitted SED and was not estimated by the model.

\FloatBarrier
\section{Spectral curvature parameter}
\label{appendix_spectralcurv}

We report here the plot of the spectral curvature parameter ($r$) as a function of the synchrotron peak frequency.
The plot highlights a significant negative Spearman correlation ($\rho=-0.542$, $p=0.011$).

\begin{figure}[h!]
    \centering
    \includegraphics[width=0.48\linewidth]{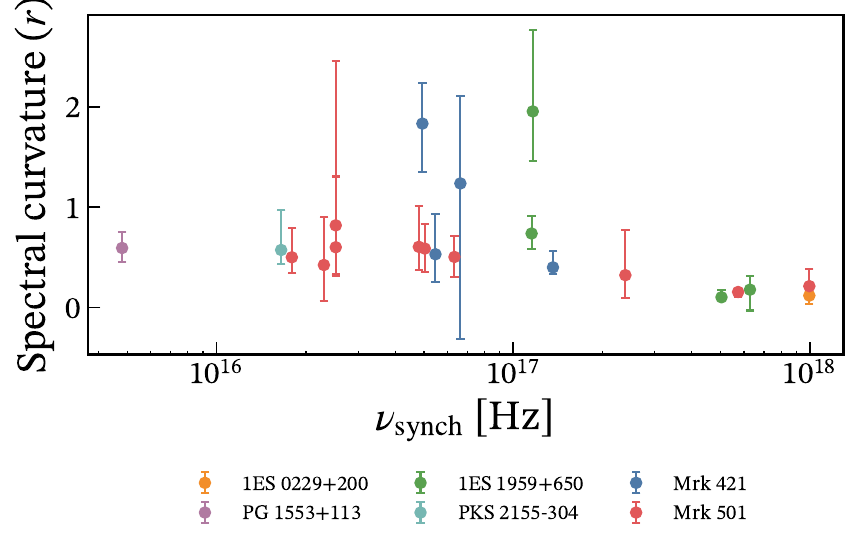}
    \caption{Spectral curvature parameter ($r$) as a function of the synchrotron peak frequency ($\nu_{\rm synch}$) for the one-zone model. Points are colour-coded by source, and the error bars indicate the 95\% MCMC confidence intervals on $r$. A significant negative Spearman correlation is found.}
    \label{fig:nu_vs_r}
\end{figure}

\clearpage
\section{Median PD ratio compared to model parameters}
\label{appendix_medianpdratio}

\begin{figure}[h!]
    \centering

    \begin{subfigure}[t]{0.32\textwidth}
        \centering
        \includegraphics[width=\linewidth]{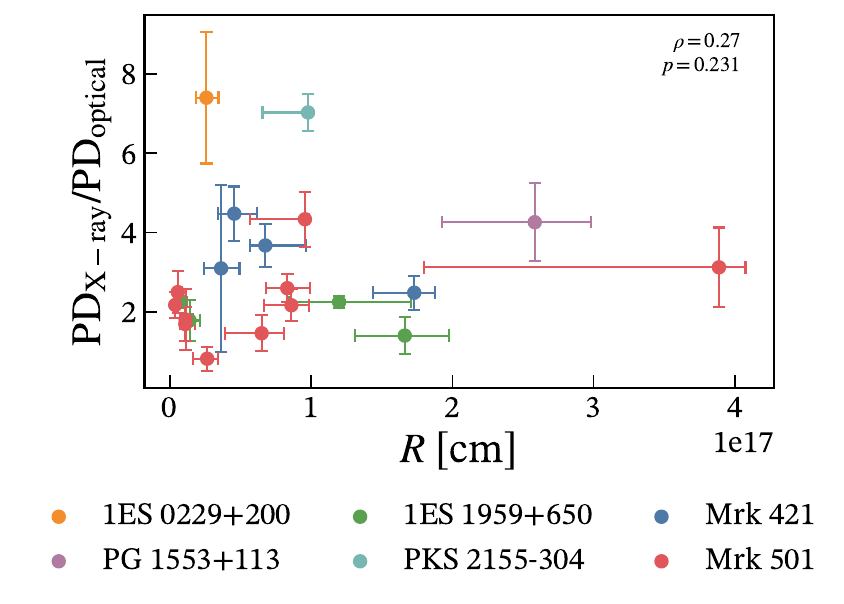}
        \caption{Radius $R$.}
    \end{subfigure}
    \hfill
    \begin{subfigure}[t]{0.32\textwidth}
        \centering
        \includegraphics[width=\linewidth]{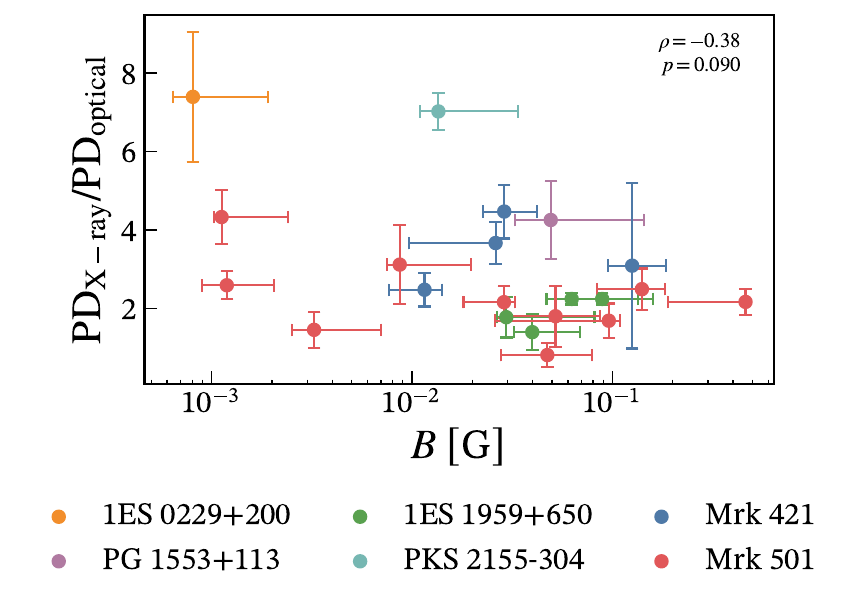}
        \caption{Magnetic field $B$.}
        \label{fig:onezone_B_ratio_opt}
    \end{subfigure}
    \hfill
    \begin{subfigure}[t]{0.32\textwidth}
        \centering
        \includegraphics[width=\linewidth]{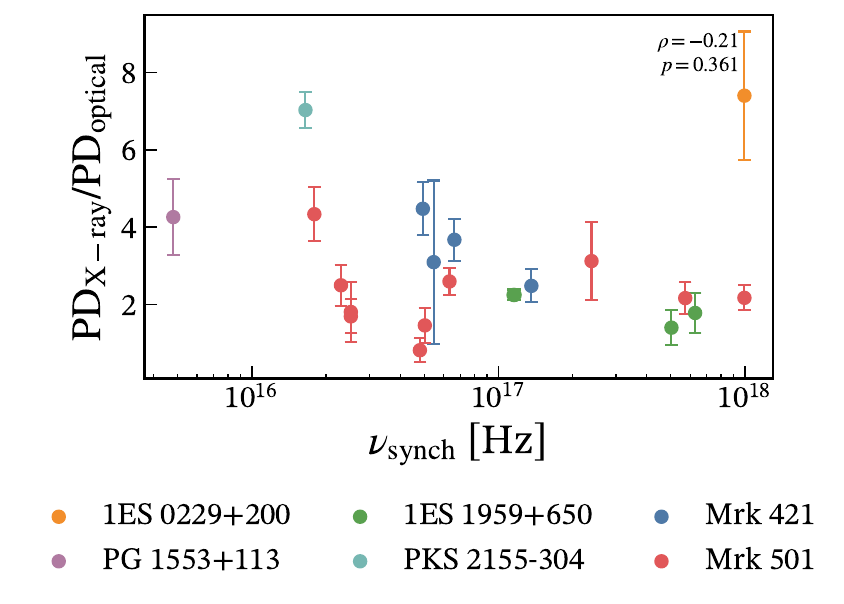}
        \caption{Synchrotron peak frequency $\nu_{\rm synch}$.}
    \end{subfigure}

    \caption{Comparison of the parameters from the one-zone model with the ratio between the median X-ray PD and the median optical PD (R band).}
    \label{fig:onezone_ratio_opt}
\end{figure}

\begin{figure}[h!]
    \centering

    \begin{subfigure}[t]{0.32\textwidth}
        \centering
        \includegraphics[width=\linewidth]{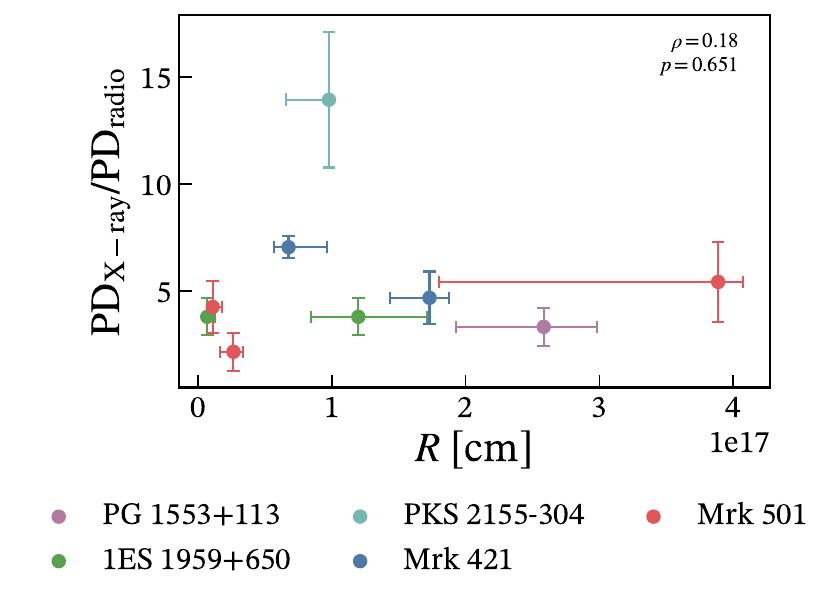}
        \caption{Radius $R$.}
    \end{subfigure}
    \hfill
    \begin{subfigure}[t]{0.32\textwidth}
        \centering
        \includegraphics[width=\linewidth]{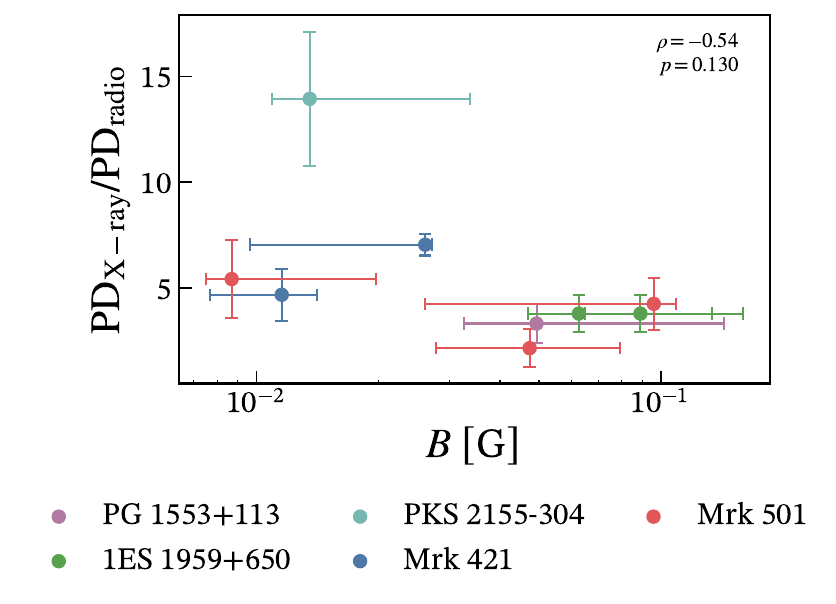}
        \caption{Magnetic field $B$.}
    \end{subfigure}
    \hfill
    \begin{subfigure}[t]{0.32\textwidth}
        \centering
        \includegraphics[width=\linewidth]{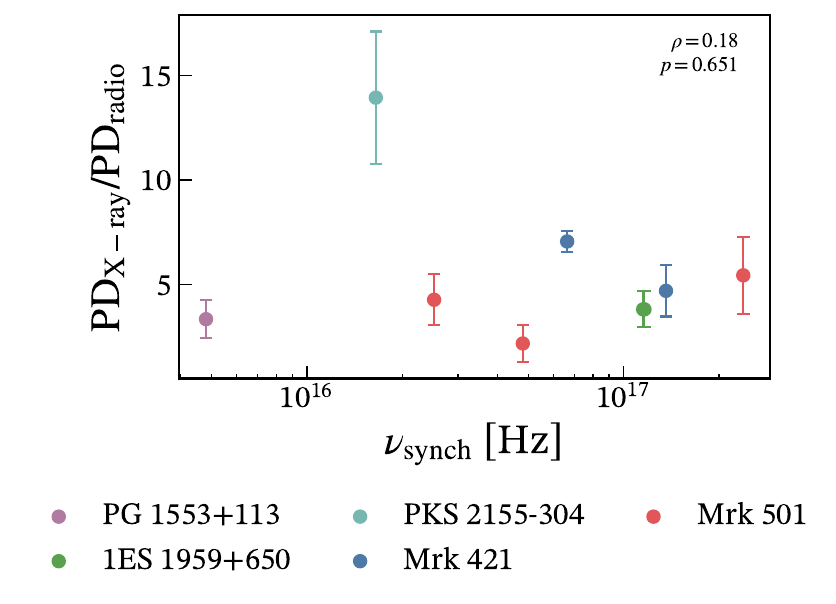}
        \caption{Synchrotron peak frequency $\nu_{\rm synch}$.}
    \end{subfigure}

    \caption{Comparison of the parameters from the one-zone model with the ratio between the median X-ray PD and the median radio PD (225 GHz).}
    \label{fig:onezone_ratio_radio}
\end{figure}

\begin{figure}[h!]
    \centering

    \begin{subfigure}[t]{0.32\textwidth}
        \centering
        \includegraphics[width=\linewidth]{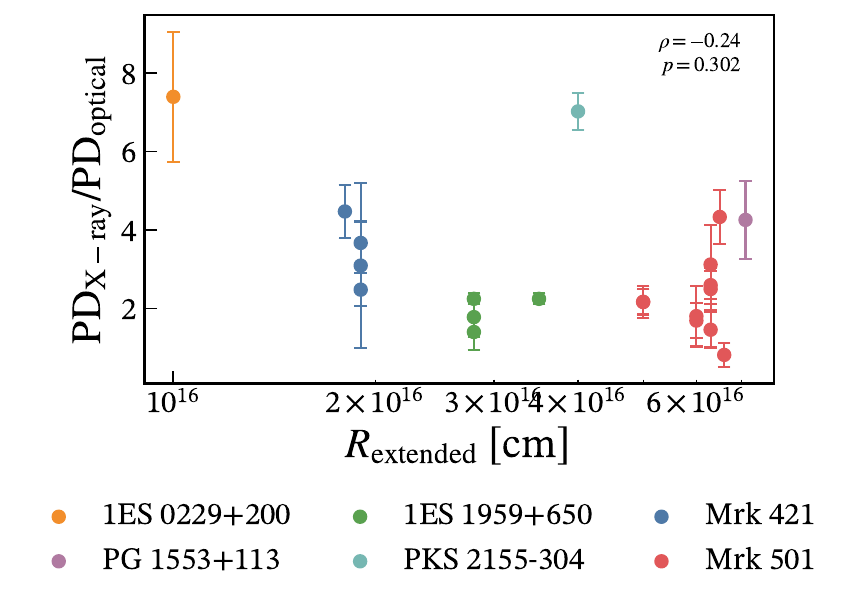}
        \caption{Radius $R$.}
    \end{subfigure}
    \hfill
    \begin{subfigure}[t]{0.32\textwidth}
        \centering
        \includegraphics[width=\linewidth]{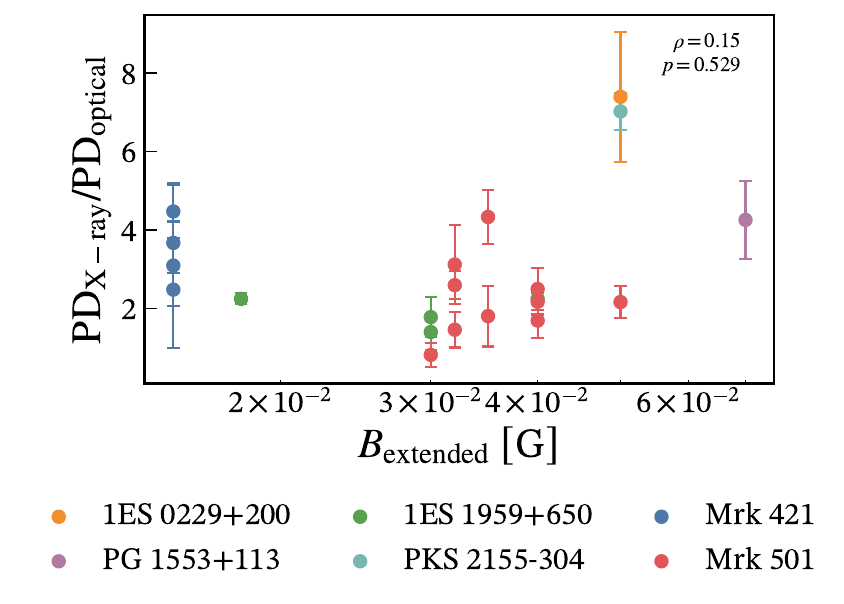}
        \caption{Magnetic field $B$.}
    \end{subfigure}
    \hfill
    \begin{subfigure}[t]{0.32\textwidth}
        \centering
        \includegraphics[width=\linewidth]{Images/nu_synch_peak_vs_X_over_R_PD_ratio.pdf}
        \caption{Synchrotron peak frequency $\nu_{\rm synch}$.}
    \end{subfigure}

    \caption{Comparison of the parameters from the two-zone model extended zone with the ratio between the median X-ray PD and the median optical PD (R band).}
    \label{fig:twozone_extended_ratio_opt}
\end{figure}

\begin{figure}[h!]
    \centering

    \begin{subfigure}[t]{0.32\textwidth}
        \centering
        \includegraphics[width=\linewidth]{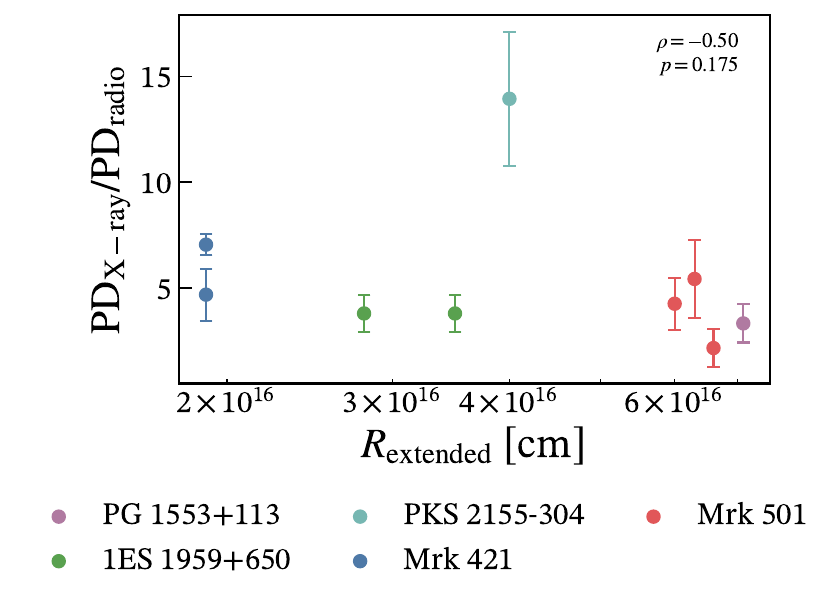}
        \caption{Radius $R$.}
    \end{subfigure}
    \hfill
    \begin{subfigure}[t]{0.32\textwidth}
        \centering
        \includegraphics[width=\linewidth]{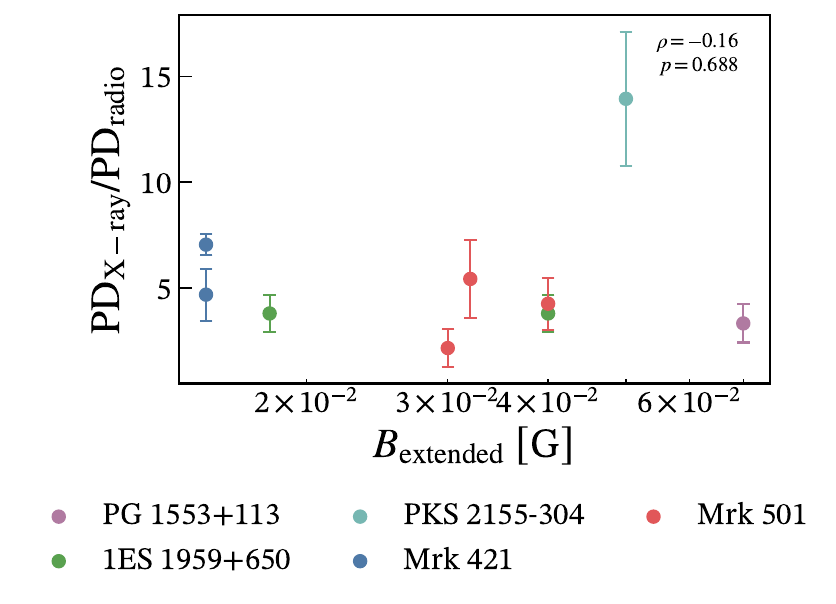}
        \caption{Magnetic field $B$.}
    \end{subfigure}
    \hfill
    \begin{subfigure}[t]{0.32\textwidth}
        \centering
        \includegraphics[width=\linewidth]{Images/nu_synch_peak_vs_X_over_225_PD_ratio.pdf}
        \caption{Synchrotron peak frequency $\nu_{\rm synch}$.}
    \end{subfigure}

    \caption{Comparison of the parameters from the two-zone model extended zone with the ratio between the median X-ray PD and the median radio PD (225 GHz).}
    \label{fig:twozone_extended_ratio_radio}
\end{figure}

\begin{figure}[h!]
    \centering

    \begin{subfigure}[t]{0.32\textwidth}
        \centering
        \includegraphics[width=\linewidth]{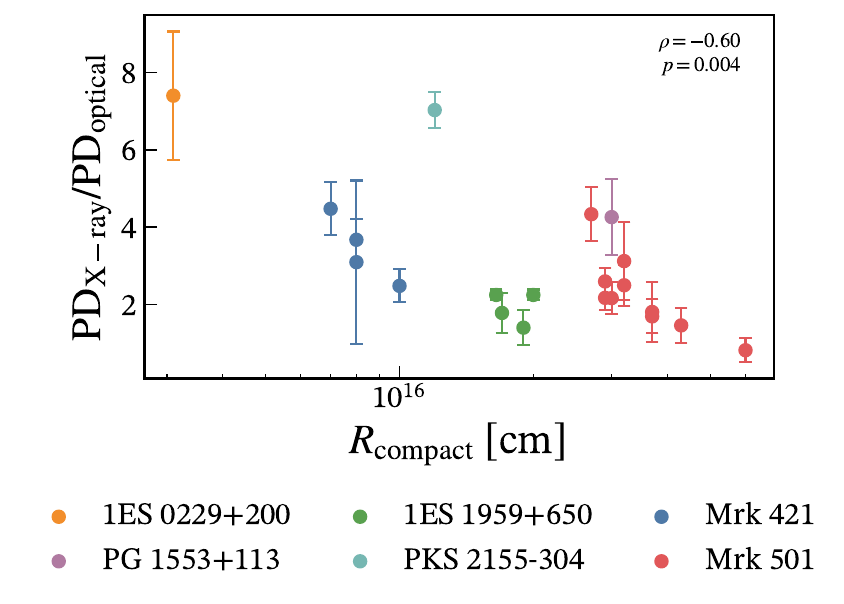}
        \caption{Radius $R$.}
    \end{subfigure}
    \begin{subfigure}[t]{0.32\textwidth}
        \centering
        \includegraphics[width=\linewidth]{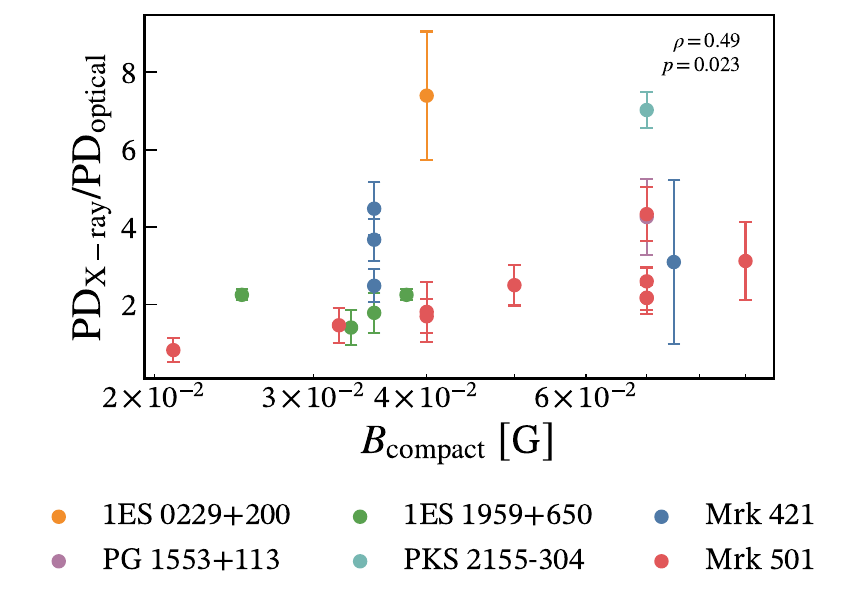}
        \caption{Magnetic field $B$.}
    \end{subfigure}

    \caption{Comparison of the parameters from the two-zone model compact zone with the ratio between the median X-ray PD and the median optical PD (R band). The synchrotron peak frequency plot is not reported, as it does not depend on the emitting zone considered.}
    \label{fig:twozone_compact_ratio_opt}
\end{figure}

\begin{figure}[h!]
    \centering

    \begin{subfigure}[t]{0.32\textwidth}
        \centering
        \includegraphics[width=\linewidth]{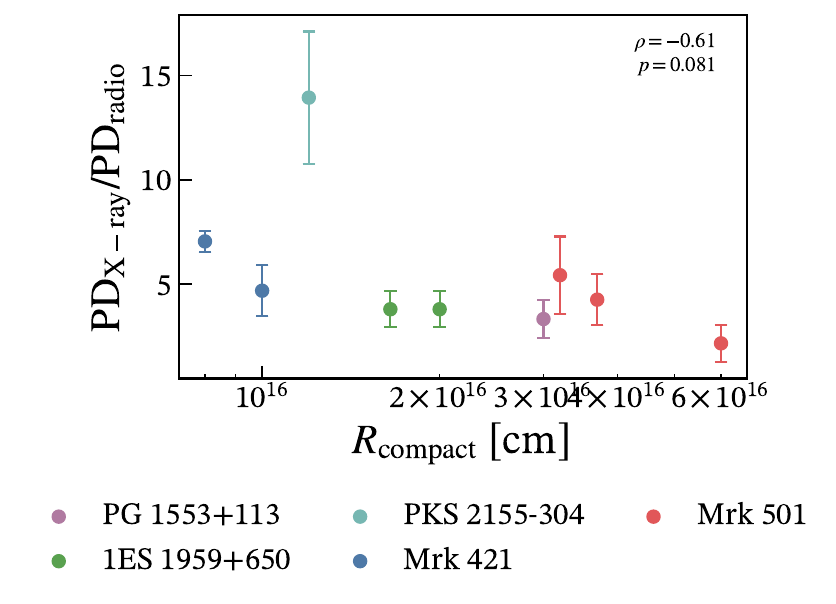}
        \caption{Radius $R$.}
    \end{subfigure}
    \begin{subfigure}[t]{0.32\textwidth}
        \centering
        \includegraphics[width=\linewidth]{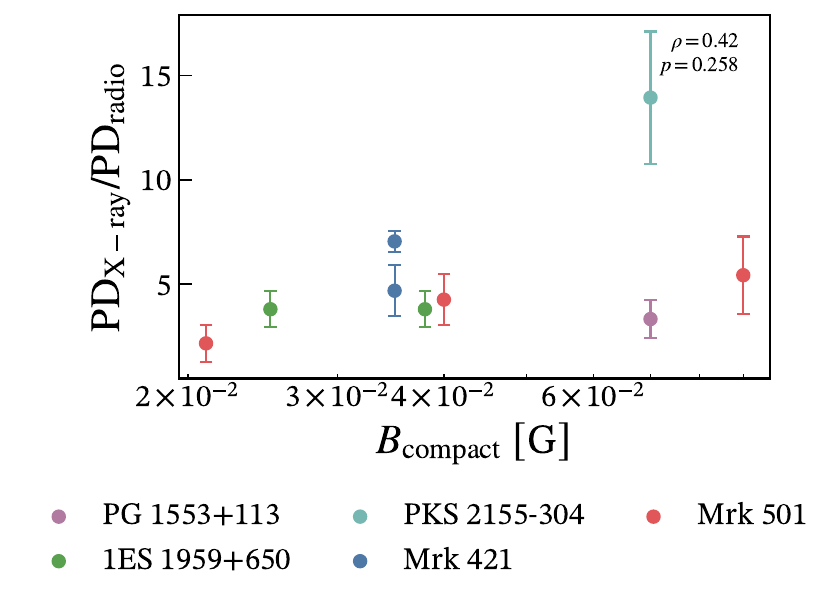}
        \caption{Magnetic field $B$.}
    \end{subfigure}

    \caption{Comparison of the parameters from the two-zone model compact zone with the ratio between the median X-ray PD and the median radio PD (225 GHz). The synchrotron peak frequency plot is not reported, as it does not depend on the emitting zone considered.}
    \label{fig:twozone_compact_ratio_radio}
\end{figure}

We report here the plots of the model parameters versus the ratio of the median PD.
Figures \ref{fig:onezone_ratio_opt} and \ref{fig:onezone_ratio_radio} report the one-zone model results.
Figures \ref{fig:twozone_extended_ratio_opt}, \ref{fig:twozone_extended_ratio_radio}, \ref{fig:twozone_compact_ratio_opt}, and \ref{fig:twozone_compact_ratio_radio} report the two-zone model results.
We report in each plot the values of $p$ and $\rho$ obtained through the Spearman statistical test.

\clearpage
\section{Two-zone model evaluation}
\label{appendix:twozone_fits}

We report here the best models obtained with the two-zone model for our observations (source and obsID reported in the caption).
The blue line represents the total SED, the green dots the data, the blue downward arrows the upper limits.
The contribution from the extended zone is represented by the dash-dotted purple line, while the one from the compact zone by the dashed black line.
The dotted pink line represents the interaction between the two zones as explained in Sect. \ref{sect_twozone}.

\begin{figure*}[h!]
\centering

\begin{minipage}{0.32\textwidth}
    \centering
    \includegraphics[width=\linewidth]{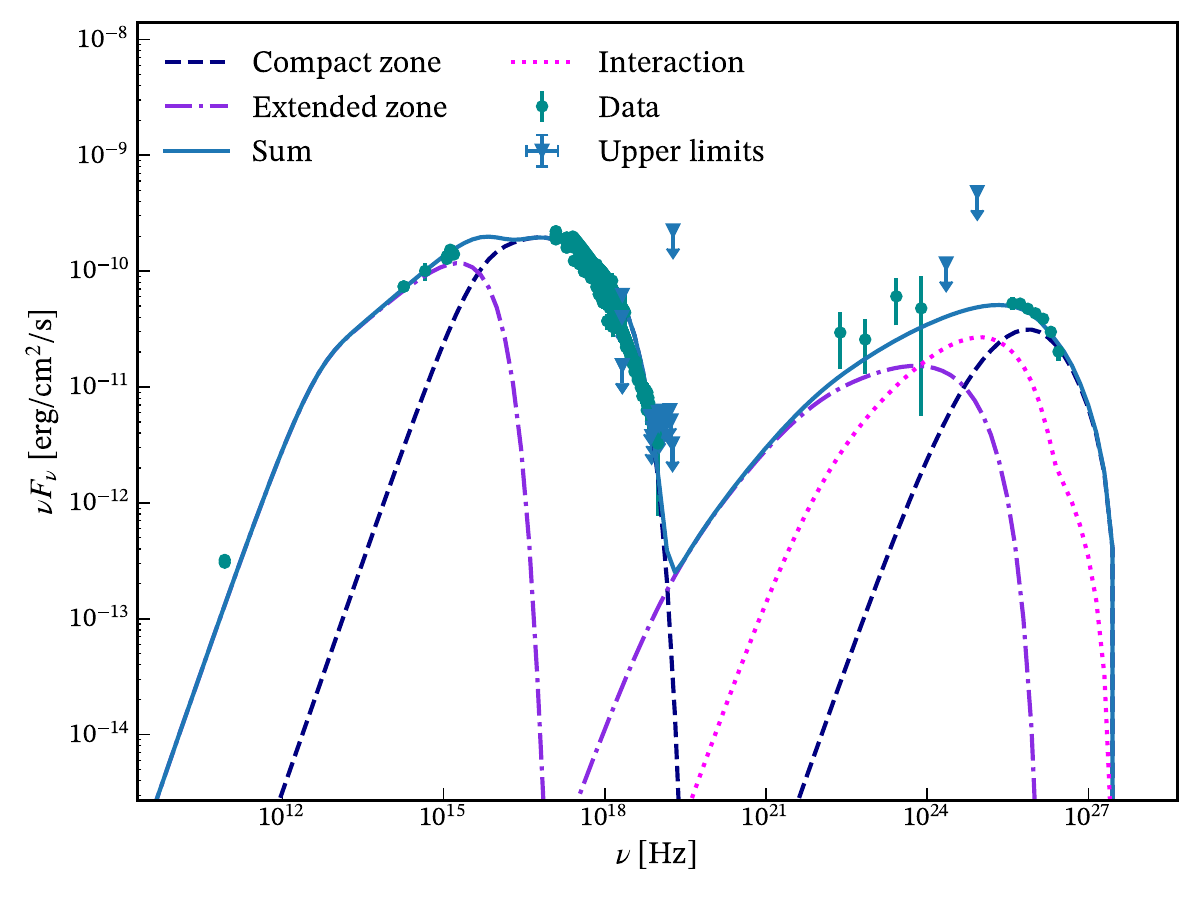}
    \small Mrk~421 -- 01003701
\end{minipage}
\hfill
\begin{minipage}{0.32\textwidth}
    \centering
    \includegraphics[width=\linewidth]{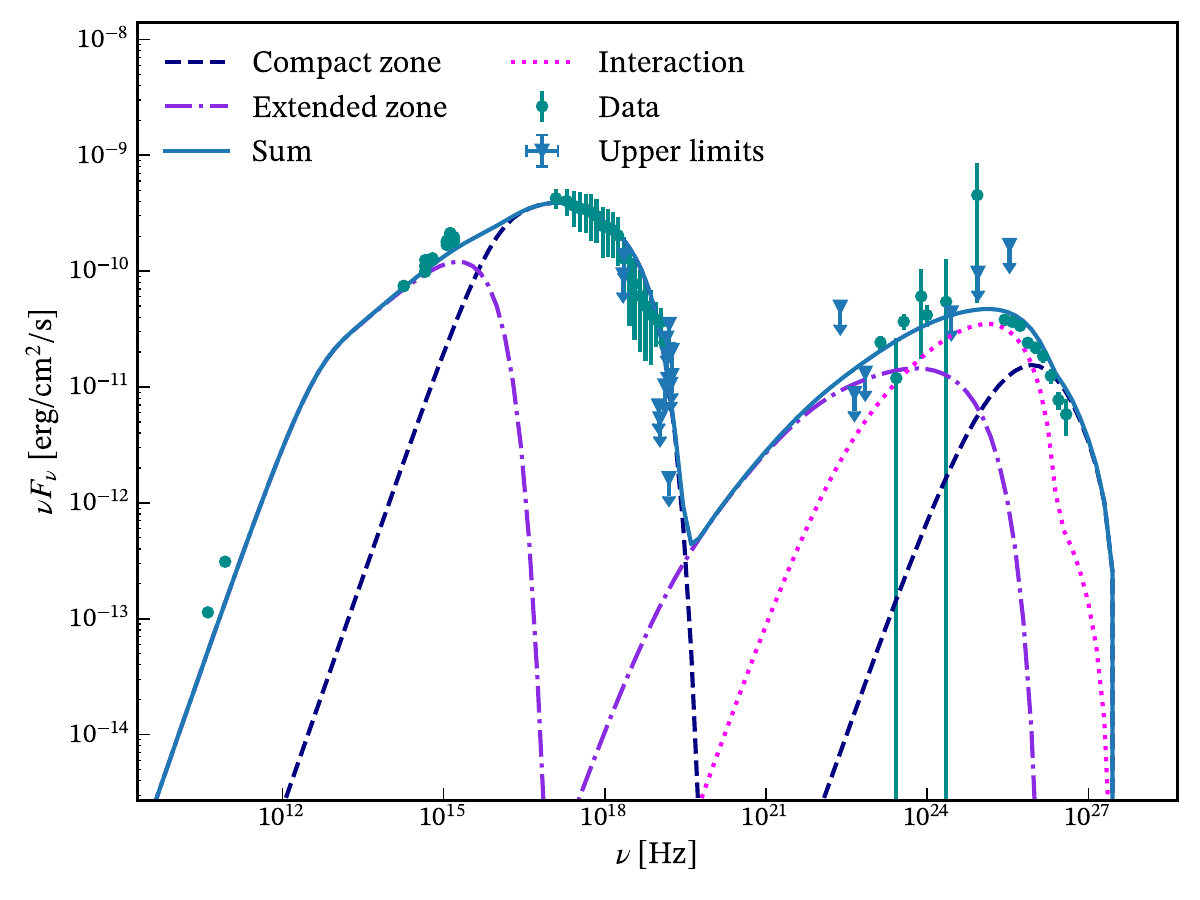}
    \small Mrk~421 -- 01003801
\end{minipage}
\hfill
\begin{minipage}{0.32\textwidth}
    \centering
    \includegraphics[width=\linewidth]{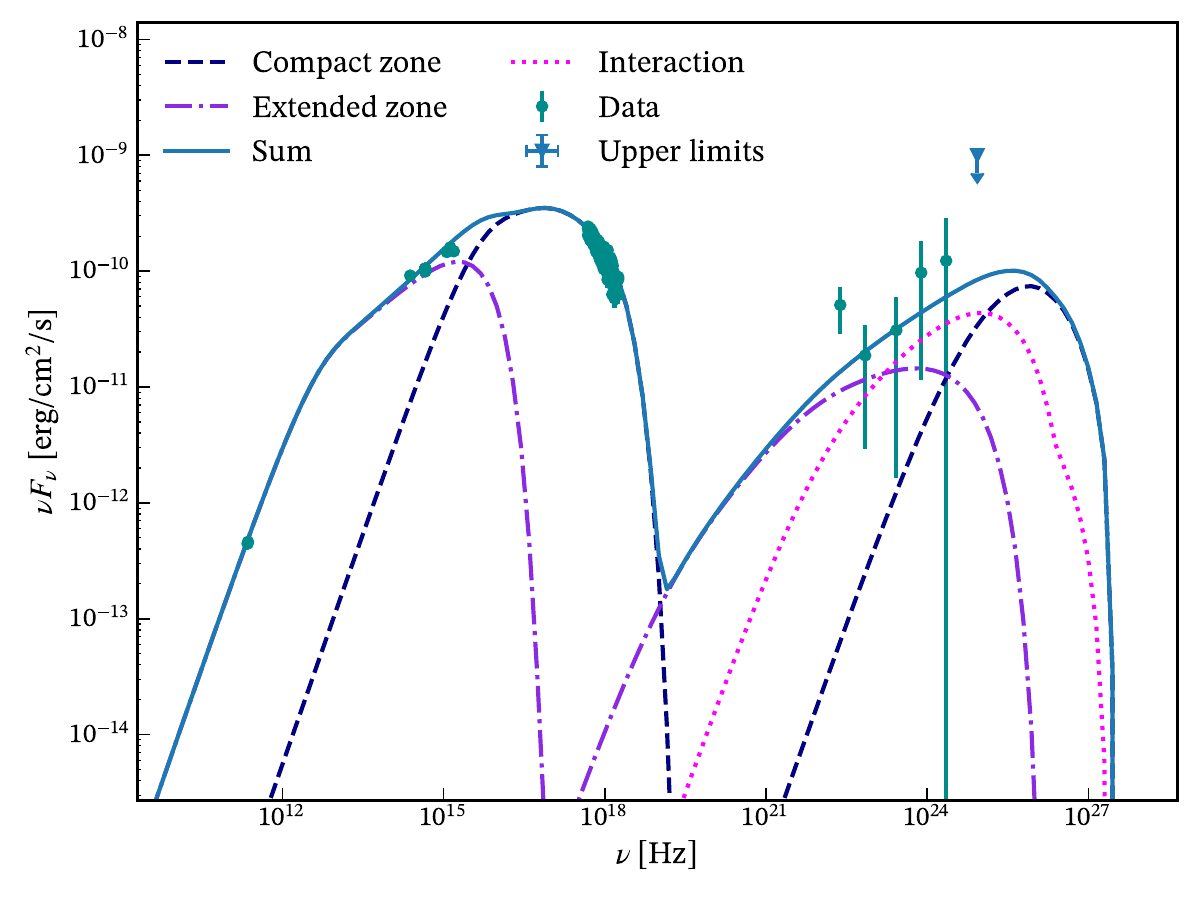}
    \small Mrk~421 -- 02004401
\end{minipage}

\vspace{0.4cm}

\begin{minipage}{0.32\textwidth}
    \centering
    \includegraphics[width=\linewidth]{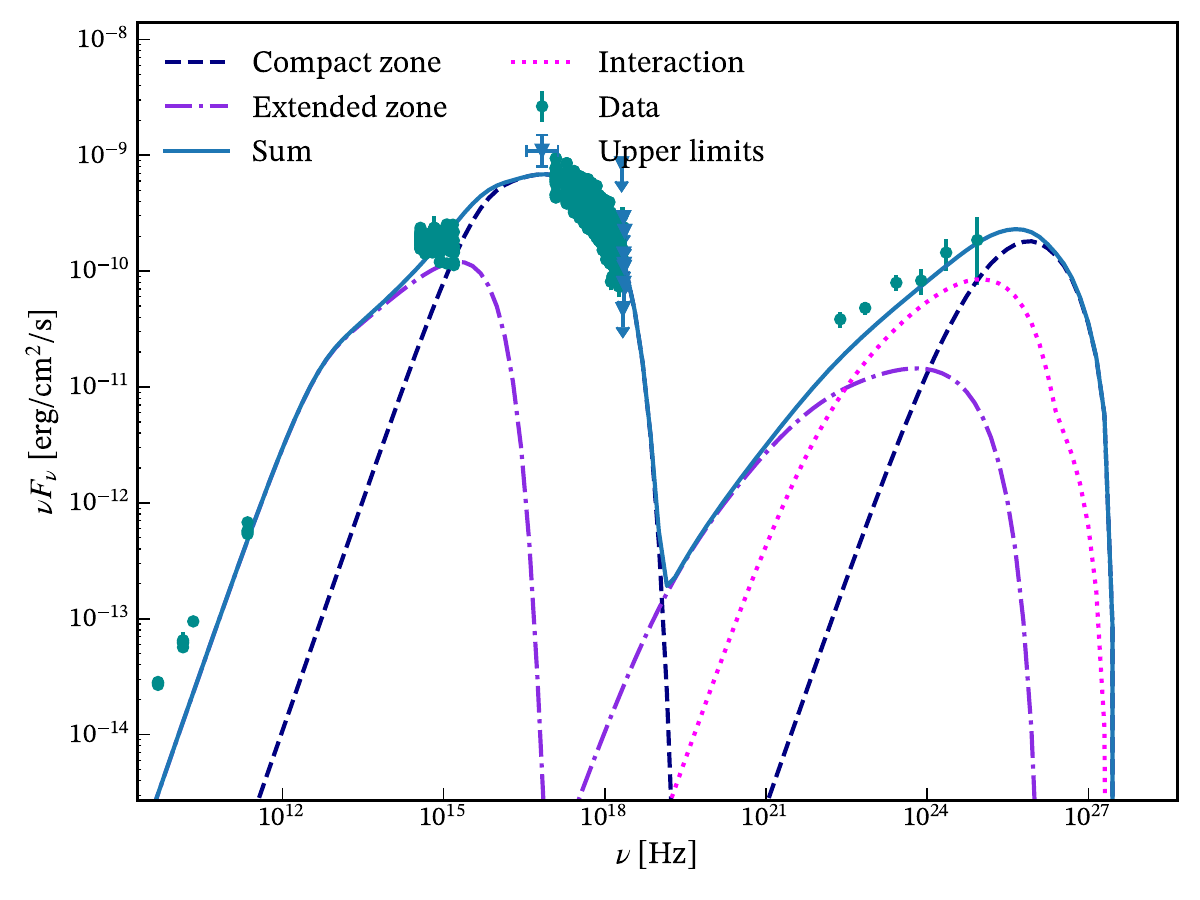}
    \small Mrk~421 -- 02008199
\end{minipage}
\hfill
\begin{minipage}{0.32\textwidth}
    \centering
    \includegraphics[width=\linewidth]{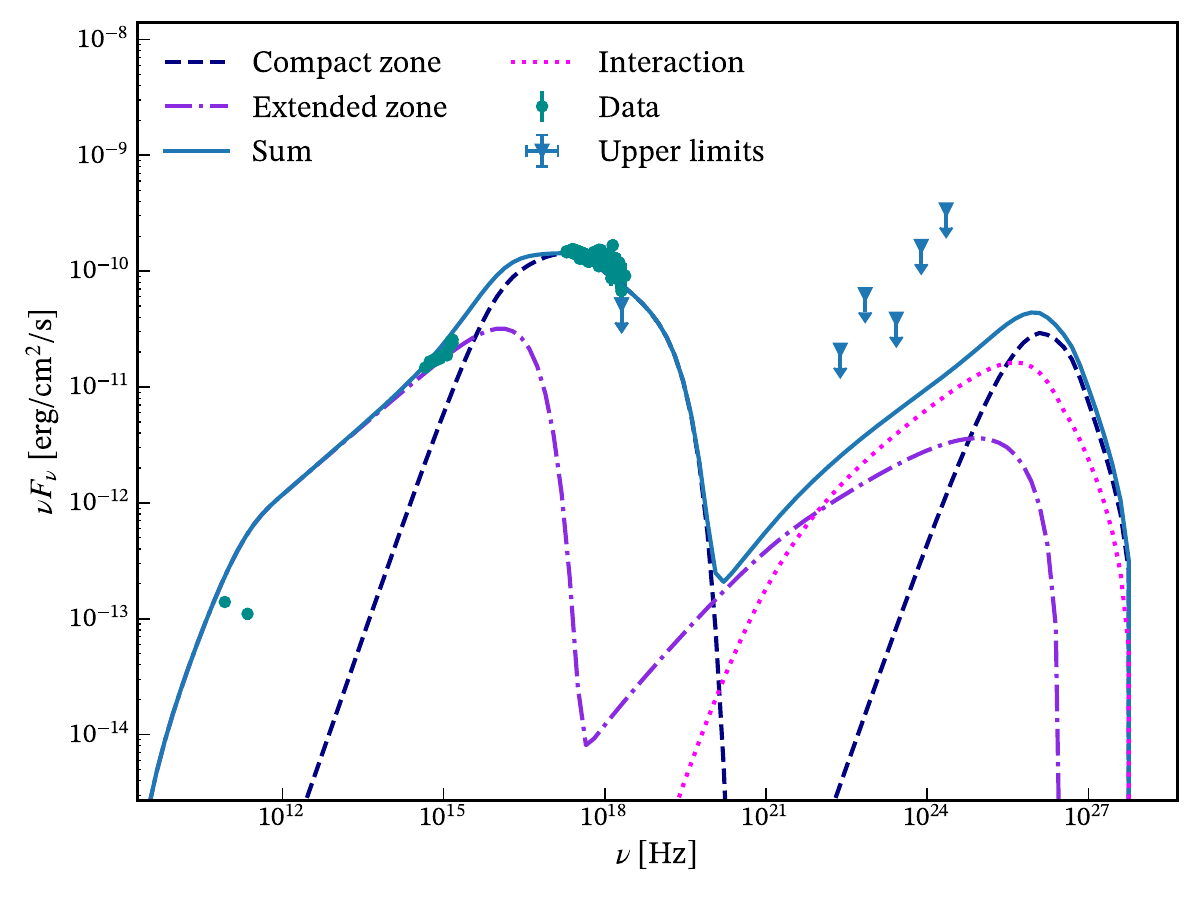}
    \small 1ES~1959$+$650 -- 01006201
\end{minipage}
\hfill
\begin{minipage}{0.32\textwidth}
    \centering
    \includegraphics[width=\linewidth]{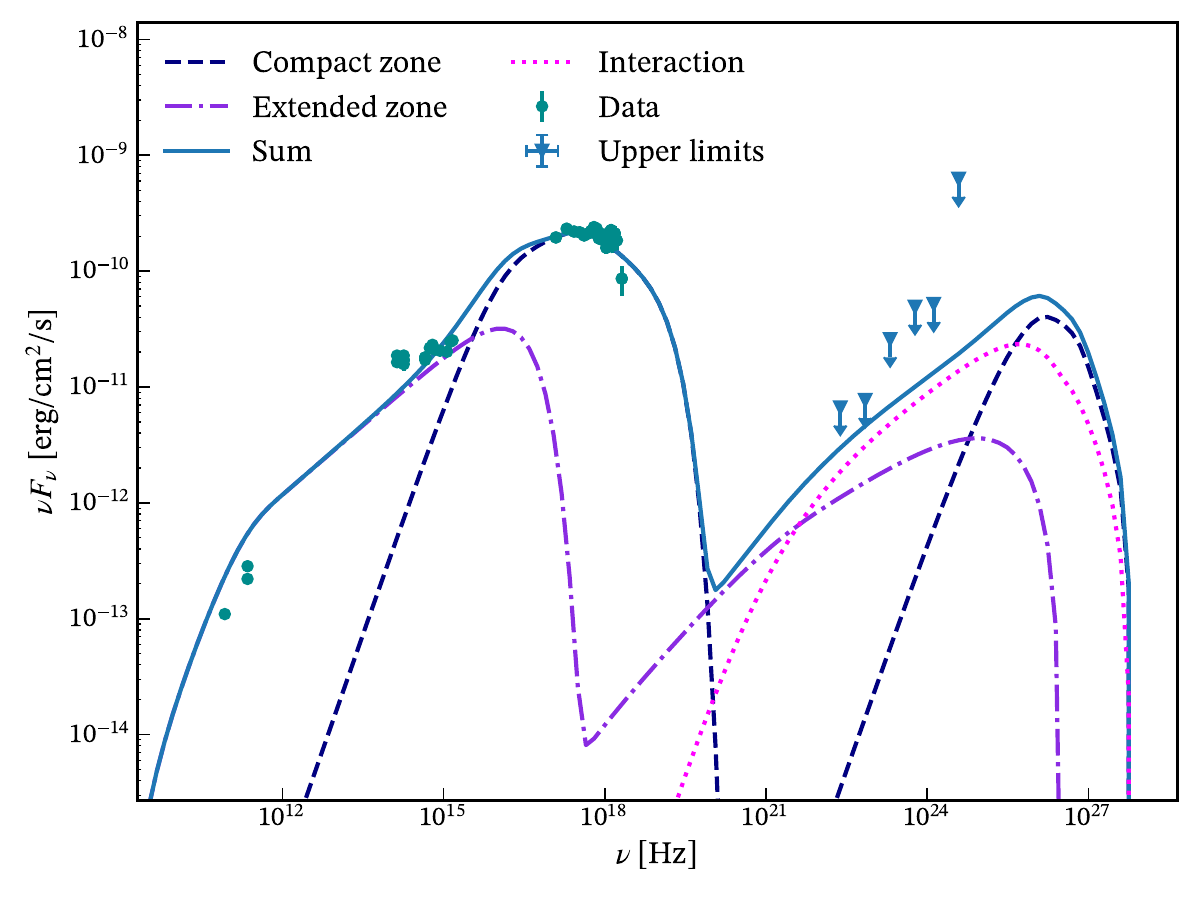}
    \small 1ES~1959$+$650 -- 01006001
\end{minipage}

\vspace{0.4cm}

\begin{minipage}{0.32\textwidth}
    \centering
    \includegraphics[width=\linewidth]{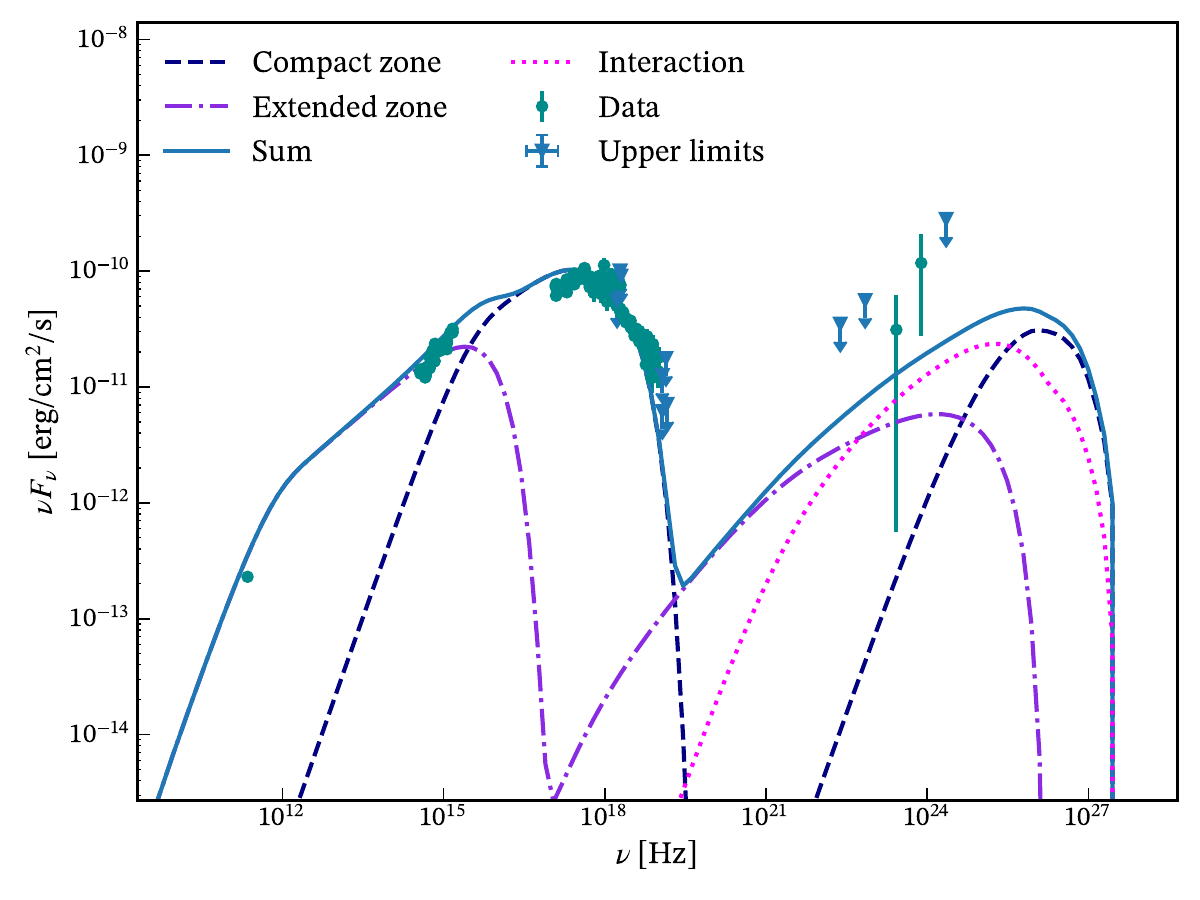}
    \small 1ES~1959$+$650 -- 02004801
\end{minipage}
\hfill
\begin{minipage}{0.32\textwidth}
    \centering
    \includegraphics[width=\linewidth]{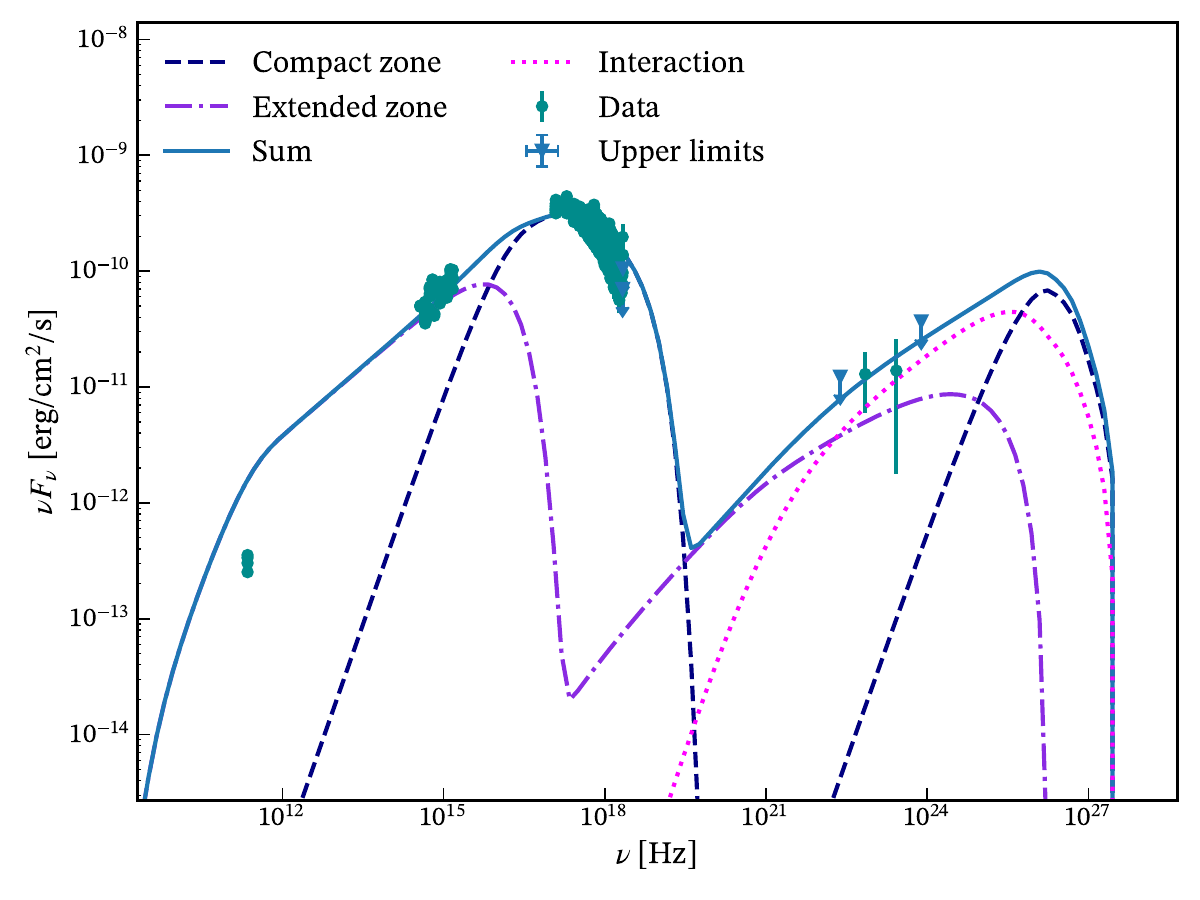}
    \small 1ES~1959$+$650 -- 02250801
\end{minipage}
\hfill
\begin{minipage}{0.32\textwidth}
    \centering
    \includegraphics[width=\linewidth]{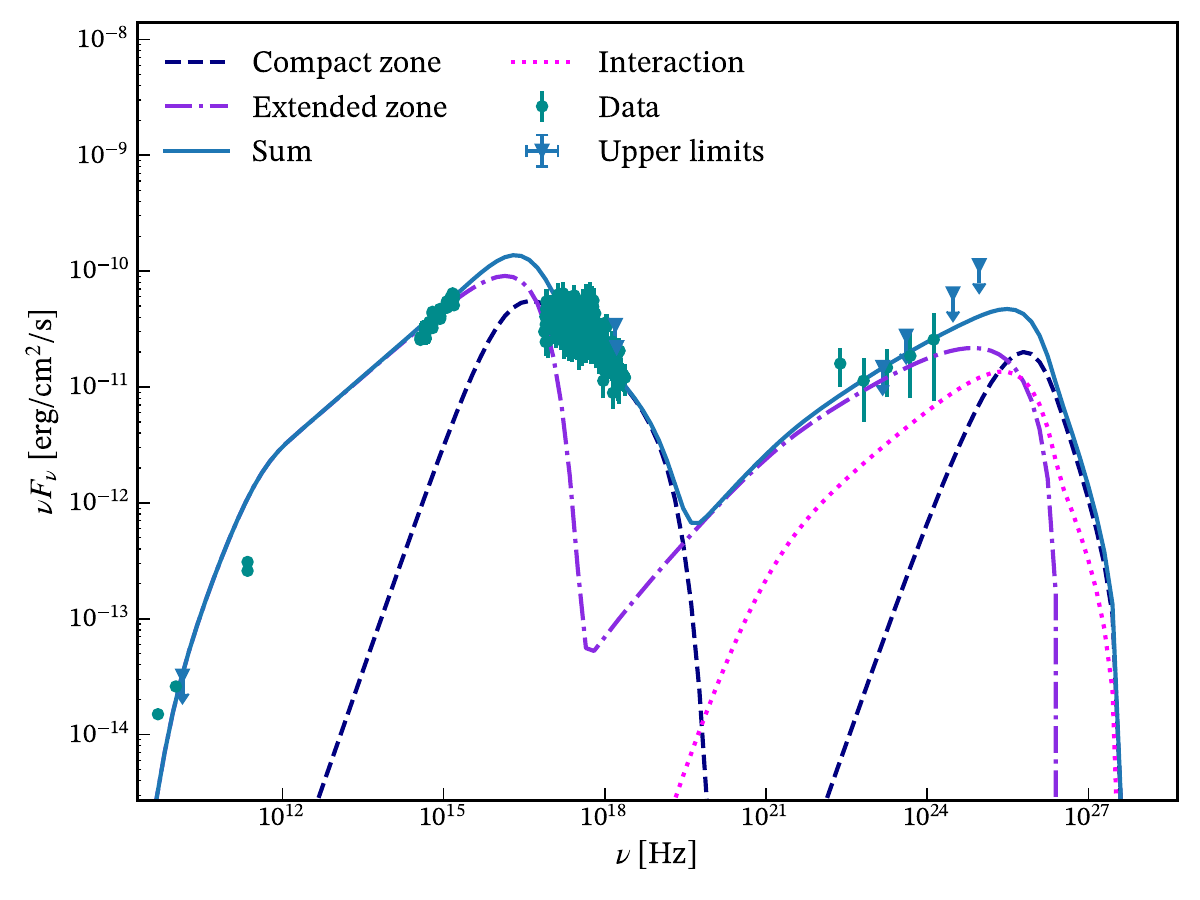}
    \small PKS~2155$-$304 -- 02005601
\end{minipage}

\vspace{0.4cm}

\begin{minipage}{0.32\textwidth}
    \centering
    \includegraphics[width=\linewidth]{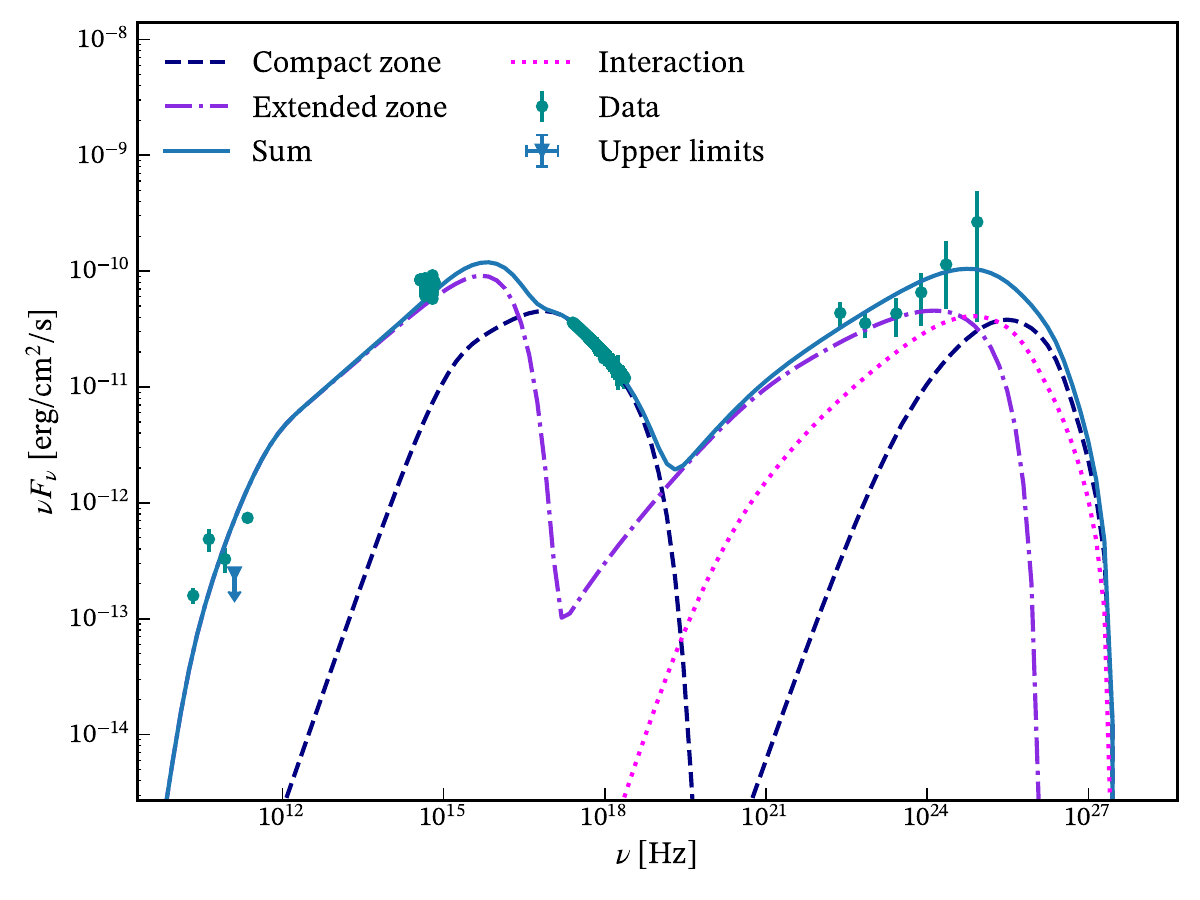}
    \small PG~1553$+$113 -- 02004999
\end{minipage}
\hfill
\begin{minipage}{0.32\textwidth}
    \centering
    \includegraphics[width=\linewidth]{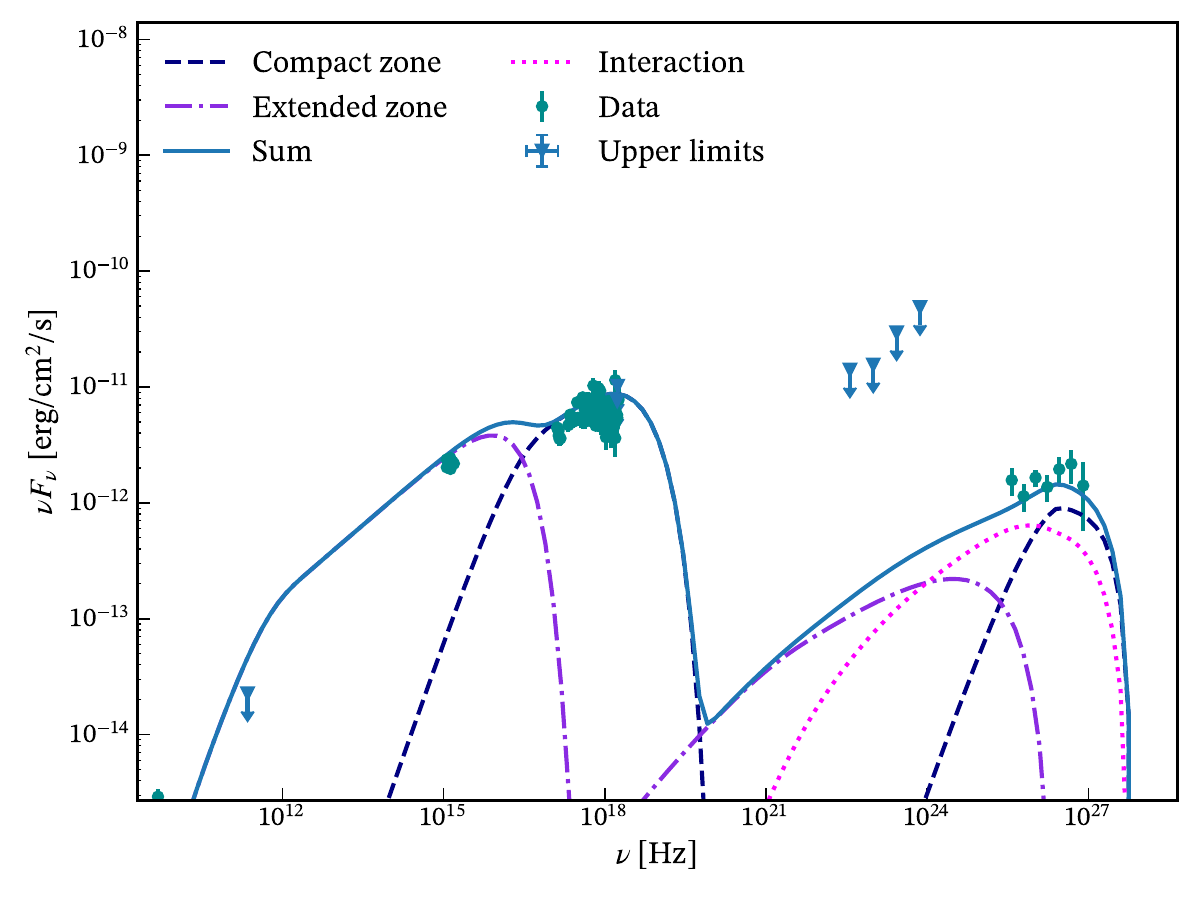}
    \small 1ES~0229$+$200 -- 01006499
\end{minipage}
\hfill
\begin{minipage}{0.32\textwidth}
    \centering
    \includegraphics[width=\linewidth]{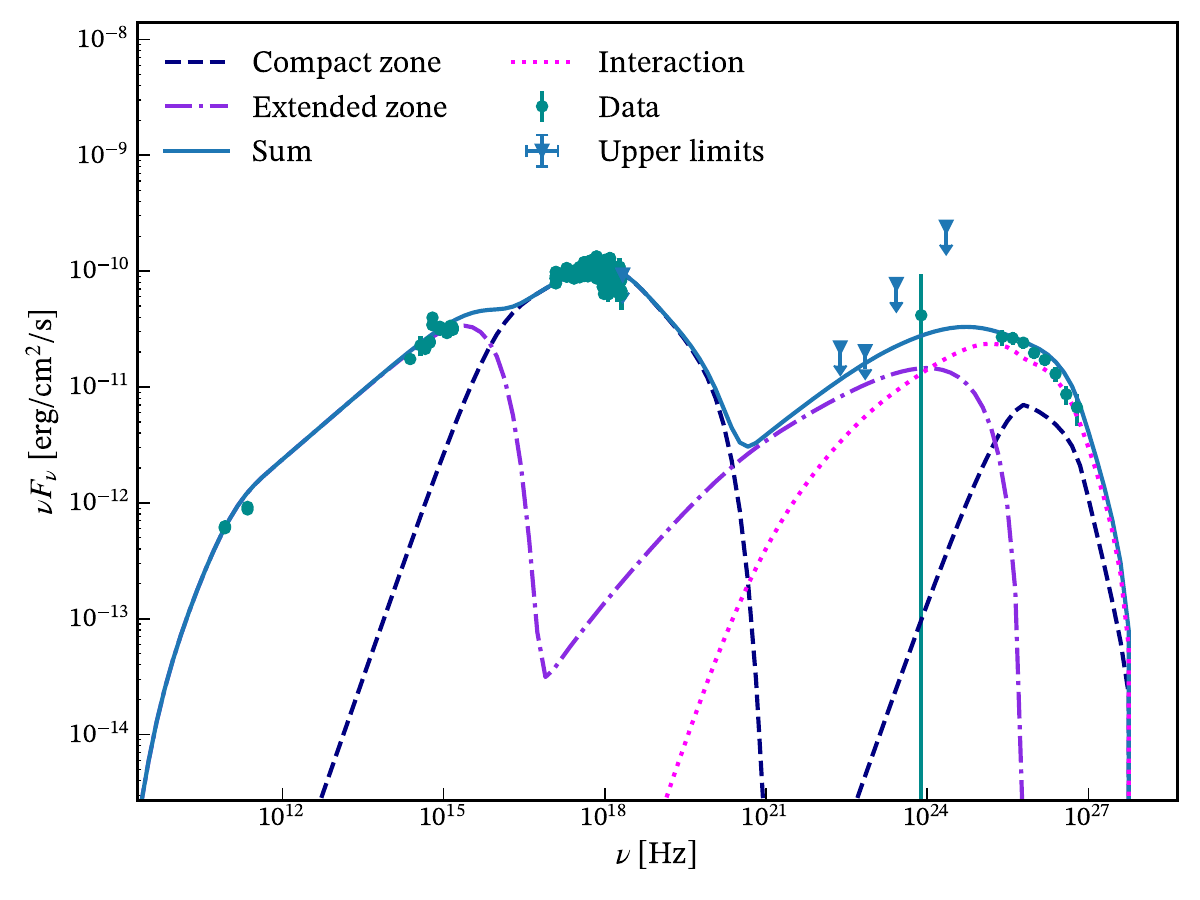}
    \small Mrk~501 -- 01004501
\end{minipage}

\end{figure*}

\begin{figure*}[h!]
\centering

\begin{minipage}{0.32\textwidth}
    \centering
    \includegraphics[width=\linewidth]{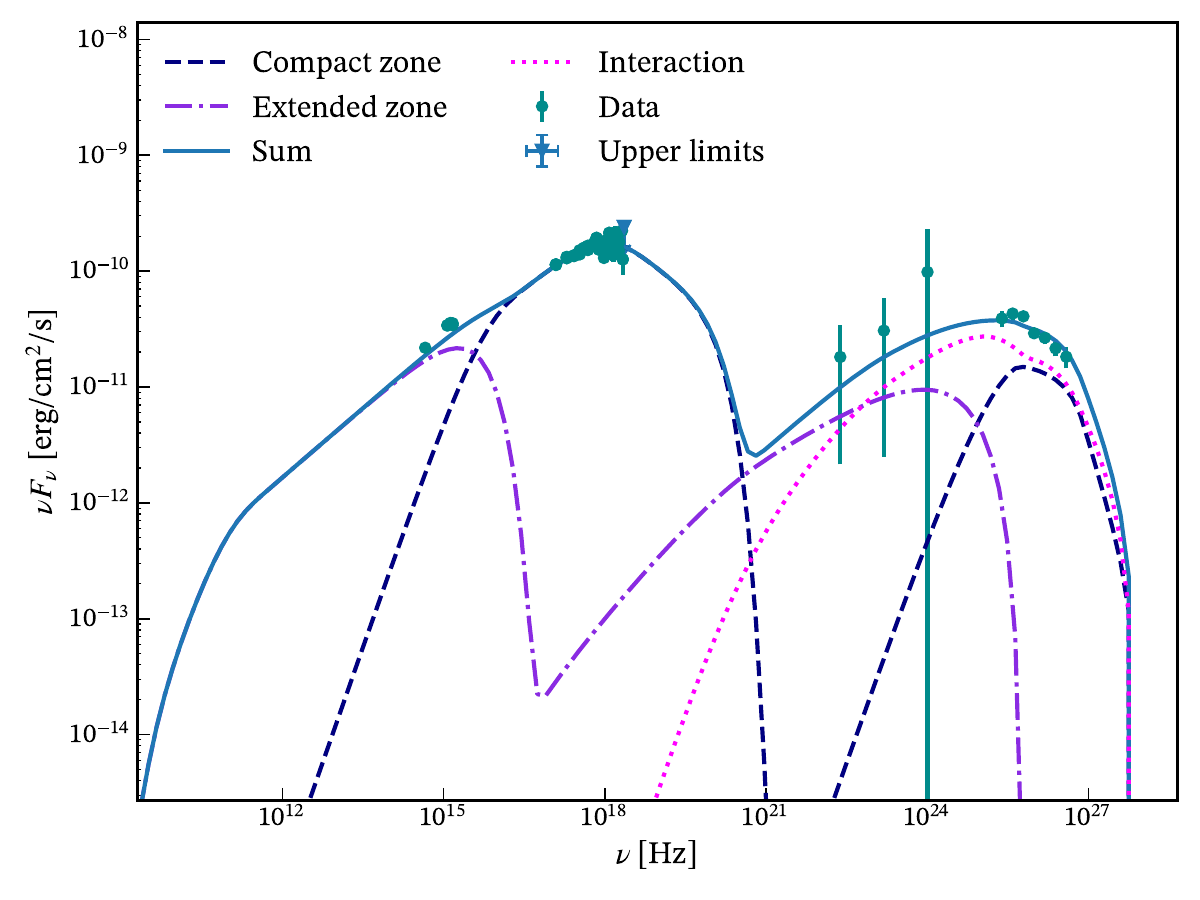}
    \small Mrk~501 -- 01004601
\end{minipage}
\hfill
\begin{minipage}{0.32\textwidth}
    \centering
    \includegraphics[width=\linewidth]{Images/501_3.pdf}
    \small Mrk~501 -- 01004701
\end{minipage}
\hfill
\begin{minipage}{0.32\textwidth}
    \centering
    \includegraphics[width=\linewidth]{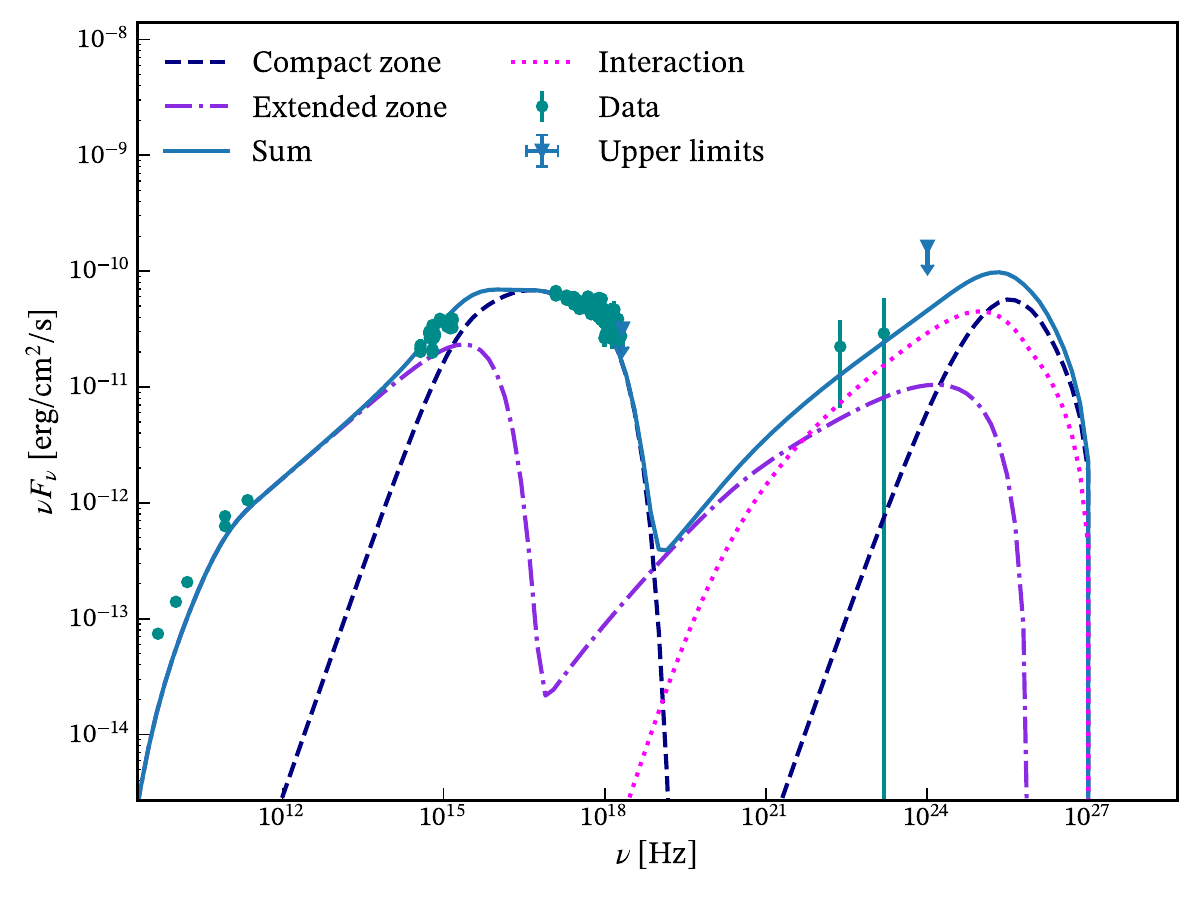}
    \small Mrk~501 -- 02004601
\end{minipage}

\vspace{0.4cm}

\begin{minipage}{0.32\textwidth}
    \centering
    \includegraphics[width=\linewidth]{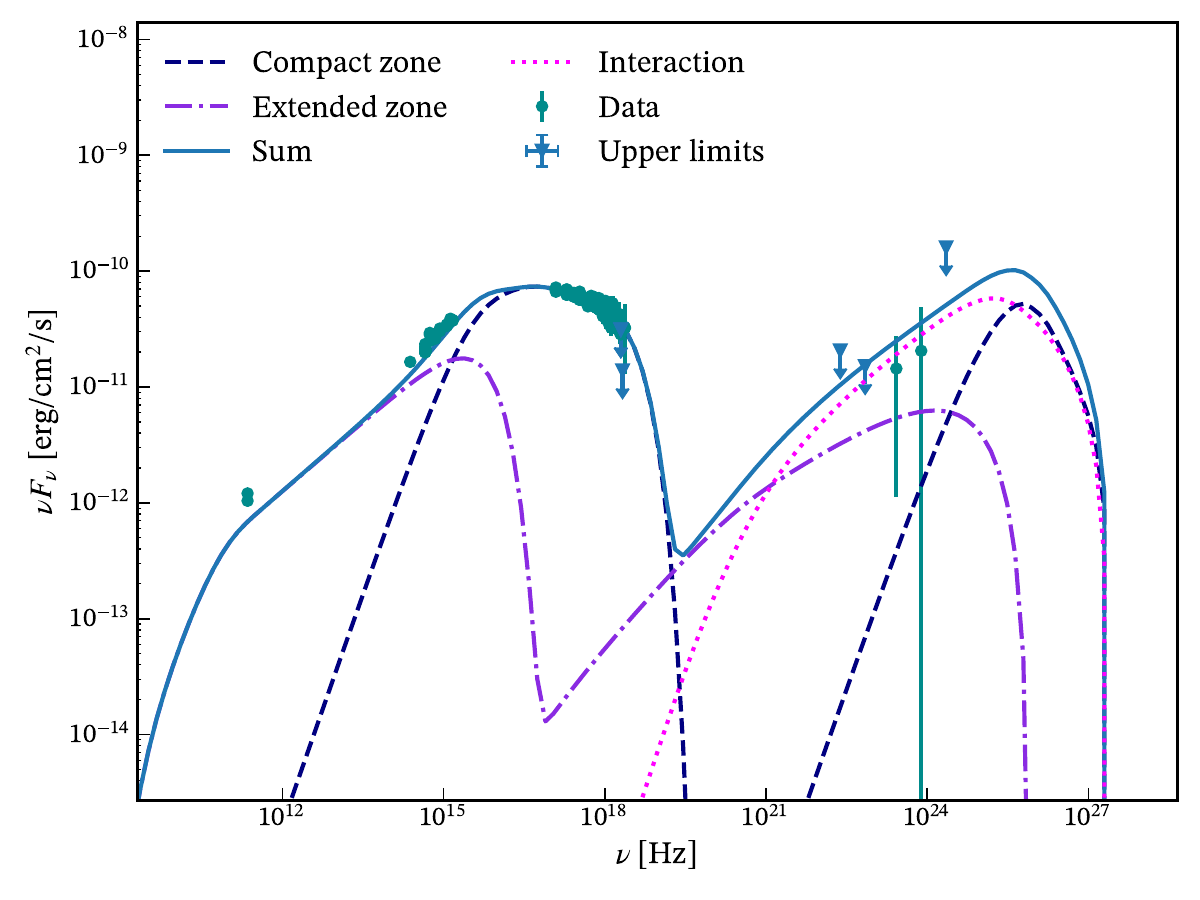}
    \small Mrk~501 -- 02004501
\end{minipage}
\hfill
\begin{minipage}{0.32\textwidth}
    \centering
    \includegraphics[width=\linewidth]{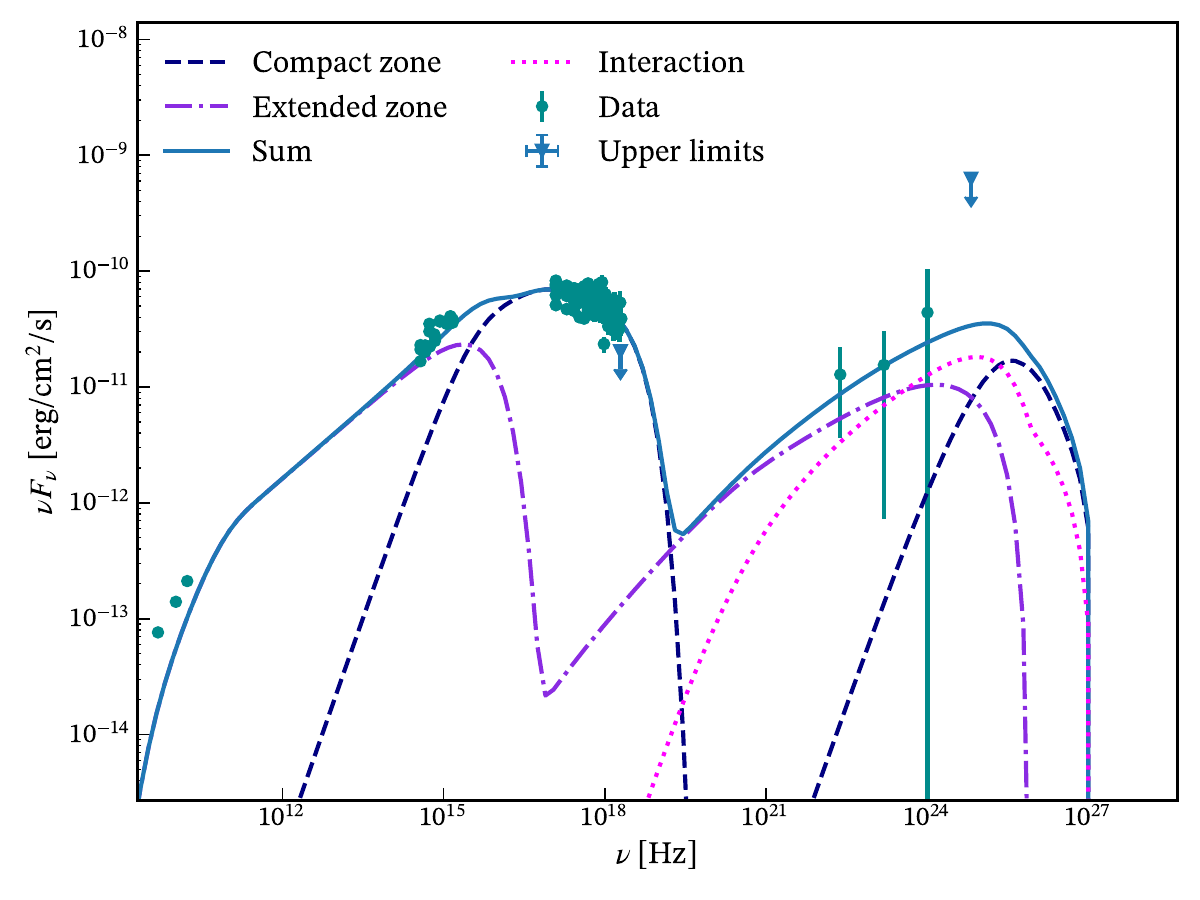}
    \small Mrk~501 -- 02004701
\end{minipage}
\hfill
\begin{minipage}{0.32\textwidth}
    \centering
    \includegraphics[width=\linewidth]{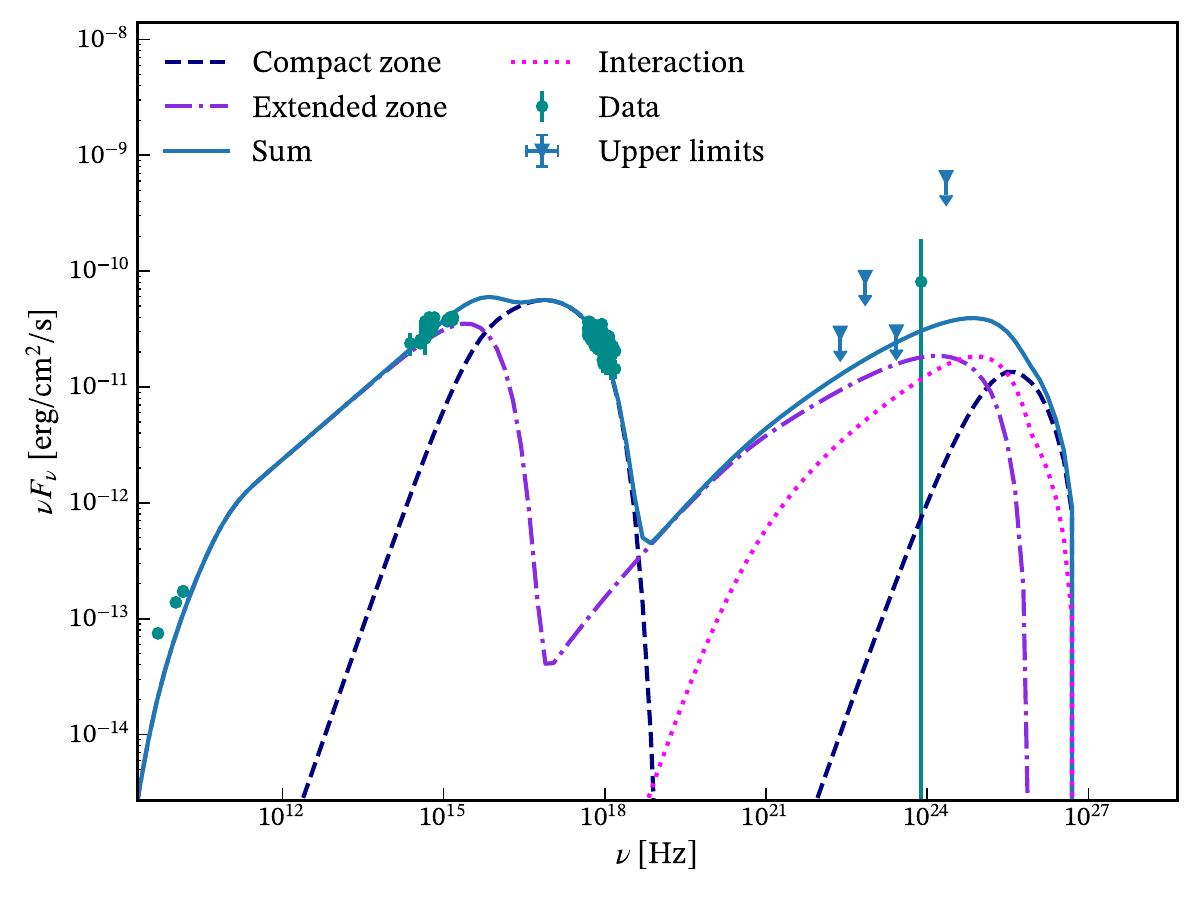}
    \small Mrk~501 -- 03008801
\end{minipage}

\vspace{0.4cm}

\begin{minipage}{0.32\textwidth}
    \centering
    \includegraphics[width=\linewidth]{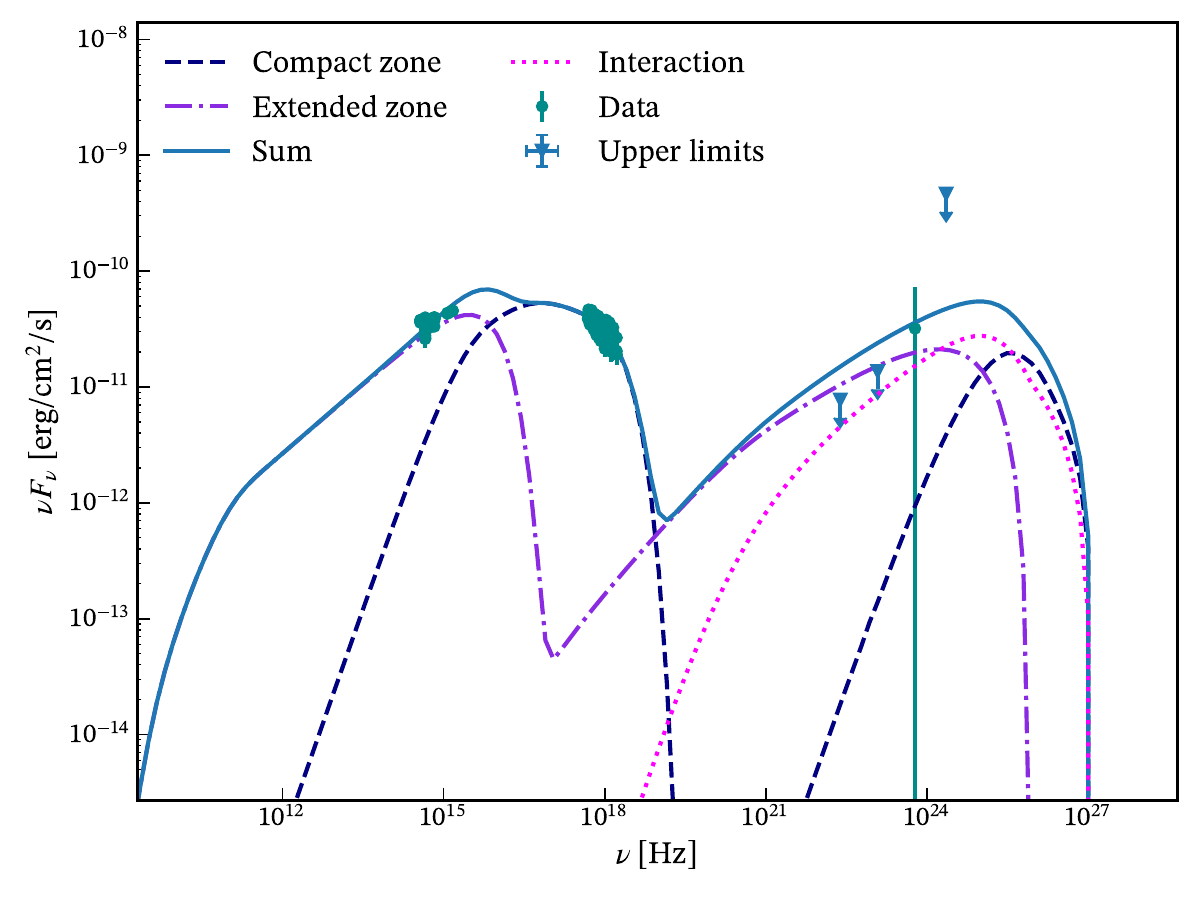}
    \small Mrk~501 -- 03008901
\end{minipage}
\hfill
\begin{minipage}{0.32\textwidth}
    \centering
    \includegraphics[width=\linewidth]{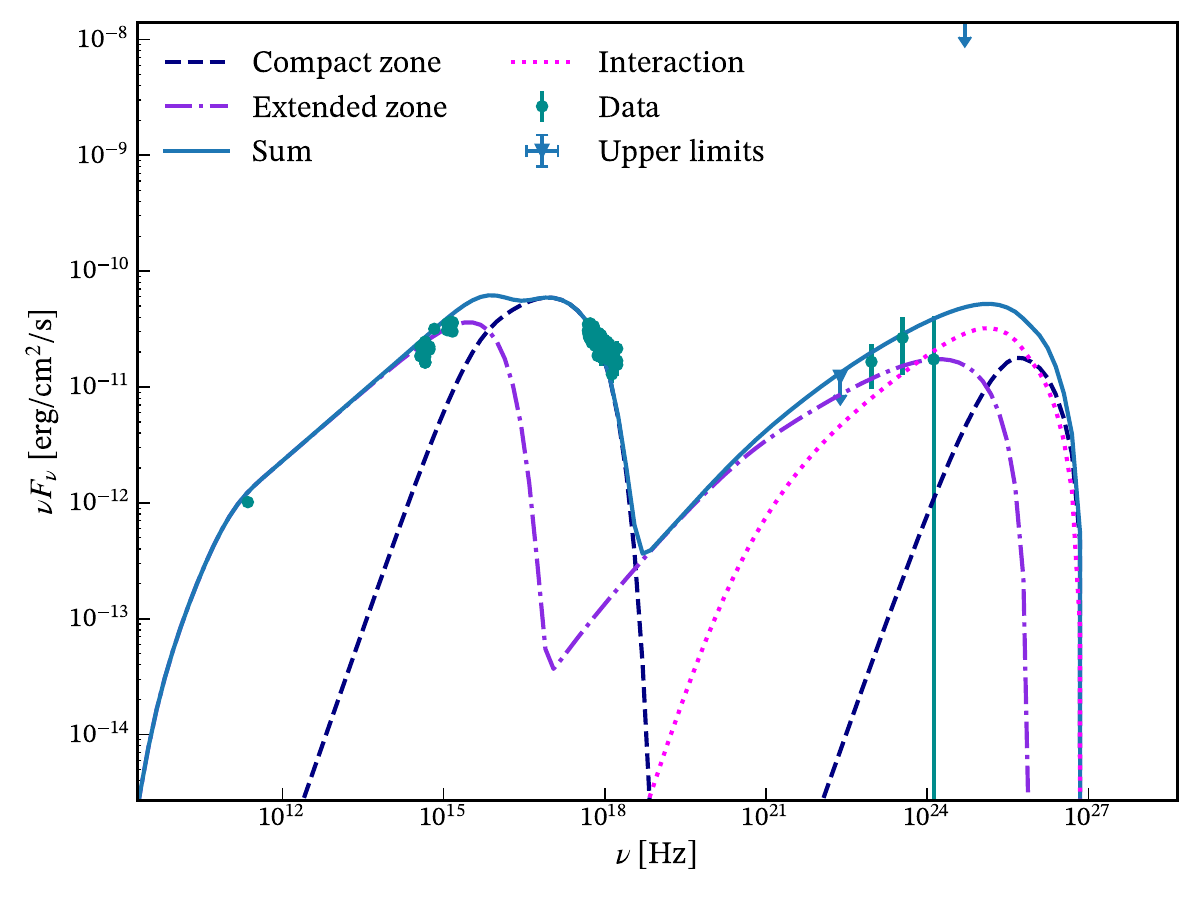}
    \small Mrk~501 -- 03009001
\end{minipage}
\hfill
\begin{minipage}{0.32\textwidth}
    \centering
    \includegraphics[width=\linewidth]{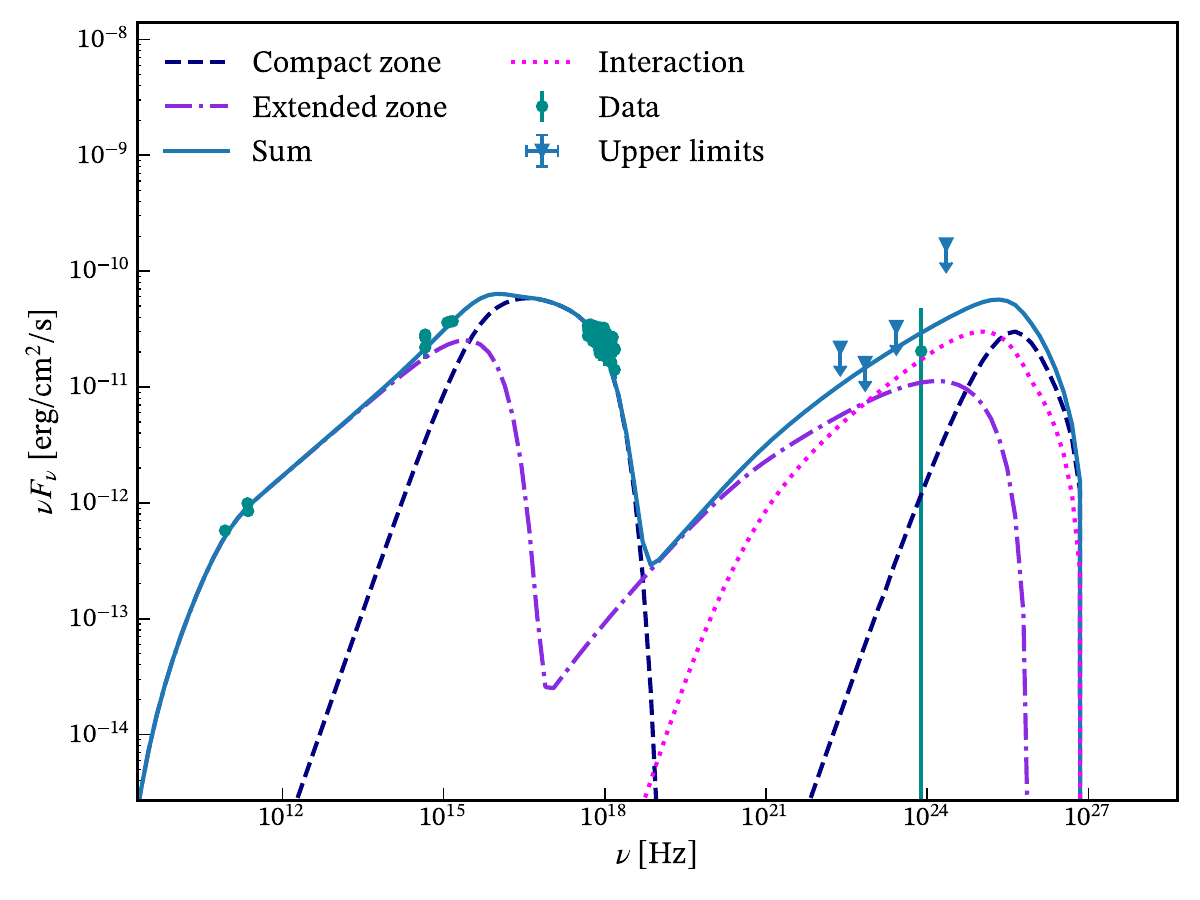}
    \small Mrk~501 -- 03009101
\end{minipage}
\caption{Best models obtained with the two-zone model for our observations. Source and obsID reported in the caption.
The blue line represents the total SED, the green dots the data, the blue downward arrows the upper limits.
The contribution from the extended zone is represented by the dash-dotted purple line, while the one from the compact zone by the dashed black line.
The dotted pink line represents the interaction between the two zones.}
\end{figure*}

\end{appendix}

\end{document}